\documentclass[twocolumn,prx,10pt,superscriptaddress, aps, amsmath, amssymb, amsfonts, floatfix]{revtex4-2}
\usepackage[T1]{fontenc}
\usepackage[utf8]{inputenc}

\usepackage{amssymb,amsmath}
\usepackage{siunitx}
\usepackage{graphicx}
\usepackage{natbib}
\usepackage{braket}

\usepackage{subfiles}

\usepackage[unicode,raiselinks=false, hyperfootnotes=false,bookmarksopen=true,hyperfigures=false,bookmarksopenlevel=2]{hyperref} 
\hypersetup{colorlinks=true,
    linkcolor=black,
    filecolor=black,      
    urlcolor=blue,
    citecolor=blue
}

\begin{document}
\title{Tailoring the band structure of twisted double bilayer graphene with pressure}
\author{Bálint Szentpéteri}
\affiliation{Department of Physics, Budapest University of Technology and Economics and Nanoelectronics Momentum Research Group of the Hungarian Academy of Sciences, Budafoki ut 8, 1111 Budapest, Hungary}
\author{Peter Rickhaus}
\affiliation{Solid State Physics Laboratory, ETH Zürich, CH-8093 Zürich, Switzerland}
\author{Folkert K. de Vries}
\affiliation{Solid State Physics Laboratory, ETH Zürich, CH-8093 Zürich, Switzerland}
\author{Albin Márffy}
\affiliation{Department of Physics, Budapest University of Technology and Economics and Correlated van der Waals Structures Momentum Research Group of the Hungarian Academy of Sciences, 1111 Budapest, Hungary}
\author{Bálint Fülöp}
\affiliation{Department of Physics, Budapest University of Technology and Economics and Nanoelectronics Momentum Research Group of the Hungarian Academy of Sciences, Budafoki ut 8, 1111 Budapest, Hungary}
\author{Endre Tóvári}
\affiliation{Department of Physics, Budapest University of Technology and Economics and Nanoelectronics Momentum Research Group of the Hungarian Academy of Sciences, Budafoki ut 8, 1111 Budapest, Hungary}
\author{Kenji Watanabe}
\affiliation{Research Center for Functional Materials, National Institute for Materials Science, 1-1 Namiki, Tsukuba 305-0044, Japan}
\author{Takashi Taniguchi}
\affiliation{International Center for Materials Nanoarchitectonics, National Institute for Materials Science, 1-1 Namiki, Tsukuba 305-0044, Japan}
\author{Andor Kormányos}
\affiliation{Department of Physics of Complex Systems, Eötvös Loránd University, Budapest, Hungary}
\author{Szabolcs Csonka}
\email{csonka.szabolcs@ttk.bme.hu}
\affiliation{Department of Physics, Budapest University of Technology and Economics and Nanoelectronics Momentum Research Group of the Hungarian Academy of Sciences, Budafoki ut 8, 1111 Budapest, Hungary}
\author{Péter Makk}
\email{makk.peter@ttk.bme.hu}
\affiliation{Department of Physics, Budapest University of Technology and Economics and Correlated van der Waals Structures Momentum Research Group of the Hungarian Academy of Sciences, 1111 Budapest, Hungary}

\date{October 11, 2021}

\begin{abstract}
Twisted two-dimensional structures open new possibilities in band structure engineering. At magic twist angles, flat bands emerge, which give a new drive to the field of strongly correlated physics. In twisted double bilayer graphene dual gating allows changing the Fermi level and hence the electron density and also allows tuning the interlayer potential, giving further control over band gaps. Here, we demonstrate that by applying hydrostatic pressure, an additional control of the band structure becomes possible due to the change of tunnel couplings between the layers. We find that the flat bands and the gaps separating them can be drastically changed by pressures up to 2\,GPa, in good agreement with our theoretical simulations. Furthermore, our measurements suggest that in finite magnetic field due to pressure a topologically non-trivial band gap opens at the charge neutrality point at zero displacement field. 
\end{abstract}
\maketitle

\section{Introduction}
Twisted van der Waals heterostructures recently opened a new platform to explore correlated electronics phases. These structures consist of two or more layers of 2D materials, e.g. graphene, placed on top of each other, with a well-defined rotation angle between their crystallographic axes. The rotation of the two layers leads to the formation of a moiré superlattice. For small angles, the hybridization between the layers becomes significant, resulting in a reconstruction of the band structure and for well defined special, so called ''magic'' angles, the bands flatten out\cite{Bistritzer2011, Jung2014a}. Due to their narrow bandwidth, the electron-electron interaction can become comparable to the kinetic energy and novel, correlated phases form\cite{Cao2018, Cao2018a, Yankowitz2019, Saito2020, Sharpe2019, Lu2019, Serlin2019, Cao2020a, Polshyn2019, Stepanov2020, Zhang2019}. The ability to control the size of the moiré unit cell size with the twist angle and implicitly the bandwidth of the low energy bands led to the discovery of various correlated phases\,\cite{Cao2018,Sharpe2019,Cao2020a,Polshyn2019,Lu2019,Serlin2019} such as correlated insulator states\,\cite{Cao2018,Cao2018a,Yankowitz2019,Saito2020,Lu2019}, ferromagnetic phase with a signature of the quantum anomalous Hall effect\,\cite{Sharpe2019,  Zhang2019, Lu2019, Serlin2019} and unconventional superconducting phases resembling high-temperature superconductors\,\cite{Cao2018a, Yankowitz2019, Stepanov2020, Saito2020} in twisted bilayer graphene (TBG). Moreover, the correlation effects are often accompanied with non-trivial topology of the bands\cite{Liu2019a, Song2019, Zhang2019, Nuckolls2020, Wu2021,Saito2021, Koshino2019,Chebrolu2019, Lee2019,Crosse2020,Wang2021, Burg2020, Ma2020}.

\begin{figure*}[!ht]
\begin{center}
\includegraphics[width=1\linewidth]{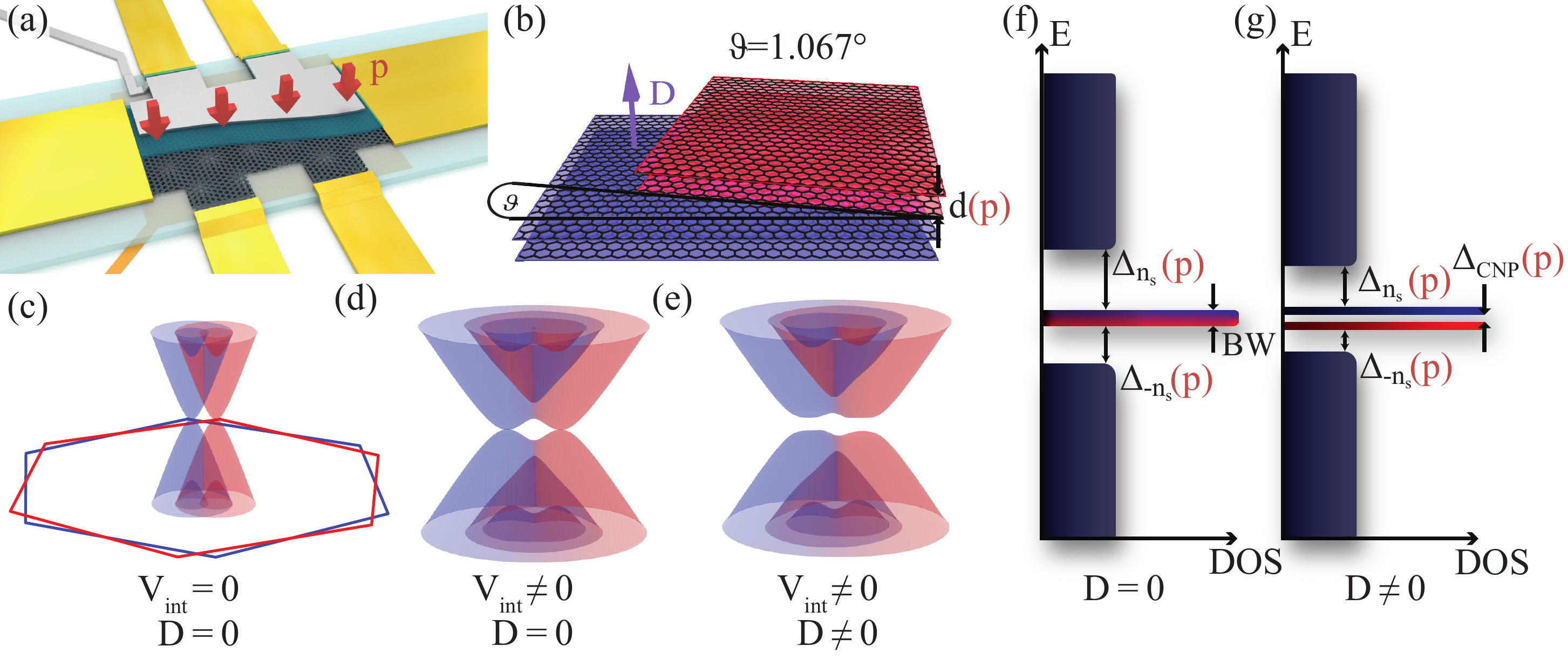}
\caption{Properties of the twisted double bilayer graphene device. (a) Schematic of the TDBG (black lattice) with a bottom graphite gate (orange) isolated by a hBN layer (light blue) and a top metallic gate (gray) isolated by a hBN layer and an AlO$_x$ layer (blue). The sample is contacted by edge contacts (yellow). The red arrows represent the pressure $p$, which modifies the distance between the layers. (b) Illustration of the twisted double bilayer structure. The purple arrow shows the direction of the transverse displacement field ($D$), $\vartheta$ is the twist angle and $d$ is the distance between the graphene layers which is tuned with the pressure.  (c-e) Illustration of the band structure of TDBG in a corner of the Brillouin zone: (c) No coupling between the top and bottom BLGs which are depicted with red and blue colors respectively, (d) a small coupling is introduced between the two BLGs which hybridizes the bands and leads to avoided crossings, (e) an external electric field can open a band gap at the Dirac points. (f) and (g) Schematic pictures of the DOS in the surrounding of the flat band in magic-angle TDBG without and with an electric field, respectively. The external electric field splits the degenerate flat bands and open a gap $\Delta_\mathrm{CNP}$ at the charge neutrality point. $\Delta_\mathrm{n_{s}}$ and  $\Delta_\mathrm{-n_{s}}$ are the band gaps separating the flat bands from the conduction and valence bands. These gaps are tunable with external pressure which is noted explicitly by the brackets.}
\label{fig1}
\end{center}
\end{figure*}

Twisted double bilayer graphene (TDBG), which consists of two Bernal stacked bilayer graphene (BLG) crystals  with a rotation angle of $\vartheta$ between them (depicted in Fig.~\ref{fig1}b.), offers a versatile platform where an external electric field can be used to control the band structure via tuning the interlayer potential. Recent experimental\,\cite{ Burg2019, Cao2020, He2020, Rickhaus2020, Burg2020, Liu2020, Shen2020} and theoretical\,\cite{Chebrolu2019,Koshino2019,Lee2019,Crosse2020,Wang2021,Lin2020, Haddadi2020} studies showed the presence of correlated insulator states and topologically non-trivial phases in TDBG. 

Since the reconstruction of the band structure and the appearance of the correlated phases depends on the interactions between the layers, these are extremely sensitive to the interlayer distance. Therefore, the tuning of the interlayer distance in these structures is of central interest. This can be achieved by applying external pressure ($p$), e.g. by using a hydrostatic pressure cell. The power of this method was demonstrated in graphene/hBN superlattices as a change of the superlattice potential\,\cite{Yankowitz2018}, in layered antiferromagnets as an antiferromagnetic to ferromagnetic transition\,\cite{Sun2018, Li2019a, Shao2021}, and in Gr/WSe$_2$ heterostructures, as an increase of induced spin\,--\,orbit coupling\cite{Gmitra2016,Fueloep2021}. In TBG the pressure changes the width of the low-energy bands\,\cite{Carr2018,Guinea2019} and it can drive the system through a superconducting phase transition\cite{Yankowitz2019}. 
Based on theoretical calculations on TDBG, the pressure is expected to have similarly drastic effects on the electronic properties\,\cite{Chebrolu2019, Lin2020} thus it gives an ideal control knob for in situ band-structure and topology engineering of TDBG. 

In order to understand the influence of pressure on correlated phases in TDBG, the first step is to study the pressure dependence of the main parameters of the band structure in the single-particle picture.
Here, we report for the first time the tuning of the band structure of twisted double bilayer graphene (TDBG) close to the magic angle ($1.05$°) by the application of hydrostatic pressure. Using bias spectroscopy and thermal activation measurements we demonstrate a strong modulation of the single-particle band gaps of the system, which can be fully closed for 2\,GPa of pressure. These findings agree well with our band-structure calculations. Moreover, our measurements indicate that pressure can lead to a topologically non-trivial gap at finite magnetic fields at the charge neutrality point.  

\section{Schematic description of the TDBG}
Our TDBG device is fabricated with the tear-and-stack and dry stacking methods\,\cite{Wang2013, Zomer2014, Kim2016a} with a twist angle of $\vartheta=1.067$° (determined from quantum oscillations in magnetoconductance shown in the Supporting Information) and is encapsulated in hexagonal Boron Nitride (hBN). It has a graphite bottom and a metallic top gate as shown schematically in Fig.~\ref{fig1}a. The details of fabrication and an optical image of the device can be found in the Methods and the Supporting Information. This dual gated geometry allows us an independent control of the charge density ($n$) and the transverse electric displacement field ($D$) in the TDBG. 

The rotation between the top and bottom BLG lattices, illustrated in Fig.~\ref{fig1}b, leads also to a rotation between their Brillouin-zones (BZ). These, along with the simplified BLG spectrum of each lattice, are presented  in Fig.~\ref{fig1}c with red and blue for the top and bottom bilayer, respectively. For small rotation angles, the spectrum of the bottom and top bilayers overlap. The coupling $V_{int}$ between the closest monolayers of the BLGs hybridizes the bands, and leads to avoided crossings as shown in Fig.~\ref{fig1}d. Moreover, at magic twist angles the low-energy moiré bands become flat\,\cite{Bistritzer2011, Santos2012, Laissardiere2012, Fang2016} due to the strong interlayer coupling driven avoided crossings. This is illustrated with the density of states (DOS) in Fig.~\ref{fig1}f. The flat bands (red and blue) have a small bandwidth (BW) and are separated from the dispersive conduction bands by a gap $\Delta_\mathrm{n_{s}}$ and from the valence bands by $\Delta_\mathrm{-n_{s}}$. Here $n_\mathrm{s}$ is the carrier density required to fill a single moiré band, either red or blue, with 4-fold degeneracy corresponding to four holes or electrons per superlattice unit cell in real space due to the spin and valley degeneracy. The indices of the gaps $\Delta_\mathrm{\pm n_{s}}$ signify that in order to fill either flat band and move the Fermi-level into either gap, the carrier density must be $\pm n_\mathrm{s}$. 

\begin{figure*}[!ht]
\begin{center}
\includegraphics[width=0.95\linewidth]{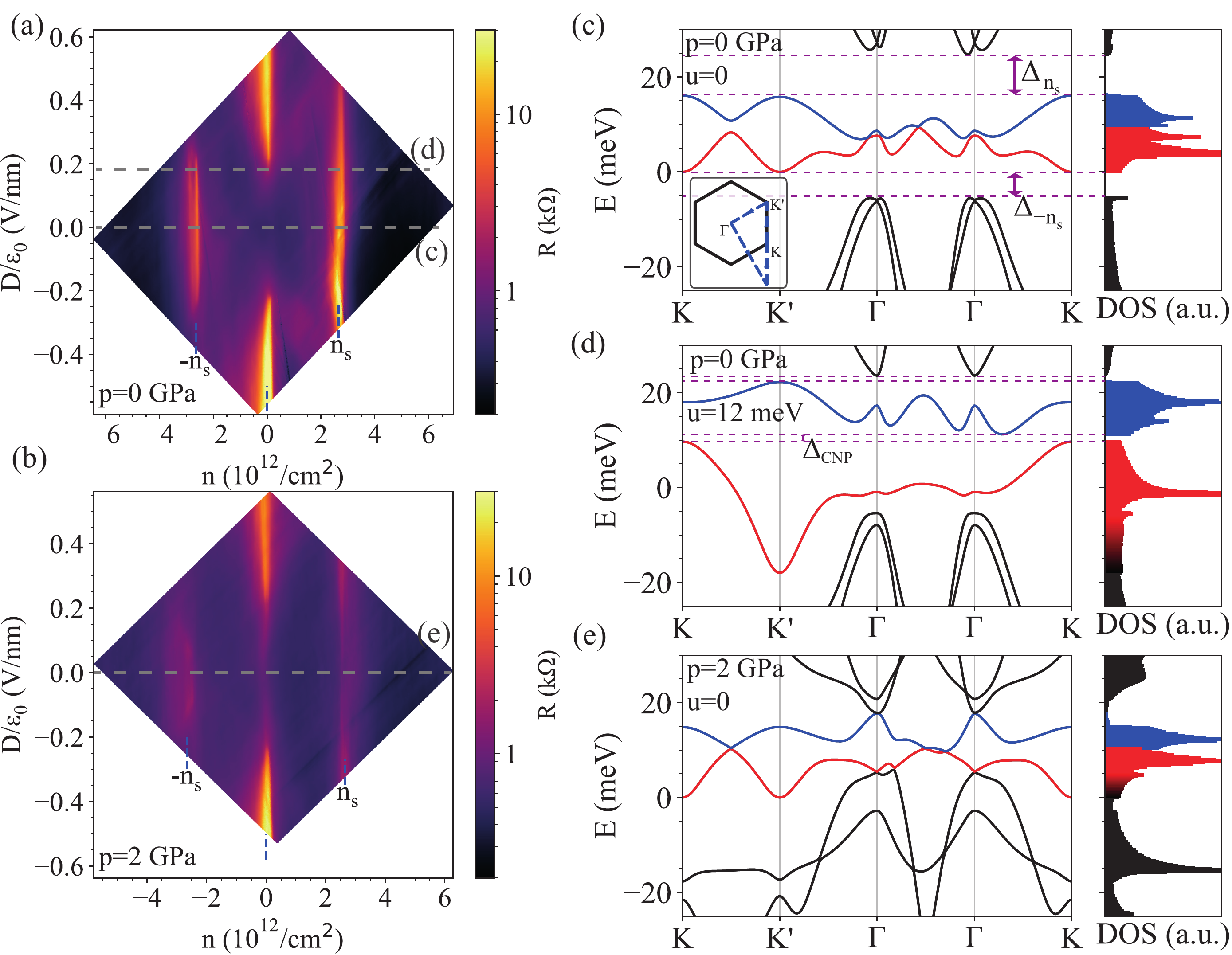}
\caption{Device characterization and band structure calculation. (a) and (b) Four-probe resistance of the TDBG as a function of  the charge density ($n$) and electric displacement field ($D$) measured in (a) at ambient pressure $p$ and in (b) at $p=2$\,GPa. Besides the charge neutrality point, there are two other high resistance regions at $\pm n_\mathrm{s}$, when the flat bands are completely filled. (c-e) Calculated band structure and DOS of the TDBG at $\vartheta=1.067$° twist angle, roughly corresponding to the dashed lines in panels (a) and (b). The flat bands are highlighted with red and blue colors. The spectra in (c) and (d) are calculated at ambient pressure for displacement fields $D=0$ and  $D/\epsilon_0\approx0.18$\,V/nm. A discussion on conversion of interlayer potential $u$ to displacement field is given later and in the Supporting Information. The band structure in (e) is calculated  for $D=0$, $p=2$\,GPa. In the inset of panel (c) the moiré-Brillouin zone is shown.}
\label{fig2}
\end{center}
\end{figure*}
In the TDBG the external perpendicular displacement field $D$ is an additional control parameter to tune the band structure and can open a gap $\Delta_\mathrm{CNP}$ at the charge neutrality point\,\cite{Koshino2019, Zhang2019, Chebrolu2019} as depicted in Fig.~\ref{fig1}e and g: the flat bands split into two fourfold degenerate bands and the gaps separating them from the dispersive bands $\Delta_\mathrm{n_{s}}$ and $\Delta_\mathrm{-n_{s}}$ are decreased.

In the following let us focus on how the pressure controls the BW, the single particle gaps and $\Delta_\mathrm{CNP}$ of TDBG.

\section{Results and Discussion}
Fig.~\ref{fig2}a shows a four-probe resistance ($R_\mathrm{xx}$) measurement as a function of top and bottom gate voltages, plotted as a function of electron density $n$ and displacement field $D$ at temperature $T=1.5$\,K. Lighter colored regions of higher resistance correspond to conditions when the Fermi energy is in a gap. If the flat bands are completely filled with electrons or holes at $n=\pm n_\mathrm{s}$ the device shows single-particle gaps which are the most prominent at $D=0$ and start to fade away for larger displacement fields. Moreover, at the charge neutrality point (CNP, $n=0$), a gap opens by increasing $|D|$ as demonstrated by the increase of the resistance with the increase of $|D|$ in Fig.~\ref{fig2}a. This can be well explained by our band structure calculations done using a continuum model.

\begin{figure*}[!ht]
\begin{center}
\includegraphics[width=1\linewidth]{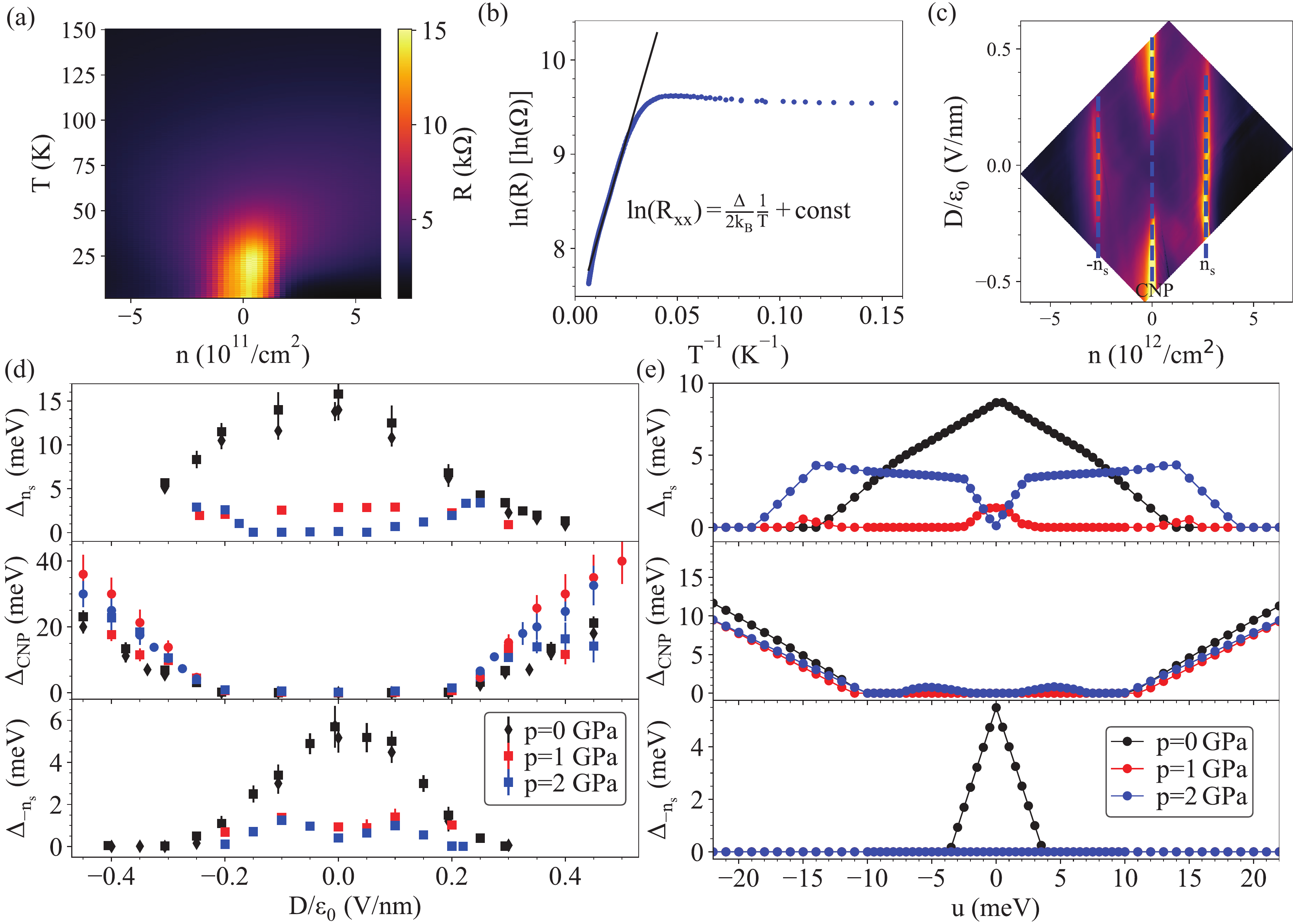}
\caption{Thermal activation measurements of band gaps and comparison to the theoretical results. (a) A typical resistance map measured at ${D}/{\epsilon_0}=0.375\,$V/nm and $p=0$ near $n=0$ as a function of the temperature $T$ and charge density $n$. At every measured temperature the highest resistance value was taken. (b) $R$ peaks extracted from panel (a) as a function of $1/T$. The black line is the fit from which the gap values were obtained according to the Arrhenius equation. (c) The same n-D map of the resistance as Fig.~\ref{fig1}a showing the lines along which gap values were estimated for several $D$ values in panel (d) using similar thermal activation measurements as presented in panel (b). (d) The measured gaps with respect to $D$. The different colors show the gaps at different pressures. The gaps obtained from different methods are shown with different markers: the squares and diamonds show gaps obtained from four-probe and two-probe thermal activation measurements, respectively, and the circles are gaps obtained from bias measurements. (e) Corresponding theoretically calculated single-particle gaps at three different pressures with respect to the on-site energy difference ($u$) of the layers.}
\label{fig3}
\end{center}
\end{figure*}

We calculated the band structure of the TDBG at $\vartheta=1.067$° using the non-interacting Bistritzer-Macdonald model\,\cite{Bistritzer2011} using the parameters from Ref.\,\citenum{Chebrolu2019} and \citenum{Jung2014}. First, in Fig.~\ref{fig2}c we show the low-energy band structure for $D=0$ and zero external pressure ($p=0$), along a path in the moire-Brillouin zone indicated by the dashed line in the inset. The flat bands are depicted in red and blue colors. We denote the gaps separating the flat bands from the dispersive ones by $\Delta_\mathrm{\pm n_{s}}$. Large resistance is expected when the Fermi energy is in the gapped regions of the band structure. A calculated spectrum for finite $D$ and zero pressure is shown in Fig.~\ref{fig2}d. On one hand, $D$ opens a band gap $\Delta_\mathrm{CNP}$ at the CNP between the two flat bands. On the other hand, it decreases and eventually closes the gaps $\Delta_\mathrm{\pm n_{s}}$. Overall, the band structure calculations are in agreement with the measurements presented in Fig.~\ref{fig2}a. 

As a next step we applied hydrostatic pressure of 2\,GPa on the sample (see Methods) and in a subsequent cooldown to $1.5$\,K $R_\mathrm{xx}$ was remeasured as a function of the gate voltages. The result is shown in Fig.~\ref{fig2}b with the same color scale as in Fig.~\ref{fig2}a. The features are similar as at $p=0$, except that the resistance values at $n=\pm n_\mathrm{s}$ are significantly smaller than in Fig.~\ref{fig2}a. For example along $n=-n_\mathrm{s}$ the resistance decreased by 70\% at $p=2$\,GPa. To understand the origin of this change we calculated the evolution of the band structure using the pressure dependence of the interlayer coupling parameters given in Ref.\,\citenum{Chebrolu2019}. These parameters increased by $\approx30$\% at $p=2$\,GPa from their value at $p=0$ while the interlayer distance decreased by $\approx5$\%. The spectrum at $p=2$\,GPa and zero displacement field is plotted in Fig.~\ref{fig2}e. As a result of pressure, the flat bands slightly narrow down and the dispersive bands shift down (up) in energy for the electron (hole) side and close the gaps at $\pm n_\mathrm{s}$. (Later, we show the $D$ dependence of the gaps at $p=2$\,GPa.) Moreover, in Fig.~\ref{fig2}a there's a sign of emerging correlated phases at half-filling, which disappears at $p=2$\,GPa (see Fig.~\ref{fig2}b.), similarly to what was observed in magic-angle bilayer graphene\cite{Yankowitz2019,Carr2018,Chittari2018} 

To quantitatively verify the pressure dependence of the band structure and to extract the gap sizes we performed thermal activation measurements. For this purpose, we measured the four-terminal resistance at a fixed $D$ in a small range of $n$ near the gapped regions as a function of the temperature ($T$). In Fig.~\ref{fig3}a we show a typical activation measurement, i.e. a resistance map as a function of $n$ and $T$. The resistance decreases with increasing temperature due to thermal activation over the gap. To determine the gap size we extract the resistance maximum for each temperature value, $R_\mathrm{xx}(T)$. Then the gap value was extracted from Arrhenius plot, where the logarithm of $R_\mathrm{xx}(T)$ was plotted as a function of $T^{-1}$ (see Fig.~\ref{fig3}b). The linear region close to zero in the x-axis is used to extract the gap energies by the Arrhenius equation ($R_{xx}\propto e^\frac{\Delta}{2k_\mathrm{B}T}$, where $k_\mathrm{B}$ is the Boltzmann constant). The slope of the linear fit (black line) provided the gap values. Error bars originate from the uncertainties in the fitting temperature range.

The gap energies were extracted in this way at several $D$ points along the three dashed lines shown in Fig.~\ref{fig3}c, which are denoted as $\pm n_\mathrm{s}$ and CNP. These data are shown in Fig.~\ref{fig3}d at three different pressures with three different colors. The middle panel presents the evolution of the gap with $D$ at the CNP ($\Delta_\mathrm{CNP}$). The system is not gapped at $D=0$ and a gap opens at $D/\epsilon_0\approx0.2$\,V/nm where $\epsilon_0$ is the vacuum permittivity. The top panel shows the $D$ dependence of the moiré gap at the electron side. A non-zero gap is present at $D=0$, which decreases with $|D|$ until it closes at $|D|/\epsilon_0\approx 0.4$\,V/nm. The bottom panel summarizes the behavior of the gap at $-n_\mathrm{s}$ ($\Delta_\mathrm{-n_{s}}$). 
 It behaves similarly to $\Delta_\mathrm{n_{s}}$ as it is finite at zero displacement field and decreases with $|D|$, but it closes at a smaller displacement field ($|D|/\epsilon_0\approx 0.3$ V/nm). 
The results at ambient pressure are in accordance with the previous findings\cite{Burg2019, Chebrolu2019, Lee2019, Shen2020, He2020}. The measurement results obtained at a pressure of $p=1$\,GPa and $p=2$\,GPa are shown in the same plots with red and blue color, respectively. At finite pressure $\Delta_\mathrm{CNP}$ behaves similar to $p=0$: the gap opens around the same displacement field (middle panel). However, for $\Delta_\mathrm{\pm n_{s}}$ the displacement field dependence strongly deviates from results at ambient pressure. At 1\,GPa (red symbol) $\Delta_\mathrm{n_{s}}$ is smaller than at $p=0$ by a factor of $4$ but follows the same tendency. On the other hand, at 2\,GPa (blue symbol) $\Delta_\mathrm{n_{s}}$ is closed at $D=0$ and opens at finite $D$. The $\Delta_\mathrm{-n_{s}}$ data at $p=1$\,GPa and $p=2$\,GPa are similar to each other: they exhibit a peak at finite $|D|$ before decaying but they are strongly reduced compared to the $p=0$ values.

We note that the gap energies remained approximately the same at the same hydrostatic pressures during different cooldowns and pressurization cycles.  We confirmed a part of the thermal activation measurements at 2\,GPa and 1\,GPa using bias spectroscopy shown by circle symbols (for details see the Supporting Information). These results are generally consistent with the thermal activation data.
 
A good qualitative understanding of the pressure dependence of the measured gaps can be obtained by comparing them to our theoretical calculations shown in Fig.~\ref{fig3}e. In these calculations we neglected the electron-electron interactions and the quantum capacitance. 
The band gaps as a function of the interlayer potential difference ($u$), which is proportional to the displacement field, are plotted in Fig.~\ref{fig3}e. The gaps in the measurements show qualitatively the same dependence on the displacement field as our calculation and the measured gap values are also comparable to the calculated ones. 
In our calculations, for $\Delta_\mathrm{n_{s}}$, the gap values at ambient pressure and at $p=1$\,GPa decrease as a function of $u$, the latter being smaller than the former. Furthermore, the $p=1$\,GPa gap closes for smaller $u$. At $p=2$\,GPa the gap opens with $u$ and closes at a higher interlayer potential difference. This tendency is similar to the experimental results. However, the theory suggests a faster opening of the gap at 2\,GPa compared to the experiment. For the hole sides, at ambient pressure a finite gap is present in the calculations, which closes by increasing $u$. In contrast, the gap is absent both for $p=1$\,GPa and $p=$\,2GPa for all interlayer potential difference. These tendencies are qualitatively the same as observed in the experiments where only small gaps are seen at large pressure. Altogether, the measured spectrum is in good agreement with the theoretical expectations considering the simplicity of our model which is discussed later.

In order to compare the interlayer potential to the displacement fields applied in the experiment, we used the relation of $u=eD\frac{d}{\epsilon_0\epsilon}$, where $d=0.33$\,nm is the interlayer distance of bilayer graphene and $\epsilon=6\pm2$ \cite{Hwang2012, Bessler2019} is the relative dielectric constant of bilayer graphene. Using $\epsilon=5$  gives a relatively good agreement for the gap at the CNP, and less good for the moiré gaps (see Supp Fig.~S9). We emphasize that as a simplification we have used equal potential drop between all four layers in the calculations. However, the potential drop between the layers can be different originating from crystal fields\cite{Rickhaus2019} and the different layer distances and dielectric constants. Moreover, we have neglected all quantum capacitance corrections, and most importantly all correlation effects in the calculations (see Supporting information).

\begin{figure}[!hbt]
\begin{center}
\includegraphics[width=7cm]{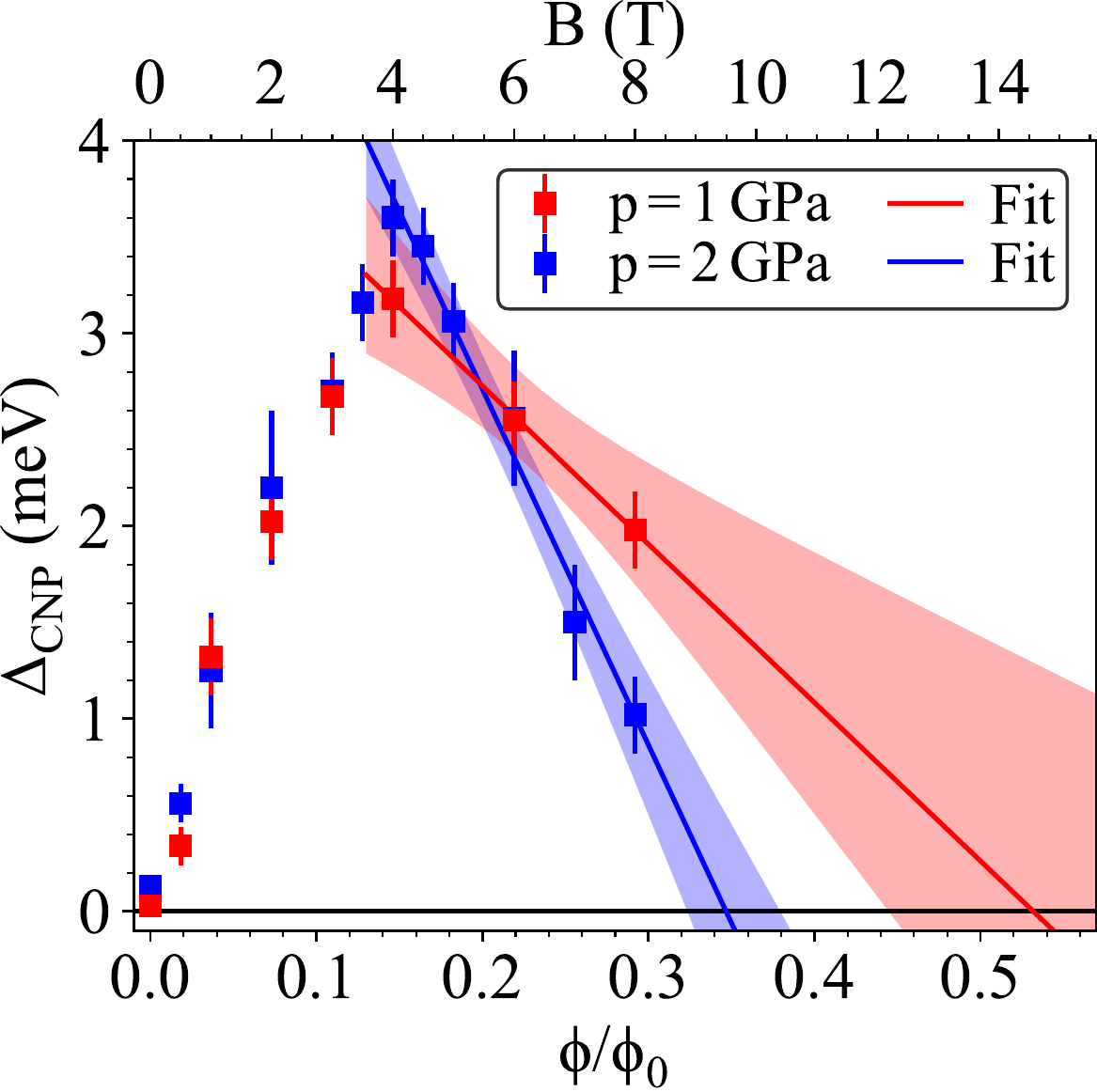}
\caption{The charge neutrality gap ($n=0$) at $D=0$ as a function of the magnetic flux in the superlattice unit cell ($\Phi=BA_\mathrm{s}$). The corresponding magnetic field $B$ is given on the top axis. The gap starts to close above $\sim\,4$\,T and closes completely approximately at $\Phi/\Phi_0=1/3$ for 2\,GPa. The blue and red lines are extrapolations for the gap at $p=2$\,GPa and $p=1$\,GPa, respectively. The shaded areas are the 95\% confidence interval of the extrapolation.}
\label{fig4}
\end{center}
\end{figure}
Finally, we also find signatures of interesting topological effects at the CNP ($D=0$) in an out-of-plane magnetic field. By increasing the magnetic field $B$, a gap opens and surprisingly above $\sim4$\,T this gap starts to close. This is shown in Fig.~\ref{fig4} for 1 and 2 GPa respectively. This finding is similar to the recent results by Burg et al. in Ref.\,\citenum{Burg2020}, where the authors argued that if the gap is non-trivial with a non-zero Chern-number ($C$) then the gap should close at $\Phi/\Phi_0=1/|C|$. Here $\Phi=BA_\mathrm{s}$ is the magnetic flux penetrating the superlattice unit cell, $A_\mathrm{s}$ is the area of the superlattice unit cell which is given by Eq.~S2 in the Supporting Information and $\phi_0=h/e$ is the flux quantum with Planck's constant $h$ and the elementary charge $e$. The measurements of Ref.\,\citenum{Burg2020} were performed at ambient pressure using a sample of $\vartheta=1.01$° twist angle. Their results suggested the presence of a non-trivial gap with $C=2$ at finite $D$, which agree with the theory \cite{Zhang2011, Wang2021, Crosse2020, Lee2019},   and similar gap opening and closing was observed for $D=0$.  Our device shows a similar behaviour, however, for $D=0$ and finite pressure as shown in Fig.~\ref{fig4}. At 2\,GPa the gap closes near $\Phi/\Phi_0=1/3$, which could suggest a band gap with a Chern-number of $C=3$. Surprisingly at 1\,GPa (red symbols and red line) the gap starts to close at the same magnetic field as for 2\,GPa, but the extrapolation suggest that it would go to zero at much higher magnetic field than accessible in our setup ($8$\,T). This suggests that the Chern-number may depend on the pressure. Another, more likely possibility stems from the decrease of correlations effects which is well visible in Fig.~\ref{fig2}, by the disappearance of correlated features at half filling at finite pressures. This will lead to a smaller value of $\Phi/\Phi_0$ where the gap closes\citep{Burg2020}. For a better understanding, further studies are required both theoretically and experimentally. 

\section{Conclusion}
In conclusion, we investigated a TDBG with $\vartheta=1.067$° under pressure at different temperatures and magnetic fields. We found that the band structure significantly changes with pressure: the single-particle moiré gaps present at zero displacement fields can be fully closed. We compared our findings with a single-particle continuum model and found a reasonable agreement with our experimental data. The changes achievable with the pressure are large and on the order of the bandwidth of the central flat band. The large tunability and the theoretical predictability suggest that the band structure of twisted structures can be precisely designed by taking into account the modified layer separation. The pressure combined with the electric field tunability allows extensive control of the band structure in situ. The exploration of the single-particle band structure is an important milestone towards tuning and understanding the emerging topological and correlated states of these systems. In addition we showed that in out-of-plane magnetic field the closing of the gap at the charge neutrality point strongly depends on pressure, suggesting pressure dependence of the Chern-number.

\section{Methods}
\subsection{Device Fabrication}
The structure of the van der Waals heterostructure is the following from bottom to the top: graphite layer, 54\,nm hBN, bilayer graphene, rotated bilayer graphene, 32\,nm hBN, aluminium oxide layer and metallic top gate. Side contacts to the TDBG are created by reactive ion etching and evaporation of 10/50\,nm Cr/Au.
\subsection{Transport measurements}
Transport measurements were made with AC voltage excitation of 0.1\,mV using lock-in technique at 177.13\,Hz. The charge density ($n$) and electric displacement field ($D$) were calculated from the gate voltages which is detailed in the Supporting Information.
\subsection{Pressurization}
Our device was measured using a special high-pressure sample holder equipped with a circuit board suitable for transport measurements of nanodevices at cryogenic temperature, and a piston-cylinder hydrostatic pressure cell. In the cell the pressure was mediated with kerosene and it was applied using a hydraulic press at room temperature. One of the unique features of our device is that we can use wire bonding to contact the device. Our pressure cell is described in more detail in Ref.\,\citenum{Fueloep2021a}. After releasing the pressure we remeasured some data at ambient pressure which was the same before the pressurization.
\subsection{Numerical calculations}
The band structure of the TDBG was calculated using a continuum model. The effect of the pressure is captured by tuning the interlayer couplings modelled in Ref.\,\citenum{Chebrolu2019}. The details of our calculations can be found in the Supporting Information.

\section{Data availability}
Source data of the measurements and the python code for the simulation are publicly available at \url{https://doi.org/10.5281/zenodo.5532887}.

\section{Author contribution}
P.R fabricated the device. F.d.V. and P.R. performed initial characterization measurements. Measurements were performed by B.Sz. with the help of B.F., E.T. and A.M. B.Sz. did the data analysis and the theoretical calculations with the help of A.K. B.Sz. and P.M. wrote the paper and all authors discussed the results and worked on the manuscript. K.W. and T.T. grew the hBN crystals. The project was guided by Sz.Cs. and P.M.

\section{Acknowledgement}
The authors thank Prof E. Tutuc, K. Ensslin and T. Ihn for useful discussions, and Márton Hajdú, Ferenc Fülöp, and Gergő Fülöp for their technical support. We thank Gergő Fülöp for helping in creating the device sketch.
This work acknowledges support from the Topograph FlagERA network, the OTKA FK-123894 and OTKA PD-134758 grants. This research was supported by the Ministry of Innovation and Technology and the National Research, Development and Innovation Office within the Quantum Information National Laboratory of Hungary and by the Quantum Technology National Excellence Program (Project Nr. 2017-1.2.1-NKP-2017-00001), by SuperTop QuantERA network, by the FET Open AndQC network and Nanocohybri COST network. P.M., E.T., and A.K. received funding from the Hungarian Academy of Sciences through the Bolyai Fellowship. A.K. acknowledges the support from the Hungarian Scientific Research Fund (OTKA)  Grant No. K134437, and the ELTE Institutional Excellence Program (TKP2020-IKA-05). K.W. and T.T. acknowledge support from the Elemental Strategy Initiative conducted by the MEXT, Japan (Grant Number JPMXP0112101001) and JSPS KAKENHI (Grant Numbers 19H05790 and JP20H00354). We acknowledge support from the Graphene Flagship and from the European Union’s Horizon 2020 research and innovation programme under grant agreement number 862660/QUANTUM E LEAPS and the Swiss National Science Foundation via NCCR Quantum Science and Technology. Low T infrastructure was provided by VEKOP-2.3.3-15-2017-00015.

\bibliographystyle{apsrev4-2}
\bibliography{Szentpeteri_etal_2021}

\begin{thebibliography}{54}%
\makeatletter
\providecommand \@ifxundefined [1]{%
 \@ifx{#1\undefined}
}%
\providecommand \@ifnum [1]{%
 \ifnum #1\expandafter \@firstoftwo
 \else \expandafter \@secondoftwo
 \fi
}%
\providecommand \@ifx [1]{%
 \ifx #1\expandafter \@firstoftwo
 \else \expandafter \@secondoftwo
 \fi
}%
\providecommand \natexlab [1]{#1}%
\providecommand \enquote  [1]{``#1''}%
\providecommand \bibnamefont  [1]{#1}%
\providecommand \bibfnamefont [1]{#1}%
\providecommand \citenamefont [1]{#1}%
\providecommand \href@noop [0]{\@secondoftwo}%
\providecommand \href [0]{\begingroup \@sanitize@url \@href}%
\providecommand \@href[1]{\@@startlink{#1}\@@href}%
\providecommand \@@href[1]{\endgroup#1\@@endlink}%
\providecommand \@sanitize@url [0]{\catcode `\\12\catcode `\$12\catcode
  `\&12\catcode `\#12\catcode `\^12\catcode `\_12\catcode `\%12\relax}%
\providecommand \@@startlink[1]{}%
\providecommand \@@endlink[0]{}%
\providecommand \url  [0]{\begingroup\@sanitize@url \@url }%
\providecommand \@url [1]{\endgroup\@href {#1}{\urlprefix }}%
\providecommand \urlprefix  [0]{URL }%
\providecommand \Eprint [0]{\href }%
\providecommand \doibase [0]{https://doi.org/}%
\providecommand \selectlanguage [0]{\@gobble}%
\providecommand \bibinfo  [0]{\@secondoftwo}%
\providecommand \bibfield  [0]{\@secondoftwo}%
\providecommand \translation [1]{[#1]}%
\providecommand \BibitemOpen [0]{}%
\providecommand \bibitemStop [0]{}%
\providecommand \bibitemNoStop [0]{.\EOS\space}%
\providecommand \EOS [0]{\spacefactor3000\relax}%
\providecommand \BibitemShut  [1]{\csname bibitem#1\endcsname}%
\let\auto@bib@innerbib\@empty
\bibitem [{\citenamefont {Bistritzer}\ and\ \citenamefont
  {MacDonald}(2011)}]{Bistritzer2011}%
  \BibitemOpen
  \bibfield  {author} {\bibinfo {author} {\bibfnamefont {R.}~\bibnamefont
  {Bistritzer}}\ and\ \bibinfo {author} {\bibfnamefont {A.~H.}\ \bibnamefont
  {MacDonald}},\ }\href {https://doi.org/10.1073/pnas.1108174108} {\bibfield
  {journal} {\bibinfo  {journal} {Proceedings of the National Academy of
  Sciences}\ }\textbf {\bibinfo {volume} {108}},\ \bibinfo {pages} {12233}
  (\bibinfo {year} {2011})}\BibitemShut {NoStop}%
\bibitem [{\citenamefont {Jung}\ \emph {et~al.}(2014)\citenamefont {Jung},
  \citenamefont {Raoux}, \citenamefont {Qiao},\ and\ \citenamefont
  {MacDonald}}]{Jung2014a}%
  \BibitemOpen
  \bibfield  {author} {\bibinfo {author} {\bibfnamefont {J.}~\bibnamefont
  {Jung}}, \bibinfo {author} {\bibfnamefont {A.}~\bibnamefont {Raoux}},
  \bibinfo {author} {\bibfnamefont {Z.}~\bibnamefont {Qiao}},\ and\ \bibinfo
  {author} {\bibfnamefont {A.~H.}\ \bibnamefont {MacDonald}},\ }\href
  {https://doi.org/10.1103/physrevb.89.205414} {\bibfield  {journal} {\bibinfo
  {journal} {Physical Review B}\ }\textbf {\bibinfo {volume} {89}},\ \bibinfo
  {pages} {205414} (\bibinfo {year} {2014})}\BibitemShut {NoStop}%
\bibitem [{\citenamefont {Cao}\ \emph {et~al.}(2018{\natexlab{a}})\citenamefont
  {Cao}, \citenamefont {Fatemi}, \citenamefont {Demir}, \citenamefont {Fang},
  \citenamefont {Tomarken}, \citenamefont {Luo}, \citenamefont
  {Sanchez-Yamagishi}, \citenamefont {Watanabe}, \citenamefont {Taniguchi},
  \citenamefont {Kaxiras}, \citenamefont {Ashoori},\ and\ \citenamefont
  {Jarillo-Herrero}}]{Cao2018}%
  \BibitemOpen
  \bibfield  {author} {\bibinfo {author} {\bibfnamefont {Y.}~\bibnamefont
  {Cao}}, \bibinfo {author} {\bibfnamefont {V.}~\bibnamefont {Fatemi}},
  \bibinfo {author} {\bibfnamefont {A.}~\bibnamefont {Demir}}, \bibinfo
  {author} {\bibfnamefont {S.}~\bibnamefont {Fang}}, \bibinfo {author}
  {\bibfnamefont {S.~L.}\ \bibnamefont {Tomarken}}, \bibinfo {author}
  {\bibfnamefont {J.~Y.}\ \bibnamefont {Luo}}, \bibinfo {author} {\bibfnamefont
  {J.~D.}\ \bibnamefont {Sanchez-Yamagishi}}, \bibinfo {author} {\bibfnamefont
  {K.}~\bibnamefont {Watanabe}}, \bibinfo {author} {\bibfnamefont
  {T.}~\bibnamefont {Taniguchi}}, \bibinfo {author} {\bibfnamefont
  {E.}~\bibnamefont {Kaxiras}}, \bibinfo {author} {\bibfnamefont {R.~C.}\
  \bibnamefont {Ashoori}},\ and\ \bibinfo {author} {\bibfnamefont
  {P.}~\bibnamefont {Jarillo-Herrero}},\ }\href
  {https://doi.org/10.1038/nature26154} {\bibfield  {journal} {\bibinfo
  {journal} {Nature}\ }\textbf {\bibinfo {volume} {556}},\ \bibinfo {pages}
  {80} (\bibinfo {year} {2018}{\natexlab{a}})}\BibitemShut {NoStop}%
\bibitem [{\citenamefont {Cao}\ \emph {et~al.}(2018{\natexlab{b}})\citenamefont
  {Cao}, \citenamefont {Fatemi}, \citenamefont {Fang}, \citenamefont
  {Watanabe}, \citenamefont {Taniguchi}, \citenamefont {Kaxiras},\ and\
  \citenamefont {Jarillo-Herrero}}]{Cao2018a}%
  \BibitemOpen
  \bibfield  {author} {\bibinfo {author} {\bibfnamefont {Y.}~\bibnamefont
  {Cao}}, \bibinfo {author} {\bibfnamefont {V.}~\bibnamefont {Fatemi}},
  \bibinfo {author} {\bibfnamefont {S.}~\bibnamefont {Fang}}, \bibinfo {author}
  {\bibfnamefont {K.}~\bibnamefont {Watanabe}}, \bibinfo {author}
  {\bibfnamefont {T.}~\bibnamefont {Taniguchi}}, \bibinfo {author}
  {\bibfnamefont {E.}~\bibnamefont {Kaxiras}},\ and\ \bibinfo {author}
  {\bibfnamefont {P.}~\bibnamefont {Jarillo-Herrero}},\ }\href
  {https://doi.org/10.1038/nature26160} {\bibfield  {journal} {\bibinfo
  {journal} {Nature}\ }\textbf {\bibinfo {volume} {556}},\ \bibinfo {pages}
  {43} (\bibinfo {year} {2018}{\natexlab{b}})}\BibitemShut {NoStop}%
\bibitem [{\citenamefont {Yankowitz}\ \emph {et~al.}(2019)\citenamefont
  {Yankowitz}, \citenamefont {Chen}, \citenamefont {Polshyn}, \citenamefont
  {Zhang}, \citenamefont {Watanabe}, \citenamefont {Taniguchi}, \citenamefont
  {Graf}, \citenamefont {Young},\ and\ \citenamefont {Dean}}]{Yankowitz2019}%
  \BibitemOpen
  \bibfield  {author} {\bibinfo {author} {\bibfnamefont {M.}~\bibnamefont
  {Yankowitz}}, \bibinfo {author} {\bibfnamefont {S.}~\bibnamefont {Chen}},
  \bibinfo {author} {\bibfnamefont {H.}~\bibnamefont {Polshyn}}, \bibinfo
  {author} {\bibfnamefont {Y.}~\bibnamefont {Zhang}}, \bibinfo {author}
  {\bibfnamefont {K.}~\bibnamefont {Watanabe}}, \bibinfo {author}
  {\bibfnamefont {T.}~\bibnamefont {Taniguchi}}, \bibinfo {author}
  {\bibfnamefont {D.}~\bibnamefont {Graf}}, \bibinfo {author} {\bibfnamefont
  {A.~F.}\ \bibnamefont {Young}},\ and\ \bibinfo {author} {\bibfnamefont
  {C.~R.}\ \bibnamefont {Dean}},\ }\href
  {https://doi.org/10.1126/science.aav1910} {\bibfield  {journal} {\bibinfo
  {journal} {Science}\ }\textbf {\bibinfo {volume} {363}},\ \bibinfo {pages}
  {1059} (\bibinfo {year} {2019})}\BibitemShut {NoStop}%
\bibitem [{\citenamefont {Saito}\ \emph {et~al.}(2020)\citenamefont {Saito},
  \citenamefont {Ge}, \citenamefont {Watanabe}, \citenamefont {Taniguchi},\
  and\ \citenamefont {Young}}]{Saito2020}%
  \BibitemOpen
  \bibfield  {author} {\bibinfo {author} {\bibfnamefont {Y.}~\bibnamefont
  {Saito}}, \bibinfo {author} {\bibfnamefont {J.}~\bibnamefont {Ge}}, \bibinfo
  {author} {\bibfnamefont {K.}~\bibnamefont {Watanabe}}, \bibinfo {author}
  {\bibfnamefont {T.}~\bibnamefont {Taniguchi}},\ and\ \bibinfo {author}
  {\bibfnamefont {A.~F.}\ \bibnamefont {Young}},\ }\href
  {https://doi.org/10.1038/s41567-020-0928-3} {\bibfield  {journal} {\bibinfo
  {journal} {Nature Physics}\ }\textbf {\bibinfo {volume} {16}},\ \bibinfo
  {pages} {926} (\bibinfo {year} {2020})}\BibitemShut {NoStop}%
\bibitem [{\citenamefont {Sharpe}\ \emph {et~al.}(2019)\citenamefont {Sharpe},
  \citenamefont {Fox}, \citenamefont {Barnard}, \citenamefont {Finney},
  \citenamefont {Watanabe}, \citenamefont {Taniguchi}, \citenamefont
  {Kastner},\ and\ \citenamefont {Goldhaber-Gordon}}]{Sharpe2019}%
  \BibitemOpen
  \bibfield  {author} {\bibinfo {author} {\bibfnamefont {A.~L.}\ \bibnamefont
  {Sharpe}}, \bibinfo {author} {\bibfnamefont {E.~J.}\ \bibnamefont {Fox}},
  \bibinfo {author} {\bibfnamefont {A.~W.}\ \bibnamefont {Barnard}}, \bibinfo
  {author} {\bibfnamefont {J.}~\bibnamefont {Finney}}, \bibinfo {author}
  {\bibfnamefont {K.}~\bibnamefont {Watanabe}}, \bibinfo {author}
  {\bibfnamefont {T.}~\bibnamefont {Taniguchi}}, \bibinfo {author}
  {\bibfnamefont {M.~A.}\ \bibnamefont {Kastner}},\ and\ \bibinfo {author}
  {\bibfnamefont {D.}~\bibnamefont {Goldhaber-Gordon}},\ }\href
  {https://doi.org/10.1126/science.aaw3780} {\bibfield  {journal} {\bibinfo
  {journal} {Science}\ }\textbf {\bibinfo {volume} {365}},\ \bibinfo {pages}
  {605} (\bibinfo {year} {2019})}\BibitemShut {NoStop}%
\bibitem [{\citenamefont {Lu}\ \emph {et~al.}(2019)\citenamefont {Lu},
  \citenamefont {Stepanov}, \citenamefont {Yang}, \citenamefont {Xie},
  \citenamefont {Aamir}, \citenamefont {Das}, \citenamefont {Urgell},
  \citenamefont {Watanabe}, \citenamefont {Taniguchi}, \citenamefont {Zhang},
  \citenamefont {Bachtold}, \citenamefont {MacDonald},\ and\ \citenamefont
  {Efetov}}]{Lu2019}%
  \BibitemOpen
  \bibfield  {author} {\bibinfo {author} {\bibfnamefont {X.}~\bibnamefont
  {Lu}}, \bibinfo {author} {\bibfnamefont {P.}~\bibnamefont {Stepanov}},
  \bibinfo {author} {\bibfnamefont {W.}~\bibnamefont {Yang}}, \bibinfo {author}
  {\bibfnamefont {M.}~\bibnamefont {Xie}}, \bibinfo {author} {\bibfnamefont
  {M.~A.}\ \bibnamefont {Aamir}}, \bibinfo {author} {\bibfnamefont
  {I.}~\bibnamefont {Das}}, \bibinfo {author} {\bibfnamefont {C.}~\bibnamefont
  {Urgell}}, \bibinfo {author} {\bibfnamefont {K.}~\bibnamefont {Watanabe}},
  \bibinfo {author} {\bibfnamefont {T.}~\bibnamefont {Taniguchi}}, \bibinfo
  {author} {\bibfnamefont {G.}~\bibnamefont {Zhang}}, \bibinfo {author}
  {\bibfnamefont {A.}~\bibnamefont {Bachtold}}, \bibinfo {author}
  {\bibfnamefont {A.~H.}\ \bibnamefont {MacDonald}},\ and\ \bibinfo {author}
  {\bibfnamefont {D.~K.}\ \bibnamefont {Efetov}},\ }\href
  {https://doi.org/10.1038/s41586-019-1695-0} {\bibfield  {journal} {\bibinfo
  {journal} {Nature}\ }\textbf {\bibinfo {volume} {574}},\ \bibinfo {pages}
  {653} (\bibinfo {year} {2019})}\BibitemShut {NoStop}%
\bibitem [{\citenamefont {Serlin}\ \emph {et~al.}(2019)\citenamefont {Serlin},
  \citenamefont {Tschirhart}, \citenamefont {Polshyn}, \citenamefont {Zhang},
  \citenamefont {Zhu}, \citenamefont {Watanabe}, \citenamefont {Taniguchi},
  \citenamefont {Balents},\ and\ \citenamefont {Young}}]{Serlin2019}%
  \BibitemOpen
  \bibfield  {author} {\bibinfo {author} {\bibfnamefont {M.}~\bibnamefont
  {Serlin}}, \bibinfo {author} {\bibfnamefont {C.~L.}\ \bibnamefont
  {Tschirhart}}, \bibinfo {author} {\bibfnamefont {H.}~\bibnamefont {Polshyn}},
  \bibinfo {author} {\bibfnamefont {Y.}~\bibnamefont {Zhang}}, \bibinfo
  {author} {\bibfnamefont {J.}~\bibnamefont {Zhu}}, \bibinfo {author}
  {\bibfnamefont {K.}~\bibnamefont {Watanabe}}, \bibinfo {author}
  {\bibfnamefont {T.}~\bibnamefont {Taniguchi}}, \bibinfo {author}
  {\bibfnamefont {L.}~\bibnamefont {Balents}},\ and\ \bibinfo {author}
  {\bibfnamefont {A.~F.}\ \bibnamefont {Young}},\ }\href
  {https://doi.org/10.1126/science.aay5533} {\bibfield  {journal} {\bibinfo
  {journal} {Science}\ }\textbf {\bibinfo {volume} {367}},\ \bibinfo {pages}
  {900} (\bibinfo {year} {2019})}\BibitemShut {NoStop}%
\bibitem [{\citenamefont {Cao}\ \emph {et~al.}(2020{\natexlab{a}})\citenamefont
  {Cao}, \citenamefont {Chowdhury}, \citenamefont {Rodan-Legrain},
  \citenamefont {Rubies-Bigorda}, \citenamefont {Watanabe}, \citenamefont
  {Taniguchi}, \citenamefont {Senthil},\ and\ \citenamefont
  {Jarillo-Herrero}}]{Cao2020a}%
  \BibitemOpen
  \bibfield  {author} {\bibinfo {author} {\bibfnamefont {Y.}~\bibnamefont
  {Cao}}, \bibinfo {author} {\bibfnamefont {D.}~\bibnamefont {Chowdhury}},
  \bibinfo {author} {\bibfnamefont {D.}~\bibnamefont {Rodan-Legrain}}, \bibinfo
  {author} {\bibfnamefont {O.}~\bibnamefont {Rubies-Bigorda}}, \bibinfo
  {author} {\bibfnamefont {K.}~\bibnamefont {Watanabe}}, \bibinfo {author}
  {\bibfnamefont {T.}~\bibnamefont {Taniguchi}}, \bibinfo {author}
  {\bibfnamefont {T.}~\bibnamefont {Senthil}},\ and\ \bibinfo {author}
  {\bibfnamefont {P.}~\bibnamefont {Jarillo-Herrero}},\ }\href
  {https://doi.org/10.1103/physrevlett.124.076801} {\bibfield  {journal}
  {\bibinfo  {journal} {Physical Review Letters}\ }\textbf {\bibinfo {volume}
  {124}},\ \bibinfo {pages} {076801} (\bibinfo {year}
  {2020}{\natexlab{a}})}\BibitemShut {NoStop}%
\bibitem [{\citenamefont {Polshyn}\ \emph {et~al.}(2019)\citenamefont
  {Polshyn}, \citenamefont {Yankowitz}, \citenamefont {Chen}, \citenamefont
  {Zhang}, \citenamefont {Watanabe}, \citenamefont {Taniguchi}, \citenamefont
  {Dean},\ and\ \citenamefont {Young}}]{Polshyn2019}%
  \BibitemOpen
  \bibfield  {author} {\bibinfo {author} {\bibfnamefont {H.}~\bibnamefont
  {Polshyn}}, \bibinfo {author} {\bibfnamefont {M.}~\bibnamefont {Yankowitz}},
  \bibinfo {author} {\bibfnamefont {S.}~\bibnamefont {Chen}}, \bibinfo {author}
  {\bibfnamefont {Y.}~\bibnamefont {Zhang}}, \bibinfo {author} {\bibfnamefont
  {K.}~\bibnamefont {Watanabe}}, \bibinfo {author} {\bibfnamefont
  {T.}~\bibnamefont {Taniguchi}}, \bibinfo {author} {\bibfnamefont {C.~R.}\
  \bibnamefont {Dean}},\ and\ \bibinfo {author} {\bibfnamefont {A.~F.}\
  \bibnamefont {Young}},\ }\href {https://doi.org/10.1038/s41567-019-0596-3}
  {\bibfield  {journal} {\bibinfo  {journal} {Nature Physics}\ }\textbf
  {\bibinfo {volume} {15}},\ \bibinfo {pages} {1011} (\bibinfo {year}
  {2019})}\BibitemShut {NoStop}%
\bibitem [{\citenamefont {Stepanov}\ \emph {et~al.}(2020)\citenamefont
  {Stepanov}, \citenamefont {Das}, \citenamefont {Lu}, \citenamefont
  {Fahimniya}, \citenamefont {Watanabe}, \citenamefont {Taniguchi},
  \citenamefont {Koppens}, \citenamefont {Lischner}, \citenamefont {Levitov},\
  and\ \citenamefont {Efetov}}]{Stepanov2020}%
  \BibitemOpen
  \bibfield  {author} {\bibinfo {author} {\bibfnamefont {P.}~\bibnamefont
  {Stepanov}}, \bibinfo {author} {\bibfnamefont {I.}~\bibnamefont {Das}},
  \bibinfo {author} {\bibfnamefont {X.}~\bibnamefont {Lu}}, \bibinfo {author}
  {\bibfnamefont {A.}~\bibnamefont {Fahimniya}}, \bibinfo {author}
  {\bibfnamefont {K.}~\bibnamefont {Watanabe}}, \bibinfo {author}
  {\bibfnamefont {T.}~\bibnamefont {Taniguchi}}, \bibinfo {author}
  {\bibfnamefont {F.~H.~L.}\ \bibnamefont {Koppens}}, \bibinfo {author}
  {\bibfnamefont {J.}~\bibnamefont {Lischner}}, \bibinfo {author}
  {\bibfnamefont {L.}~\bibnamefont {Levitov}},\ and\ \bibinfo {author}
  {\bibfnamefont {D.~K.}\ \bibnamefont {Efetov}},\ }\href
  {https://doi.org/10.1038/s41586-020-2459-6} {\bibfield  {journal} {\bibinfo
  {journal} {Nature}\ }\textbf {\bibinfo {volume} {583}},\ \bibinfo {pages}
  {375} (\bibinfo {year} {2020})}\BibitemShut {NoStop}%
\bibitem [{\citenamefont {Zhang}\ \emph {et~al.}(2019)\citenamefont {Zhang},
  \citenamefont {Mao}, \citenamefont {Cao}, \citenamefont {Jarillo-Herrero},\
  and\ \citenamefont {Senthil}}]{Zhang2019}%
  \BibitemOpen
  \bibfield  {author} {\bibinfo {author} {\bibfnamefont {Y.-H.}\ \bibnamefont
  {Zhang}}, \bibinfo {author} {\bibfnamefont {D.}~\bibnamefont {Mao}}, \bibinfo
  {author} {\bibfnamefont {Y.}~\bibnamefont {Cao}}, \bibinfo {author}
  {\bibfnamefont {P.}~\bibnamefont {Jarillo-Herrero}},\ and\ \bibinfo {author}
  {\bibfnamefont {T.}~\bibnamefont {Senthil}},\ }\href
  {https://doi.org/10.1103/physrevb.99.075127} {\bibfield  {journal} {\bibinfo
  {journal} {Physical Review B}\ }\textbf {\bibinfo {volume} {99}},\ \bibinfo
  {pages} {075127} (\bibinfo {year} {2019})}\BibitemShut {NoStop}%
\bibitem [{\citenamefont {Liu}\ \emph {et~al.}(2019)\citenamefont {Liu},
  \citenamefont {Liu},\ and\ \citenamefont {Dai}}]{Liu2019a}%
  \BibitemOpen
  \bibfield  {author} {\bibinfo {author} {\bibfnamefont {J.}~\bibnamefont
  {Liu}}, \bibinfo {author} {\bibfnamefont {J.}~\bibnamefont {Liu}},\ and\
  \bibinfo {author} {\bibfnamefont {X.}~\bibnamefont {Dai}},\ }\href
  {https://doi.org/10.1103/physrevb.99.155415} {\bibfield  {journal} {\bibinfo
  {journal} {Physical Review B}\ }\textbf {\bibinfo {volume} {99}},\ \bibinfo
  {pages} {155415} (\bibinfo {year} {2019})}\BibitemShut {NoStop}%
\bibitem [{\citenamefont {Song}\ \emph {et~al.}(2019)\citenamefont {Song},
  \citenamefont {Wang}, \citenamefont {Shi}, \citenamefont {Li}, \citenamefont
  {Fang},\ and\ \citenamefont {Bernevig}}]{Song2019}%
  \BibitemOpen
  \bibfield  {author} {\bibinfo {author} {\bibfnamefont {Z.}~\bibnamefont
  {Song}}, \bibinfo {author} {\bibfnamefont {Z.}~\bibnamefont {Wang}}, \bibinfo
  {author} {\bibfnamefont {W.}~\bibnamefont {Shi}}, \bibinfo {author}
  {\bibfnamefont {G.}~\bibnamefont {Li}}, \bibinfo {author} {\bibfnamefont
  {C.}~\bibnamefont {Fang}},\ and\ \bibinfo {author} {\bibfnamefont {B.~A.}\
  \bibnamefont {Bernevig}},\ }\href
  {https://doi.org/10.1103/physrevlett.123.036401} {\bibfield  {journal}
  {\bibinfo  {journal} {Physical Review Letters}\ }\textbf {\bibinfo {volume}
  {123}},\ \bibinfo {pages} {036401} (\bibinfo {year} {2019})}\BibitemShut
  {NoStop}%
\bibitem [{\citenamefont {Nuckolls}\ \emph {et~al.}(2020)\citenamefont
  {Nuckolls}, \citenamefont {Oh}, \citenamefont {Wong}, \citenamefont {Lian},
  \citenamefont {Watanabe}, \citenamefont {Taniguchi}, \citenamefont
  {Bernevig},\ and\ \citenamefont {Yazdani}}]{Nuckolls2020}%
  \BibitemOpen
  \bibfield  {author} {\bibinfo {author} {\bibfnamefont {K.~P.}\ \bibnamefont
  {Nuckolls}}, \bibinfo {author} {\bibfnamefont {M.}~\bibnamefont {Oh}},
  \bibinfo {author} {\bibfnamefont {D.}~\bibnamefont {Wong}}, \bibinfo {author}
  {\bibfnamefont {B.}~\bibnamefont {Lian}}, \bibinfo {author} {\bibfnamefont
  {K.}~\bibnamefont {Watanabe}}, \bibinfo {author} {\bibfnamefont
  {T.}~\bibnamefont {Taniguchi}}, \bibinfo {author} {\bibfnamefont {B.~A.}\
  \bibnamefont {Bernevig}},\ and\ \bibinfo {author} {\bibfnamefont
  {A.}~\bibnamefont {Yazdani}},\ }\href
  {https://doi.org/10.1038/s41586-020-3028-8} {\bibfield  {journal} {\bibinfo
  {journal} {Nature}\ }\textbf {\bibinfo {volume} {588}},\ \bibinfo {pages}
  {610} (\bibinfo {year} {2020})}\BibitemShut {NoStop}%
\bibitem [{\citenamefont {Wu}\ \emph {et~al.}(2021)\citenamefont {Wu},
  \citenamefont {Zhang}, \citenamefont {Watanabe}, \citenamefont {Taniguchi},\
  and\ \citenamefont {Andrei}}]{Wu2021}%
  \BibitemOpen
  \bibfield  {author} {\bibinfo {author} {\bibfnamefont {S.}~\bibnamefont
  {Wu}}, \bibinfo {author} {\bibfnamefont {Z.}~\bibnamefont {Zhang}}, \bibinfo
  {author} {\bibfnamefont {K.}~\bibnamefont {Watanabe}}, \bibinfo {author}
  {\bibfnamefont {T.}~\bibnamefont {Taniguchi}},\ and\ \bibinfo {author}
  {\bibfnamefont {E.~Y.}\ \bibnamefont {Andrei}},\ }\href
  {https://doi.org/10.1038/s41563-020-00911-2} {\bibfield  {journal} {\bibinfo
  {journal} {Nature Materials}\ }\textbf {\bibinfo {volume} {20}},\ \bibinfo
  {pages} {488} (\bibinfo {year} {2021})}\BibitemShut {NoStop}%
\bibitem [{\citenamefont {Saito}\ \emph {et~al.}(2021)\citenamefont {Saito},
  \citenamefont {Ge}, \citenamefont {Rademaker}, \citenamefont {Watanabe},
  \citenamefont {Taniguchi}, \citenamefont {Abanin},\ and\ \citenamefont
  {Young}}]{Saito2021}%
  \BibitemOpen
  \bibfield  {author} {\bibinfo {author} {\bibfnamefont {Y.}~\bibnamefont
  {Saito}}, \bibinfo {author} {\bibfnamefont {J.}~\bibnamefont {Ge}}, \bibinfo
  {author} {\bibfnamefont {L.}~\bibnamefont {Rademaker}}, \bibinfo {author}
  {\bibfnamefont {K.}~\bibnamefont {Watanabe}}, \bibinfo {author}
  {\bibfnamefont {T.}~\bibnamefont {Taniguchi}}, \bibinfo {author}
  {\bibfnamefont {D.~A.}\ \bibnamefont {Abanin}},\ and\ \bibinfo {author}
  {\bibfnamefont {A.~F.}\ \bibnamefont {Young}},\ }\href
  {https://doi.org/10.1038/s41567-020-01129-4} {\bibfield  {journal} {\bibinfo
  {journal} {Nature Physics}\ }\textbf {\bibinfo {volume} {17}},\ \bibinfo
  {pages} {478} (\bibinfo {year} {2021})}\BibitemShut {NoStop}%
\bibitem [{\citenamefont {Koshino}(2019)}]{Koshino2019}%
  \BibitemOpen
  \bibfield  {author} {\bibinfo {author} {\bibfnamefont {M.}~\bibnamefont
  {Koshino}},\ }\href {https://doi.org/10.1103/physrevb.99.235406} {\bibfield
  {journal} {\bibinfo  {journal} {Physical Review B}\ }\textbf {\bibinfo
  {volume} {99}},\ \bibinfo {pages} {235406} (\bibinfo {year}
  {2019})}\BibitemShut {NoStop}%
\bibitem [{\citenamefont {Chebrolu}\ \emph {et~al.}(2019)\citenamefont
  {Chebrolu}, \citenamefont {Chittari},\ and\ \citenamefont
  {Jung}}]{Chebrolu2019}%
  \BibitemOpen
  \bibfield  {author} {\bibinfo {author} {\bibfnamefont {N.~R.}\ \bibnamefont
  {Chebrolu}}, \bibinfo {author} {\bibfnamefont {B.~L.}\ \bibnamefont
  {Chittari}},\ and\ \bibinfo {author} {\bibfnamefont {J.}~\bibnamefont
  {Jung}},\ }\href {https://doi.org/10.1103/physrevb.99.235417} {\bibfield
  {journal} {\bibinfo  {journal} {Physical Review B}\ }\textbf {\bibinfo
  {volume} {99}},\ \bibinfo {pages} {235417} (\bibinfo {year}
  {2019})}\BibitemShut {NoStop}%
\bibitem [{\citenamefont {Lee}\ \emph {et~al.}(2019)\citenamefont {Lee},
  \citenamefont {Khalaf}, \citenamefont {Liu}, \citenamefont {Liu},
  \citenamefont {Hao}, \citenamefont {Kim},\ and\ \citenamefont
  {Vishwanath}}]{Lee2019}%
  \BibitemOpen
  \bibfield  {author} {\bibinfo {author} {\bibfnamefont {J.~Y.}\ \bibnamefont
  {Lee}}, \bibinfo {author} {\bibfnamefont {E.}~\bibnamefont {Khalaf}},
  \bibinfo {author} {\bibfnamefont {S.}~\bibnamefont {Liu}}, \bibinfo {author}
  {\bibfnamefont {X.}~\bibnamefont {Liu}}, \bibinfo {author} {\bibfnamefont
  {Z.}~\bibnamefont {Hao}}, \bibinfo {author} {\bibfnamefont {P.}~\bibnamefont
  {Kim}},\ and\ \bibinfo {author} {\bibfnamefont {A.}~\bibnamefont
  {Vishwanath}},\ }\href {https://doi.org/10.1038/s41467-019-12981-1}
  {\bibfield  {journal} {\bibinfo  {journal} {Nature Communications}\ }\textbf
  {\bibinfo {volume} {10}},\ \bibinfo {pages} {5333} (\bibinfo {year}
  {2019})}\BibitemShut {NoStop}%
\bibitem [{\citenamefont {Crosse}\ \emph {et~al.}(2020)\citenamefont {Crosse},
  \citenamefont {Nakatsuji}, \citenamefont {Koshino},\ and\ \citenamefont
  {Moon}}]{Crosse2020}%
  \BibitemOpen
  \bibfield  {author} {\bibinfo {author} {\bibfnamefont {J.~A.}\ \bibnamefont
  {Crosse}}, \bibinfo {author} {\bibfnamefont {N.}~\bibnamefont {Nakatsuji}},
  \bibinfo {author} {\bibfnamefont {M.}~\bibnamefont {Koshino}},\ and\ \bibinfo
  {author} {\bibfnamefont {P.}~\bibnamefont {Moon}},\ }\href
  {https://doi.org/10.1103/physrevb.102.035421} {\bibfield  {journal} {\bibinfo
   {journal} {Physical Review B}\ }\textbf {\bibinfo {volume} {102}},\ \bibinfo
  {pages} {035421} (\bibinfo {year} {2020})}\BibitemShut {NoStop}%
\bibitem [{\citenamefont {Wang}\ \emph {et~al.}(2021)\citenamefont {Wang},
  \citenamefont {Li},\ and\ \citenamefont {Zhang}}]{Wang2021}%
  \BibitemOpen
  \bibfield  {author} {\bibinfo {author} {\bibfnamefont {Y.-X.}\ \bibnamefont
  {Wang}}, \bibinfo {author} {\bibfnamefont {F.}~\bibnamefont {Li}},\ and\
  \bibinfo {author} {\bibfnamefont {Z.-Y.}\ \bibnamefont {Zhang}},\ }\href
  {https://doi.org/10.1103/physrevb.103.115201} {\bibfield  {journal} {\bibinfo
   {journal} {Physical Review B}\ }\textbf {\bibinfo {volume} {103}},\ \bibinfo
  {pages} {115201} (\bibinfo {year} {2021})}\BibitemShut {NoStop}%
\bibitem [{\citenamefont {Burg}\ \emph {et~al.}(2020)\citenamefont {Burg},
  \citenamefont {Lian}, \citenamefont {Taniguchi}, \citenamefont {Watanabe},
  \citenamefont {Bernevig},\ and\ \citenamefont {Tutuc}}]{Burg2020}%
  \BibitemOpen
  \bibfield  {author} {\bibinfo {author} {\bibfnamefont {G.~W.}\ \bibnamefont
  {Burg}}, \bibinfo {author} {\bibfnamefont {B.}~\bibnamefont {Lian}}, \bibinfo
  {author} {\bibfnamefont {T.}~\bibnamefont {Taniguchi}}, \bibinfo {author}
  {\bibfnamefont {K.}~\bibnamefont {Watanabe}}, \bibinfo {author}
  {\bibfnamefont {B.~A.}\ \bibnamefont {Bernevig}},\ and\ \bibinfo {author}
  {\bibfnamefont {E.}~\bibnamefont {Tutuc}},\ }\href@noop {} {\bibinfo {title}
  {Evidence of emergent symmetry and valley chern number in twisted
  double-bilayer graphene}} (\bibinfo {year} {2020}),\ \bibinfo {note}
  {2006.14000. arXiv. \url{https://arxiv.org/abs/2006.14000} (date accessed:
  09/10/2021)},\ \Eprint {https://arxiv.org/abs/2006.14000} {arXiv:2006.14000
  [cond-mat.mes-hall]} \BibitemShut {NoStop}%
\bibitem [{\citenamefont {Ma}\ \emph {et~al.}(2020)\citenamefont {Ma},
  \citenamefont {Wang}, \citenamefont {Mills}, \citenamefont {Chen},
  \citenamefont {Deng}, \citenamefont {Yuan}, \citenamefont {Li}, \citenamefont
  {Watanabe}, \citenamefont {Taniguchi}, \citenamefont {Du}, \citenamefont
  {Zhang},\ and\ \citenamefont {Xia}}]{Ma2020}%
  \BibitemOpen
  \bibfield  {author} {\bibinfo {author} {\bibfnamefont {C.}~\bibnamefont
  {Ma}}, \bibinfo {author} {\bibfnamefont {Q.}~\bibnamefont {Wang}}, \bibinfo
  {author} {\bibfnamefont {S.}~\bibnamefont {Mills}}, \bibinfo {author}
  {\bibfnamefont {X.}~\bibnamefont {Chen}}, \bibinfo {author} {\bibfnamefont
  {B.}~\bibnamefont {Deng}}, \bibinfo {author} {\bibfnamefont {S.}~\bibnamefont
  {Yuan}}, \bibinfo {author} {\bibfnamefont {C.}~\bibnamefont {Li}}, \bibinfo
  {author} {\bibfnamefont {K.}~\bibnamefont {Watanabe}}, \bibinfo {author}
  {\bibfnamefont {T.}~\bibnamefont {Taniguchi}}, \bibinfo {author}
  {\bibfnamefont {X.}~\bibnamefont {Du}}, \bibinfo {author} {\bibfnamefont
  {F.}~\bibnamefont {Zhang}},\ and\ \bibinfo {author} {\bibfnamefont
  {F.}~\bibnamefont {Xia}},\ }\href
  {https://doi.org/10.1021/acs.nanolett.0c02131} {\bibfield  {journal}
  {\bibinfo  {journal} {Nano Letters}\ }\textbf {\bibinfo {volume} {20}},\
  \bibinfo {pages} {6076} (\bibinfo {year} {2020})}\BibitemShut {NoStop}%
\bibitem [{\citenamefont {Burg}\ \emph {et~al.}(2019)\citenamefont {Burg},
  \citenamefont {Zhu}, \citenamefont {Taniguchi}, \citenamefont {Watanabe},
  \citenamefont {MacDonald},\ and\ \citenamefont {Tutuc}}]{Burg2019}%
  \BibitemOpen
  \bibfield  {author} {\bibinfo {author} {\bibfnamefont {G.~W.}\ \bibnamefont
  {Burg}}, \bibinfo {author} {\bibfnamefont {J.}~\bibnamefont {Zhu}}, \bibinfo
  {author} {\bibfnamefont {T.}~\bibnamefont {Taniguchi}}, \bibinfo {author}
  {\bibfnamefont {K.}~\bibnamefont {Watanabe}}, \bibinfo {author}
  {\bibfnamefont {A.~H.}\ \bibnamefont {MacDonald}},\ and\ \bibinfo {author}
  {\bibfnamefont {E.}~\bibnamefont {Tutuc}},\ }\href
  {https://doi.org/10.1103/physrevlett.123.197702} {\bibfield  {journal}
  {\bibinfo  {journal} {Physical Review Letters}\ }\textbf {\bibinfo {volume}
  {123}},\ \bibinfo {pages} {197702} (\bibinfo {year} {2019})}\BibitemShut
  {NoStop}%
\bibitem [{\citenamefont {Cao}\ \emph {et~al.}(2020{\natexlab{b}})\citenamefont
  {Cao}, \citenamefont {Rodan-Legrain}, \citenamefont {Rubies-Bigorda},
  \citenamefont {Park}, \citenamefont {Watanabe}, \citenamefont {Taniguchi},\
  and\ \citenamefont {Jarillo-Herrero}}]{Cao2020}%
  \BibitemOpen
  \bibfield  {author} {\bibinfo {author} {\bibfnamefont {Y.}~\bibnamefont
  {Cao}}, \bibinfo {author} {\bibfnamefont {D.}~\bibnamefont {Rodan-Legrain}},
  \bibinfo {author} {\bibfnamefont {O.}~\bibnamefont {Rubies-Bigorda}},
  \bibinfo {author} {\bibfnamefont {J.~M.}\ \bibnamefont {Park}}, \bibinfo
  {author} {\bibfnamefont {K.}~\bibnamefont {Watanabe}}, \bibinfo {author}
  {\bibfnamefont {T.}~\bibnamefont {Taniguchi}},\ and\ \bibinfo {author}
  {\bibfnamefont {P.}~\bibnamefont {Jarillo-Herrero}},\ }\href
  {https://doi.org/10.1038/s41586-020-2260-6} {\bibfield  {journal} {\bibinfo
  {journal} {Nature}\ }\textbf {\bibinfo {volume} {583}},\ \bibinfo {pages}
  {215} (\bibinfo {year} {2020}{\natexlab{b}})}\BibitemShut {NoStop}%
\bibitem [{\citenamefont {He}\ \emph {et~al.}(2020)\citenamefont {He},
  \citenamefont {Li}, \citenamefont {Cai}, \citenamefont {Liu}, \citenamefont
  {Watanabe}, \citenamefont {Taniguchi}, \citenamefont {Xu},\ and\
  \citenamefont {Yankowitz}}]{He2020}%
  \BibitemOpen
  \bibfield  {author} {\bibinfo {author} {\bibfnamefont {M.}~\bibnamefont
  {He}}, \bibinfo {author} {\bibfnamefont {Y.}~\bibnamefont {Li}}, \bibinfo
  {author} {\bibfnamefont {J.}~\bibnamefont {Cai}}, \bibinfo {author}
  {\bibfnamefont {Y.}~\bibnamefont {Liu}}, \bibinfo {author} {\bibfnamefont
  {K.}~\bibnamefont {Watanabe}}, \bibinfo {author} {\bibfnamefont
  {T.}~\bibnamefont {Taniguchi}}, \bibinfo {author} {\bibfnamefont
  {X.}~\bibnamefont {Xu}},\ and\ \bibinfo {author} {\bibfnamefont
  {M.}~\bibnamefont {Yankowitz}},\ }\href
  {https://doi.org/10.1038/s41567-020-1030-6} {\bibfield  {journal} {\bibinfo
  {journal} {Nature Physics}\ }\textbf {\bibinfo {volume} {17}},\ \bibinfo
  {pages} {26} (\bibinfo {year} {2020})}\BibitemShut {NoStop}%
\bibitem [{\citenamefont {Rickhaus}\ \emph {et~al.}(2021)\citenamefont
  {Rickhaus}, \citenamefont {de~Vries}, \citenamefont {Zhu}, \citenamefont
  {Portoles}, \citenamefont {Zheng}, \citenamefont {Masseroni}, \citenamefont
  {Kurzmann}, \citenamefont {Taniguchi}, \citenamefont {Watanabe},
  \citenamefont {MacDonald}, \citenamefont {Ihn},\ and\ \citenamefont
  {Ensslin}}]{Rickhaus2020}%
  \BibitemOpen
  \bibfield  {author} {\bibinfo {author} {\bibfnamefont {P.}~\bibnamefont
  {Rickhaus}}, \bibinfo {author} {\bibfnamefont {F.~K.}\ \bibnamefont
  {de~Vries}}, \bibinfo {author} {\bibfnamefont {J.}~\bibnamefont {Zhu}},
  \bibinfo {author} {\bibfnamefont {E.}~\bibnamefont {Portoles}}, \bibinfo
  {author} {\bibfnamefont {G.}~\bibnamefont {Zheng}}, \bibinfo {author}
  {\bibfnamefont {M.}~\bibnamefont {Masseroni}}, \bibinfo {author}
  {\bibfnamefont {A.}~\bibnamefont {Kurzmann}}, \bibinfo {author}
  {\bibfnamefont {T.}~\bibnamefont {Taniguchi}}, \bibinfo {author}
  {\bibfnamefont {K.}~\bibnamefont {Watanabe}}, \bibinfo {author}
  {\bibfnamefont {A.~H.}\ \bibnamefont {MacDonald}}, \bibinfo {author}
  {\bibfnamefont {T.}~\bibnamefont {Ihn}},\ and\ \bibinfo {author}
  {\bibfnamefont {K.}~\bibnamefont {Ensslin}},\ }\href
  {https://doi.org/10.1126/science.abc3534} {\bibfield  {journal} {\bibinfo
  {journal} {Science}\ }\textbf {\bibinfo {volume} {373}},\ \bibinfo {pages}
  {1257} (\bibinfo {year} {2021})}\BibitemShut {NoStop}%
\bibitem [{\citenamefont {Liu}\ \emph {et~al.}(2020)\citenamefont {Liu},
  \citenamefont {Hao}, \citenamefont {Khalaf}, \citenamefont {Lee},
  \citenamefont {Ronen}, \citenamefont {Yoo}, \citenamefont {Najafabadi},
  \citenamefont {Watanabe}, \citenamefont {Taniguchi}, \citenamefont
  {Vishwanath},\ and\ \citenamefont {Kim}}]{Liu2020}%
  \BibitemOpen
  \bibfield  {author} {\bibinfo {author} {\bibfnamefont {X.}~\bibnamefont
  {Liu}}, \bibinfo {author} {\bibfnamefont {Z.}~\bibnamefont {Hao}}, \bibinfo
  {author} {\bibfnamefont {E.}~\bibnamefont {Khalaf}}, \bibinfo {author}
  {\bibfnamefont {J.~Y.}\ \bibnamefont {Lee}}, \bibinfo {author} {\bibfnamefont
  {Y.}~\bibnamefont {Ronen}}, \bibinfo {author} {\bibfnamefont
  {H.}~\bibnamefont {Yoo}}, \bibinfo {author} {\bibfnamefont {D.~H.}\
  \bibnamefont {Najafabadi}}, \bibinfo {author} {\bibfnamefont
  {K.}~\bibnamefont {Watanabe}}, \bibinfo {author} {\bibfnamefont
  {T.}~\bibnamefont {Taniguchi}}, \bibinfo {author} {\bibfnamefont
  {A.}~\bibnamefont {Vishwanath}},\ and\ \bibinfo {author} {\bibfnamefont
  {P.}~\bibnamefont {Kim}},\ }\href {https://doi.org/10.1038/s41586-020-2458-7}
  {\bibfield  {journal} {\bibinfo  {journal} {Nature}\ }\textbf {\bibinfo
  {volume} {583}},\ \bibinfo {pages} {221} (\bibinfo {year}
  {2020})}\BibitemShut {NoStop}%
\bibitem [{\citenamefont {Shen}\ \emph {et~al.}(2020)\citenamefont {Shen},
  \citenamefont {Chu}, \citenamefont {Wu}, \citenamefont {Li}, \citenamefont
  {Wang}, \citenamefont {Zhao}, \citenamefont {Tang}, \citenamefont {Liu},
  \citenamefont {Tian}, \citenamefont {Watanabe}, \citenamefont {Taniguchi},
  \citenamefont {Yang}, \citenamefont {Meng}, \citenamefont {Shi},
  \citenamefont {Yazyev},\ and\ \citenamefont {Zhang}}]{Shen2020}%
  \BibitemOpen
  \bibfield  {author} {\bibinfo {author} {\bibfnamefont {C.}~\bibnamefont
  {Shen}}, \bibinfo {author} {\bibfnamefont {Y.}~\bibnamefont {Chu}}, \bibinfo
  {author} {\bibfnamefont {Q.}~\bibnamefont {Wu}}, \bibinfo {author}
  {\bibfnamefont {N.}~\bibnamefont {Li}}, \bibinfo {author} {\bibfnamefont
  {S.}~\bibnamefont {Wang}}, \bibinfo {author} {\bibfnamefont {Y.}~\bibnamefont
  {Zhao}}, \bibinfo {author} {\bibfnamefont {J.}~\bibnamefont {Tang}}, \bibinfo
  {author} {\bibfnamefont {J.}~\bibnamefont {Liu}}, \bibinfo {author}
  {\bibfnamefont {J.}~\bibnamefont {Tian}}, \bibinfo {author} {\bibfnamefont
  {K.}~\bibnamefont {Watanabe}}, \bibinfo {author} {\bibfnamefont
  {T.}~\bibnamefont {Taniguchi}}, \bibinfo {author} {\bibfnamefont
  {R.}~\bibnamefont {Yang}}, \bibinfo {author} {\bibfnamefont {Z.~Y.}\
  \bibnamefont {Meng}}, \bibinfo {author} {\bibfnamefont {D.}~\bibnamefont
  {Shi}}, \bibinfo {author} {\bibfnamefont {O.~V.}\ \bibnamefont {Yazyev}},\
  and\ \bibinfo {author} {\bibfnamefont {G.}~\bibnamefont {Zhang}},\ }\href
  {https://doi.org/10.1038/s41567-020-0825-9} {\bibfield  {journal} {\bibinfo
  {journal} {Nature Physics}\ }\textbf {\bibinfo {volume} {16}},\ \bibinfo
  {pages} {520} (\bibinfo {year} {2020})}\BibitemShut {NoStop}%
\bibitem [{\citenamefont {Lin}\ \emph {et~al.}(2020)\citenamefont {Lin},
  \citenamefont {Zhu},\ and\ \citenamefont {Ni}}]{Lin2020}%
  \BibitemOpen
  \bibfield  {author} {\bibinfo {author} {\bibfnamefont {X.}~\bibnamefont
  {Lin}}, \bibinfo {author} {\bibfnamefont {H.}~\bibnamefont {Zhu}},\ and\
  \bibinfo {author} {\bibfnamefont {J.}~\bibnamefont {Ni}},\ }\href
  {https://doi.org/10.1103/physrevb.101.155405} {\bibfield  {journal} {\bibinfo
   {journal} {Physical Review B}\ }\textbf {\bibinfo {volume} {101}},\ \bibinfo
  {pages} {155405} (\bibinfo {year} {2020})}\BibitemShut {NoStop}%
\bibitem [{\citenamefont {Haddadi}\ \emph {et~al.}(2020)\citenamefont
  {Haddadi}, \citenamefont {Wu}, \citenamefont {Kruchkov},\ and\ \citenamefont
  {Yazyev}}]{Haddadi2020}%
  \BibitemOpen
  \bibfield  {author} {\bibinfo {author} {\bibfnamefont {F.}~\bibnamefont
  {Haddadi}}, \bibinfo {author} {\bibfnamefont {Q.}~\bibnamefont {Wu}},
  \bibinfo {author} {\bibfnamefont {A.~J.}\ \bibnamefont {Kruchkov}},\ and\
  \bibinfo {author} {\bibfnamefont {O.~V.}\ \bibnamefont {Yazyev}},\ }\href
  {https://doi.org/10.1021/acs.nanolett.9b05117} {\bibfield  {journal}
  {\bibinfo  {journal} {Nano Letters}\ }\textbf {\bibinfo {volume} {20}},\
  \bibinfo {pages} {2410} (\bibinfo {year} {2020})}\BibitemShut {NoStop}%
\bibitem [{\citenamefont {Yankowitz}\ \emph {et~al.}(2018)\citenamefont
  {Yankowitz}, \citenamefont {Jung}, \citenamefont {Laksono}, \citenamefont
  {Leconte}, \citenamefont {Chittari}, \citenamefont {Watanabe}, \citenamefont
  {Taniguchi}, \citenamefont {Adam}, \citenamefont {Graf},\ and\ \citenamefont
  {Dean}}]{Yankowitz2018}%
  \BibitemOpen
  \bibfield  {author} {\bibinfo {author} {\bibfnamefont {M.}~\bibnamefont
  {Yankowitz}}, \bibinfo {author} {\bibfnamefont {J.}~\bibnamefont {Jung}},
  \bibinfo {author} {\bibfnamefont {E.}~\bibnamefont {Laksono}}, \bibinfo
  {author} {\bibfnamefont {N.}~\bibnamefont {Leconte}}, \bibinfo {author}
  {\bibfnamefont {B.~L.}\ \bibnamefont {Chittari}}, \bibinfo {author}
  {\bibfnamefont {K.}~\bibnamefont {Watanabe}}, \bibinfo {author}
  {\bibfnamefont {T.}~\bibnamefont {Taniguchi}}, \bibinfo {author}
  {\bibfnamefont {S.}~\bibnamefont {Adam}}, \bibinfo {author} {\bibfnamefont
  {D.}~\bibnamefont {Graf}},\ and\ \bibinfo {author} {\bibfnamefont {C.~R.}\
  \bibnamefont {Dean}},\ }\href {https://doi.org/10.1038/s41586-018-0107-1}
  {\bibfield  {journal} {\bibinfo  {journal} {Nature}\ }\textbf {\bibinfo
  {volume} {557}},\ \bibinfo {pages} {404} (\bibinfo {year}
  {2018})}\BibitemShut {NoStop}%
\bibitem [{\citenamefont {Sun}\ \emph {et~al.}(2018)\citenamefont {Sun},
  \citenamefont {Xiao}, \citenamefont {Lin}, \citenamefont {Zhang},
  \citenamefont {Ling}, \citenamefont {Ma}, \citenamefont {Luo}, \citenamefont
  {Lu}, \citenamefont {Sun},\ and\ \citenamefont {Sheng}}]{Sun2018}%
  \BibitemOpen
  \bibfield  {author} {\bibinfo {author} {\bibfnamefont {Y.}~\bibnamefont
  {Sun}}, \bibinfo {author} {\bibfnamefont {R.~C.}\ \bibnamefont {Xiao}},
  \bibinfo {author} {\bibfnamefont {G.~T.}\ \bibnamefont {Lin}}, \bibinfo
  {author} {\bibfnamefont {R.~R.}\ \bibnamefont {Zhang}}, \bibinfo {author}
  {\bibfnamefont {L.~S.}\ \bibnamefont {Ling}}, \bibinfo {author}
  {\bibfnamefont {Z.~W.}\ \bibnamefont {Ma}}, \bibinfo {author} {\bibfnamefont
  {X.}~\bibnamefont {Luo}}, \bibinfo {author} {\bibfnamefont {W.~J.}\
  \bibnamefont {Lu}}, \bibinfo {author} {\bibfnamefont {Y.~P.}\ \bibnamefont
  {Sun}},\ and\ \bibinfo {author} {\bibfnamefont {Z.~G.}\ \bibnamefont
  {Sheng}},\ }\href {https://doi.org/10.1063/1.5016568} {\bibfield  {journal}
  {\bibinfo  {journal} {Applied Physics Letters}\ }\textbf {\bibinfo {volume}
  {112}},\ \bibinfo {pages} {072409} (\bibinfo {year} {2018})}\BibitemShut
  {NoStop}%
\bibitem [{\citenamefont {Li}\ \emph {et~al.}(2019)\citenamefont {Li},
  \citenamefont {Jiang}, \citenamefont {Sivadas}, \citenamefont {Wang},
  \citenamefont {Xu}, \citenamefont {Weber}, \citenamefont {Goldberger},
  \citenamefont {Watanabe}, \citenamefont {Taniguchi}, \citenamefont {Fennie},
  \citenamefont {Mak},\ and\ \citenamefont {Shan}}]{Li2019a}%
  \BibitemOpen
  \bibfield  {author} {\bibinfo {author} {\bibfnamefont {T.}~\bibnamefont
  {Li}}, \bibinfo {author} {\bibfnamefont {S.}~\bibnamefont {Jiang}}, \bibinfo
  {author} {\bibfnamefont {N.}~\bibnamefont {Sivadas}}, \bibinfo {author}
  {\bibfnamefont {Z.}~\bibnamefont {Wang}}, \bibinfo {author} {\bibfnamefont
  {Y.}~\bibnamefont {Xu}}, \bibinfo {author} {\bibfnamefont {D.}~\bibnamefont
  {Weber}}, \bibinfo {author} {\bibfnamefont {J.~E.}\ \bibnamefont
  {Goldberger}}, \bibinfo {author} {\bibfnamefont {K.}~\bibnamefont
  {Watanabe}}, \bibinfo {author} {\bibfnamefont {T.}~\bibnamefont {Taniguchi}},
  \bibinfo {author} {\bibfnamefont {C.~J.}\ \bibnamefont {Fennie}}, \bibinfo
  {author} {\bibfnamefont {K.~F.}\ \bibnamefont {Mak}},\ and\ \bibinfo {author}
  {\bibfnamefont {J.}~\bibnamefont {Shan}},\ }\href
  {https://doi.org/10.1038/s41563-019-0506-1} {\bibfield  {journal} {\bibinfo
  {journal} {Nature Materials}\ }\textbf {\bibinfo {volume} {18}},\ \bibinfo
  {pages} {1303} (\bibinfo {year} {2019})}\BibitemShut {NoStop}%
\bibitem [{\citenamefont {Shao}\ \emph {et~al.}(2021)\citenamefont {Shao},
  \citenamefont {Liu}, \citenamefont {Zeng}, \citenamefont {Li}, \citenamefont
  {Wu}, \citenamefont {Ma}, \citenamefont {Jin}, \citenamefont {Lu},
  \citenamefont {Sun}, \citenamefont {Gu}, \citenamefont {Wang}, \citenamefont
  {Wu}, \citenamefont {Wu}, \citenamefont {Liu}, \citenamefont {Liu},\ and\
  \citenamefont {Zhao}}]{Shao2021}%
  \BibitemOpen
  \bibfield  {author} {\bibinfo {author} {\bibfnamefont {J.}~\bibnamefont
  {Shao}}, \bibinfo {author} {\bibfnamefont {Y.}~\bibnamefont {Liu}}, \bibinfo
  {author} {\bibfnamefont {M.}~\bibnamefont {Zeng}}, \bibinfo {author}
  {\bibfnamefont {J.}~\bibnamefont {Li}}, \bibinfo {author} {\bibfnamefont
  {X.}~\bibnamefont {Wu}}, \bibinfo {author} {\bibfnamefont {X.-M.}\
  \bibnamefont {Ma}}, \bibinfo {author} {\bibfnamefont {F.}~\bibnamefont
  {Jin}}, \bibinfo {author} {\bibfnamefont {R.}~\bibnamefont {Lu}}, \bibinfo
  {author} {\bibfnamefont {Y.}~\bibnamefont {Sun}}, \bibinfo {author}
  {\bibfnamefont {M.}~\bibnamefont {Gu}}, \bibinfo {author} {\bibfnamefont
  {K.}~\bibnamefont {Wang}}, \bibinfo {author} {\bibfnamefont {W.}~\bibnamefont
  {Wu}}, \bibinfo {author} {\bibfnamefont {L.}~\bibnamefont {Wu}}, \bibinfo
  {author} {\bibfnamefont {C.}~\bibnamefont {Liu}}, \bibinfo {author}
  {\bibfnamefont {Q.}~\bibnamefont {Liu}},\ and\ \bibinfo {author}
  {\bibfnamefont {Y.}~\bibnamefont {Zhao}},\ }\href
  {https://doi.org/10.1021/acs.nanolett.1c01874} {\bibfield  {journal}
  {\bibinfo  {journal} {Nano Letters}\ }\textbf {\bibinfo {volume} {21}},\
  \bibinfo {pages} {5874} (\bibinfo {year} {2021})}\BibitemShut {NoStop}%
\bibitem [{\citenamefont {Gmitra}\ \emph {et~al.}(2016)\citenamefont {Gmitra},
  \citenamefont {Kochan}, \citenamefont {Högl},\ and\ \citenamefont
  {Fabian}}]{Gmitra2016}%
  \BibitemOpen
  \bibfield  {author} {\bibinfo {author} {\bibfnamefont {M.}~\bibnamefont
  {Gmitra}}, \bibinfo {author} {\bibfnamefont {D.}~\bibnamefont {Kochan}},
  \bibinfo {author} {\bibfnamefont {P.}~\bibnamefont {Högl}},\ and\ \bibinfo
  {author} {\bibfnamefont {J.}~\bibnamefont {Fabian}},\ }\href
  {https://doi.org/10.1103/physrevb.93.155104} {\bibfield  {journal} {\bibinfo
  {journal} {Physical Review B}\ }\textbf {\bibinfo {volume} {93}},\ \bibinfo
  {pages} {155104} (\bibinfo {year} {2016})}\BibitemShut {NoStop}%
\bibitem [{\citenamefont {Fülöp}\ \emph
  {et~al.}(2021{\natexlab{a}})\citenamefont {Fülöp}, \citenamefont
  {M{\'{a}}rffy}, \citenamefont {Zihlmann}, \citenamefont {Gmitra},
  \citenamefont {T{\'{o}}v{\'{a}}ri}, \citenamefont {Szentp{\'{e}}teri},
  \citenamefont {Kedves}, \citenamefont {Watanabe}, \citenamefont {Taniguchi},
  \citenamefont {Fabian}, \citenamefont {Schönenberger}, \citenamefont
  {Makk},\ and\ \citenamefont {Csonka}}]{Fueloep2021}%
  \BibitemOpen
  \bibfield  {author} {\bibinfo {author} {\bibfnamefont {B.}~\bibnamefont
  {Fülöp}}, \bibinfo {author} {\bibfnamefont {A.}~\bibnamefont
  {M{\'{a}}rffy}}, \bibinfo {author} {\bibfnamefont {S.}~\bibnamefont
  {Zihlmann}}, \bibinfo {author} {\bibfnamefont {M.}~\bibnamefont {Gmitra}},
  \bibinfo {author} {\bibfnamefont {E.}~\bibnamefont {T{\'{o}}v{\'{a}}ri}},
  \bibinfo {author} {\bibfnamefont {B.}~\bibnamefont {Szentp{\'{e}}teri}},
  \bibinfo {author} {\bibfnamefont {M.}~\bibnamefont {Kedves}}, \bibinfo
  {author} {\bibfnamefont {K.}~\bibnamefont {Watanabe}}, \bibinfo {author}
  {\bibfnamefont {T.}~\bibnamefont {Taniguchi}}, \bibinfo {author}
  {\bibfnamefont {J.}~\bibnamefont {Fabian}}, \bibinfo {author} {\bibfnamefont
  {C.}~\bibnamefont {Schönenberger}}, \bibinfo {author} {\bibfnamefont
  {P.}~\bibnamefont {Makk}},\ and\ \bibinfo {author} {\bibfnamefont
  {S.}~\bibnamefont {Csonka}},\ }\bibfield  {journal} {\bibinfo  {journal} {npj
  2D Materials and Applications}\ }\textbf {\bibinfo {volume} {5}},\ \href
  {https://doi.org/10.1038/s41699-021-00262-9} {10.1038/s41699-021-00262-9}
  (\bibinfo {year} {2021}{\natexlab{a}}),\ \Eprint
  {https://arxiv.org/abs/2103.13325} {2103.13325} \BibitemShut {NoStop}%
\bibitem [{\citenamefont {Carr}\ \emph {et~al.}(2018)\citenamefont {Carr},
  \citenamefont {Fang}, \citenamefont {Jarillo-Herrero},\ and\ \citenamefont
  {Kaxiras}}]{Carr2018}%
  \BibitemOpen
  \bibfield  {author} {\bibinfo {author} {\bibfnamefont {S.}~\bibnamefont
  {Carr}}, \bibinfo {author} {\bibfnamefont {S.}~\bibnamefont {Fang}}, \bibinfo
  {author} {\bibfnamefont {P.}~\bibnamefont {Jarillo-Herrero}},\ and\ \bibinfo
  {author} {\bibfnamefont {E.}~\bibnamefont {Kaxiras}},\ }\href
  {https://doi.org/10.1103/physrevb.98.085144} {\bibfield  {journal} {\bibinfo
  {journal} {Physical Review B}\ }\textbf {\bibinfo {volume} {98}},\ \bibinfo
  {pages} {085144} (\bibinfo {year} {2018})}\BibitemShut {NoStop}%
\bibitem [{\citenamefont {Guinea}\ and\ \citenamefont
  {Walet}(2019)}]{Guinea2019}%
  \BibitemOpen
  \bibfield  {author} {\bibinfo {author} {\bibfnamefont {F.}~\bibnamefont
  {Guinea}}\ and\ \bibinfo {author} {\bibfnamefont {N.~R.}\ \bibnamefont
  {Walet}},\ }\href {https://doi.org/10.1103/physrevb.99.205134} {\bibfield
  {journal} {\bibinfo  {journal} {Physical Review B}\ }\textbf {\bibinfo
  {volume} {99}},\ \bibinfo {pages} {205134} (\bibinfo {year}
  {2019})}\BibitemShut {NoStop}%
\bibitem [{\citenamefont {Wang}\ \emph {et~al.}(2013)\citenamefont {Wang},
  \citenamefont {Meric}, \citenamefont {Huang}, \citenamefont {Gao},
  \citenamefont {Gao}, \citenamefont {Tran}, \citenamefont {Taniguchi},
  \citenamefont {Watanabe}, \citenamefont {Campos}, \citenamefont {Muller},
  \citenamefont {Guo}, \citenamefont {Kim}, \citenamefont {Hone}, \citenamefont
  {Shepard},\ and\ \citenamefont {Dean}}]{Wang2013}%
  \BibitemOpen
  \bibfield  {author} {\bibinfo {author} {\bibfnamefont {L.}~\bibnamefont
  {Wang}}, \bibinfo {author} {\bibfnamefont {I.}~\bibnamefont {Meric}},
  \bibinfo {author} {\bibfnamefont {P.~Y.}\ \bibnamefont {Huang}}, \bibinfo
  {author} {\bibfnamefont {Q.}~\bibnamefont {Gao}}, \bibinfo {author}
  {\bibfnamefont {Y.}~\bibnamefont {Gao}}, \bibinfo {author} {\bibfnamefont
  {H.}~\bibnamefont {Tran}}, \bibinfo {author} {\bibfnamefont {T.}~\bibnamefont
  {Taniguchi}}, \bibinfo {author} {\bibfnamefont {K.}~\bibnamefont {Watanabe}},
  \bibinfo {author} {\bibfnamefont {L.~M.}\ \bibnamefont {Campos}}, \bibinfo
  {author} {\bibfnamefont {D.~A.}\ \bibnamefont {Muller}}, \bibinfo {author}
  {\bibfnamefont {J.}~\bibnamefont {Guo}}, \bibinfo {author} {\bibfnamefont
  {P.}~\bibnamefont {Kim}}, \bibinfo {author} {\bibfnamefont {J.}~\bibnamefont
  {Hone}}, \bibinfo {author} {\bibfnamefont {K.~L.}\ \bibnamefont {Shepard}},\
  and\ \bibinfo {author} {\bibfnamefont {C.~R.}\ \bibnamefont {Dean}},\ }\href
  {https://doi.org/10.1126/science.1244358} {\bibfield  {journal} {\bibinfo
  {journal} {Science}\ }\textbf {\bibinfo {volume} {342}},\ \bibinfo {pages}
  {614} (\bibinfo {year} {2013})}\BibitemShut {NoStop}%
\bibitem [{\citenamefont {Zomer}\ \emph {et~al.}(2014)\citenamefont {Zomer},
  \citenamefont {Guimar{\~{a}}es}, \citenamefont {Brant}, \citenamefont
  {Tombros},\ and\ \citenamefont {van Wees}}]{Zomer2014}%
  \BibitemOpen
  \bibfield  {author} {\bibinfo {author} {\bibfnamefont {P.~J.}\ \bibnamefont
  {Zomer}}, \bibinfo {author} {\bibfnamefont {M.~H.~D.}\ \bibnamefont
  {Guimar{\~{a}}es}}, \bibinfo {author} {\bibfnamefont {J.~C.}\ \bibnamefont
  {Brant}}, \bibinfo {author} {\bibfnamefont {N.}~\bibnamefont {Tombros}},\
  and\ \bibinfo {author} {\bibfnamefont {B.~J.}\ \bibnamefont {van Wees}},\
  }\href {https://doi.org/10.1063/1.4886096} {\bibfield  {journal} {\bibinfo
  {journal} {Applied Physics Letters}\ }\textbf {\bibinfo {volume} {105}},\
  \bibinfo {pages} {013101} (\bibinfo {year} {2014})}\BibitemShut {NoStop}%
\bibitem [{\citenamefont {Kim}\ \emph {et~al.}(2016)\citenamefont {Kim},
  \citenamefont {Yankowitz}, \citenamefont {Fallahazad}, \citenamefont {Kang},
  \citenamefont {Movva}, \citenamefont {Huang}, \citenamefont {Larentis},
  \citenamefont {Corbet}, \citenamefont {Taniguchi}, \citenamefont {Watanabe},
  \citenamefont {Banerjee}, \citenamefont {LeRoy},\ and\ \citenamefont
  {Tutuc}}]{Kim2016a}%
  \BibitemOpen
  \bibfield  {author} {\bibinfo {author} {\bibfnamefont {K.}~\bibnamefont
  {Kim}}, \bibinfo {author} {\bibfnamefont {M.}~\bibnamefont {Yankowitz}},
  \bibinfo {author} {\bibfnamefont {B.}~\bibnamefont {Fallahazad}}, \bibinfo
  {author} {\bibfnamefont {S.}~\bibnamefont {Kang}}, \bibinfo {author}
  {\bibfnamefont {H.~C.~P.}\ \bibnamefont {Movva}}, \bibinfo {author}
  {\bibfnamefont {S.}~\bibnamefont {Huang}}, \bibinfo {author} {\bibfnamefont
  {S.}~\bibnamefont {Larentis}}, \bibinfo {author} {\bibfnamefont {C.~M.}\
  \bibnamefont {Corbet}}, \bibinfo {author} {\bibfnamefont {T.}~\bibnamefont
  {Taniguchi}}, \bibinfo {author} {\bibfnamefont {K.}~\bibnamefont {Watanabe}},
  \bibinfo {author} {\bibfnamefont {S.~K.}\ \bibnamefont {Banerjee}}, \bibinfo
  {author} {\bibfnamefont {B.~J.}\ \bibnamefont {LeRoy}},\ and\ \bibinfo
  {author} {\bibfnamefont {E.}~\bibnamefont {Tutuc}},\ }\href
  {https://doi.org/10.1021/acs.nanolett.5b05263} {\bibfield  {journal}
  {\bibinfo  {journal} {Nano Letters}\ }\textbf {\bibinfo {volume} {16}},\
  \bibinfo {pages} {1989} (\bibinfo {year} {2016})}\BibitemShut {NoStop}%
\bibitem [{\citenamefont {dos Santos}\ \emph {et~al.}(2012)\citenamefont {dos
  Santos}, \citenamefont {Peres},\ and\ \citenamefont {Neto}}]{Santos2012}%
  \BibitemOpen
  \bibfield  {author} {\bibinfo {author} {\bibfnamefont {J.~M. B.~L.}\
  \bibnamefont {dos Santos}}, \bibinfo {author} {\bibfnamefont {N.~M.~R.}\
  \bibnamefont {Peres}},\ and\ \bibinfo {author} {\bibfnamefont {A.~H.~C.}\
  \bibnamefont {Neto}},\ }\href {https://doi.org/10.1103/physrevb.86.155449}
  {\bibfield  {journal} {\bibinfo  {journal} {Physical Review B}\ }\textbf
  {\bibinfo {volume} {86}},\ \bibinfo {pages} {155449} (\bibinfo {year}
  {2012})}\BibitemShut {NoStop}%
\bibitem [{\citenamefont {de~Laissardi{\`{e}}re}\ \emph
  {et~al.}(2012)\citenamefont {de~Laissardi{\`{e}}re}, \citenamefont {Mayou},\
  and\ \citenamefont {Magaud}}]{Laissardiere2012}%
  \BibitemOpen
  \bibfield  {author} {\bibinfo {author} {\bibfnamefont {G.~T.}\ \bibnamefont
  {de~Laissardi{\`{e}}re}}, \bibinfo {author} {\bibfnamefont {D.}~\bibnamefont
  {Mayou}},\ and\ \bibinfo {author} {\bibfnamefont {L.}~\bibnamefont
  {Magaud}},\ }\href {https://doi.org/10.1103/physrevb.86.125413} {\bibfield
  {journal} {\bibinfo  {journal} {Physical Review B}\ }\textbf {\bibinfo
  {volume} {86}},\ \bibinfo {pages} {125413} (\bibinfo {year}
  {2012})}\BibitemShut {NoStop}%
\bibitem [{\citenamefont {Fang}\ and\ \citenamefont
  {Kaxiras}(2016)}]{Fang2016}%
  \BibitemOpen
  \bibfield  {author} {\bibinfo {author} {\bibfnamefont {S.}~\bibnamefont
  {Fang}}\ and\ \bibinfo {author} {\bibfnamefont {E.}~\bibnamefont {Kaxiras}},\
  }\href {https://doi.org/10.1103/physrevb.93.235153} {\bibfield  {journal}
  {\bibinfo  {journal} {Physical Review B}\ }\textbf {\bibinfo {volume} {93}},\
  \bibinfo {pages} {235153} (\bibinfo {year} {2016})}\BibitemShut {NoStop}%
\bibitem [{\citenamefont {Jung}\ and\ \citenamefont
  {MacDonald}(2014)}]{Jung2014}%
  \BibitemOpen
  \bibfield  {author} {\bibinfo {author} {\bibfnamefont {J.}~\bibnamefont
  {Jung}}\ and\ \bibinfo {author} {\bibfnamefont {A.~H.}\ \bibnamefont
  {MacDonald}},\ }\href {https://doi.org/10.1103/physrevb.89.035405} {\bibfield
   {journal} {\bibinfo  {journal} {Physical Review B}\ }\textbf {\bibinfo
  {volume} {89}},\ \bibinfo {pages} {035405} (\bibinfo {year}
  {2014})}\BibitemShut {NoStop}%
\bibitem [{\citenamefont {Chittari}\ \emph {et~al.}(2018)\citenamefont
  {Chittari}, \citenamefont {Leconte}, \citenamefont {Javvaji},\ and\
  \citenamefont {Jung}}]{Chittari2018}%
  \BibitemOpen
  \bibfield  {author} {\bibinfo {author} {\bibfnamefont {B.~L.}\ \bibnamefont
  {Chittari}}, \bibinfo {author} {\bibfnamefont {N.}~\bibnamefont {Leconte}},
  \bibinfo {author} {\bibfnamefont {S.}~\bibnamefont {Javvaji}},\ and\ \bibinfo
  {author} {\bibfnamefont {J.}~\bibnamefont {Jung}},\ }\href
  {https://doi.org/10.1088/2516-1075/aaead3} {\bibfield  {journal} {\bibinfo
  {journal} {Electronic Structure}\ }\textbf {\bibinfo {volume} {1}},\ \bibinfo
  {pages} {015001} (\bibinfo {year} {2018})}\BibitemShut {NoStop}%
\bibitem [{\citenamefont {Hwang}\ \emph {et~al.}(2012)\citenamefont {Hwang},
  \citenamefont {Siegel}, \citenamefont {Mo}, \citenamefont {Regan},
  \citenamefont {Ismach}, \citenamefont {Zhang}, \citenamefont {Zettl},\ and\
  \citenamefont {Lanzara}}]{Hwang2012}%
  \BibitemOpen
  \bibfield  {author} {\bibinfo {author} {\bibfnamefont {C.}~\bibnamefont
  {Hwang}}, \bibinfo {author} {\bibfnamefont {D.~A.}\ \bibnamefont {Siegel}},
  \bibinfo {author} {\bibfnamefont {S.-K.}\ \bibnamefont {Mo}}, \bibinfo
  {author} {\bibfnamefont {W.}~\bibnamefont {Regan}}, \bibinfo {author}
  {\bibfnamefont {A.}~\bibnamefont {Ismach}}, \bibinfo {author} {\bibfnamefont
  {Y.}~\bibnamefont {Zhang}}, \bibinfo {author} {\bibfnamefont
  {A.}~\bibnamefont {Zettl}},\ and\ \bibinfo {author} {\bibfnamefont
  {A.}~\bibnamefont {Lanzara}},\ }\href {https://doi.org/10.1038/srep00590}
  {\bibfield  {journal} {\bibinfo  {journal} {Scientific Reports}\ }\textbf
  {\bibinfo {volume} {2}},\ \bibinfo {pages} {590} (\bibinfo {year}
  {2012})}\BibitemShut {NoStop}%
\bibitem [{\citenamefont {Bessler}\ \emph {et~al.}(2019)\citenamefont
  {Bessler}, \citenamefont {Duerig},\ and\ \citenamefont
  {Koren}}]{Bessler2019}%
  \BibitemOpen
  \bibfield  {author} {\bibinfo {author} {\bibfnamefont {R.}~\bibnamefont
  {Bessler}}, \bibinfo {author} {\bibfnamefont {U.}~\bibnamefont {Duerig}},\
  and\ \bibinfo {author} {\bibfnamefont {E.}~\bibnamefont {Koren}},\ }\href
  {https://doi.org/10.1039/c8na00350e} {\bibfield  {journal} {\bibinfo
  {journal} {Nanoscale Advances}\ }\textbf {\bibinfo {volume} {1}},\ \bibinfo
  {pages} {1702} (\bibinfo {year} {2019})}\BibitemShut {NoStop}%
\bibitem [{\citenamefont {Rickhaus}\ \emph {et~al.}(2019)\citenamefont
  {Rickhaus}, \citenamefont {Zheng}, \citenamefont {Lado}, \citenamefont {Lee},
  \citenamefont {Kurzmann}, \citenamefont {Eich}, \citenamefont {Pisoni},
  \citenamefont {Tong}, \citenamefont {Garreis}, \citenamefont {Gold},
  \citenamefont {Masseroni}, \citenamefont {Taniguchi}, \citenamefont
  {Wantanabe}, \citenamefont {Ihn},\ and\ \citenamefont
  {Ensslin}}]{Rickhaus2019}%
  \BibitemOpen
  \bibfield  {author} {\bibinfo {author} {\bibfnamefont {P.}~\bibnamefont
  {Rickhaus}}, \bibinfo {author} {\bibfnamefont {G.}~\bibnamefont {Zheng}},
  \bibinfo {author} {\bibfnamefont {J.~L.}\ \bibnamefont {Lado}}, \bibinfo
  {author} {\bibfnamefont {Y.}~\bibnamefont {Lee}}, \bibinfo {author}
  {\bibfnamefont {A.}~\bibnamefont {Kurzmann}}, \bibinfo {author}
  {\bibfnamefont {M.}~\bibnamefont {Eich}}, \bibinfo {author} {\bibfnamefont
  {R.}~\bibnamefont {Pisoni}}, \bibinfo {author} {\bibfnamefont
  {C.}~\bibnamefont {Tong}}, \bibinfo {author} {\bibfnamefont {R.}~\bibnamefont
  {Garreis}}, \bibinfo {author} {\bibfnamefont {C.}~\bibnamefont {Gold}},
  \bibinfo {author} {\bibfnamefont {M.}~\bibnamefont {Masseroni}}, \bibinfo
  {author} {\bibfnamefont {T.}~\bibnamefont {Taniguchi}}, \bibinfo {author}
  {\bibfnamefont {K.}~\bibnamefont {Wantanabe}}, \bibinfo {author}
  {\bibfnamefont {T.}~\bibnamefont {Ihn}},\ and\ \bibinfo {author}
  {\bibfnamefont {K.}~\bibnamefont {Ensslin}},\ }\href
  {https://doi.org/10.1021/acs.nanolett.9b03660} {\bibfield  {journal}
  {\bibinfo  {journal} {Nano Letters}\ }\textbf {\bibinfo {volume} {19}},\
  \bibinfo {pages} {8821} (\bibinfo {year} {2019})}\BibitemShut {NoStop}%
\bibitem [{\citenamefont {Zhang}\ \emph {et~al.}(2011)\citenamefont {Zhang},
  \citenamefont {Jung}, \citenamefont {Fiete}, \citenamefont {Niu},\ and\
  \citenamefont {MacDonald}}]{Zhang2011}%
  \BibitemOpen
  \bibfield  {author} {\bibinfo {author} {\bibfnamefont {F.}~\bibnamefont
  {Zhang}}, \bibinfo {author} {\bibfnamefont {J.}~\bibnamefont {Jung}},
  \bibinfo {author} {\bibfnamefont {G.~A.}\ \bibnamefont {Fiete}}, \bibinfo
  {author} {\bibfnamefont {Q.}~\bibnamefont {Niu}},\ and\ \bibinfo {author}
  {\bibfnamefont {A.~H.}\ \bibnamefont {MacDonald}},\ }\href
  {https://doi.org/10.1103/physrevlett.106.156801} {\bibfield  {journal}
  {\bibinfo  {journal} {Physical Review Letters}\ }\textbf {\bibinfo {volume}
  {106}},\ \bibinfo {pages} {156801} (\bibinfo {year} {2011})}\BibitemShut
  {NoStop}%
\bibitem [{\citenamefont {Fülöp}\ \emph
  {et~al.}(2021{\natexlab{b}})\citenamefont {Fülöp}, \citenamefont
  {M{\'{a}}rffy}, \citenamefont {T{\'{o}}v{\'{a}}ri}, \citenamefont {Kedves},
  \citenamefont {Zihlmann}, \citenamefont {Indolese}, \citenamefont
  {Kov{\'{a}}cs-Krausz}, \citenamefont {Watanabe}, \citenamefont {Taniguchi},
  \citenamefont {Schönenberger}, \citenamefont {K{\'{e}}zsm{\'{a}}rki},
  \citenamefont {Makk},\ and\ \citenamefont {Csonka}}]{Fueloep2021a}%
  \BibitemOpen
  \bibfield  {author} {\bibinfo {author} {\bibfnamefont {B.}~\bibnamefont
  {Fülöp}}, \bibinfo {author} {\bibfnamefont {A.}~\bibnamefont
  {M{\'{a}}rffy}}, \bibinfo {author} {\bibfnamefont {E.}~\bibnamefont
  {T{\'{o}}v{\'{a}}ri}}, \bibinfo {author} {\bibfnamefont {M.}~\bibnamefont
  {Kedves}}, \bibinfo {author} {\bibfnamefont {S.}~\bibnamefont {Zihlmann}},
  \bibinfo {author} {\bibfnamefont {D.}~\bibnamefont {Indolese}}, \bibinfo
  {author} {\bibfnamefont {Z.}~\bibnamefont {Kov{\'{a}}cs-Krausz}}, \bibinfo
  {author} {\bibfnamefont {K.}~\bibnamefont {Watanabe}}, \bibinfo {author}
  {\bibfnamefont {T.}~\bibnamefont {Taniguchi}}, \bibinfo {author}
  {\bibfnamefont {C.}~\bibnamefont {Schönenberger}}, \bibinfo {author}
  {\bibfnamefont {I.}~\bibnamefont {K{\'{e}}zsm{\'{a}}rki}}, \bibinfo {author}
  {\bibfnamefont {P.}~\bibnamefont {Makk}},\ and\ \bibinfo {author}
  {\bibfnamefont {S.}~\bibnamefont {Csonka}},\ }\href
  {https://doi.org/10.1063/5.0058583} {\bibfield  {journal} {\bibinfo
  {journal} {Journal of Applied Physics}\ }\textbf {\bibinfo {volume} {130}},\
  \bibinfo {pages} {064303} (\bibinfo {year} {2021}{\natexlab{b}})}\BibitemShut
  {NoStop}%
\end{thebibliography}%


\begin{thebibliography}{14}%
\makeatletter
\providecommand \@ifxundefined [1]{%
 \@ifx{#1\undefined}
}%
\providecommand \@ifnum [1]{%
 \ifnum #1\expandafter \@firstoftwo
 \else \expandafter \@secondoftwo
 \fi
}%
\providecommand \@ifx [1]{%
 \ifx #1\expandafter \@firstoftwo
 \else \expandafter \@secondoftwo
 \fi
}%
\providecommand \natexlab [1]{#1}%
\providecommand \enquote  [1]{``#1''}%
\providecommand \bibnamefont  [1]{#1}%
\providecommand \bibfnamefont [1]{#1}%
\providecommand \citenamefont [1]{#1}%
\providecommand \href@noop [0]{\@secondoftwo}%
\providecommand \href [0]{\begingroup \@sanitize@url \@href}%
\providecommand \@href[1]{\@@startlink{#1}\@@href}%
\providecommand \@@href[1]{\endgroup#1\@@endlink}%
\providecommand \@sanitize@url [0]{\catcode `\\12\catcode `\$12\catcode
  `\&12\catcode `\#12\catcode `\^12\catcode `\_12\catcode `\%12\relax}%
\providecommand \@@startlink[1]{}%
\providecommand \@@endlink[0]{}%
\providecommand \url  [0]{\begingroup\@sanitize@url \@url }%
\providecommand \@url [1]{\endgroup\@href {#1}{\urlprefix }}%
\providecommand \urlprefix  [0]{URL }%
\providecommand \Eprint [0]{\href }%
\providecommand \doibase [0]{https://doi.org/}%
\providecommand \selectlanguage [0]{\@gobble}%
\providecommand \bibinfo  [0]{\@secondoftwo}%
\providecommand \bibfield  [0]{\@secondoftwo}%
\providecommand \translation [1]{[#1]}%
\providecommand \BibitemOpen [0]{}%
\providecommand \bibitemStop [0]{}%
\providecommand \bibitemNoStop [0]{.\EOS\space}%
\providecommand \EOS [0]{\spacefactor3000\relax}%
\providecommand \BibitemShut  [1]{\csname bibitem#1\endcsname}%
\let\auto@bib@innerbib\@empty
\bibitem [{\citenamefont {Kim}\ \emph {et~al.}(2016)\citenamefont {Kim},
  \citenamefont {Yankowitz}, \citenamefont {Fallahazad}, \citenamefont {Kang},
  \citenamefont {Movva}, \citenamefont {Huang}, \citenamefont {Larentis},
  \citenamefont {Corbet}, \citenamefont {Taniguchi}, \citenamefont {Watanabe},
  \citenamefont {Banerjee}, \citenamefont {LeRoy},\ and\ \citenamefont
  {Tutuc}}]{Kim2016as}%
  \BibitemOpen
  \bibfield  {author} {\bibinfo {author} {\bibfnamefont {K.}~\bibnamefont
  {Kim}}, \bibinfo {author} {\bibfnamefont {M.}~\bibnamefont {Yankowitz}},
  \bibinfo {author} {\bibfnamefont {B.}~\bibnamefont {Fallahazad}}, \bibinfo
  {author} {\bibfnamefont {S.}~\bibnamefont {Kang}}, \bibinfo {author}
  {\bibfnamefont {H.~C.~P.}\ \bibnamefont {Movva}}, \bibinfo {author}
  {\bibfnamefont {S.}~\bibnamefont {Huang}}, \bibinfo {author} {\bibfnamefont
  {S.}~\bibnamefont {Larentis}}, \bibinfo {author} {\bibfnamefont {C.~M.}\
  \bibnamefont {Corbet}}, \bibinfo {author} {\bibfnamefont {T.}~\bibnamefont
  {Taniguchi}}, \bibinfo {author} {\bibfnamefont {K.}~\bibnamefont {Watanabe}},
  \bibinfo {author} {\bibfnamefont {S.~K.}\ \bibnamefont {Banerjee}}, \bibinfo
  {author} {\bibfnamefont {B.~J.}\ \bibnamefont {LeRoy}},\ and\ \bibinfo
  {author} {\bibfnamefont {E.}~\bibnamefont {Tutuc}},\ }\bibfield  {title}
  {\bibinfo {title} {van der waals heterostructures with high accuracy
  rotational alignment},\ }\href {https://doi.org/10.1021/acs.nanolett.5b05263}
  {\bibfield  {journal} {\bibinfo  {journal} {Nano Letters}\ }\textbf {\bibinfo
  {volume} {16}},\ \bibinfo {pages} {1989} (\bibinfo {year}
  {2016})}\BibitemShut {NoStop}%
\bibitem [{\citenamefont {Solozhenko}\ \emph {et~al.}(1995)\citenamefont
  {Solozhenko}, \citenamefont {Will},\ and\ \citenamefont
  {Elf}}]{Solozhenko1995s}%
  \BibitemOpen
  \bibfield  {author} {\bibinfo {author} {\bibfnamefont {V.}~\bibnamefont
  {Solozhenko}}, \bibinfo {author} {\bibfnamefont {G.}~\bibnamefont {Will}},\
  and\ \bibinfo {author} {\bibfnamefont {F.}~\bibnamefont {Elf}},\ }\bibfield
  {title} {\bibinfo {title} {Isothermal compression of hexagonal graphite-like
  boron nitride up to 12 {GPa}},\ }\href
  {https://doi.org/10.1016/0038-1098(95)00381-9} {\bibfield  {journal}
  {\bibinfo  {journal} {Solid State Communications}\ }\textbf {\bibinfo
  {volume} {96}},\ \bibinfo {pages} {1} (\bibinfo {year} {1995})}\BibitemShut
  {NoStop}%
\bibitem [{\citenamefont {Yankowitz}\ \emph {et~al.}(2018)\citenamefont
  {Yankowitz}, \citenamefont {Jung}, \citenamefont {Laksono}, \citenamefont
  {Leconte}, \citenamefont {Chittari}, \citenamefont {Watanabe}, \citenamefont
  {Taniguchi}, \citenamefont {Adam}, \citenamefont {Graf},\ and\ \citenamefont
  {Dean}}]{Yankowitz2018s}%
  \BibitemOpen
  \bibfield  {author} {\bibinfo {author} {\bibfnamefont {M.}~\bibnamefont
  {Yankowitz}}, \bibinfo {author} {\bibfnamefont {J.}~\bibnamefont {Jung}},
  \bibinfo {author} {\bibfnamefont {E.}~\bibnamefont {Laksono}}, \bibinfo
  {author} {\bibfnamefont {N.}~\bibnamefont {Leconte}}, \bibinfo {author}
  {\bibfnamefont {B.~L.}\ \bibnamefont {Chittari}}, \bibinfo {author}
  {\bibfnamefont {K.}~\bibnamefont {Watanabe}}, \bibinfo {author}
  {\bibfnamefont {T.}~\bibnamefont {Taniguchi}}, \bibinfo {author}
  {\bibfnamefont {S.}~\bibnamefont {Adam}}, \bibinfo {author} {\bibfnamefont
  {D.}~\bibnamefont {Graf}},\ and\ \bibinfo {author} {\bibfnamefont {C.~R.}\
  \bibnamefont {Dean}},\ }\bibfield  {title} {\bibinfo {title} {Dynamic
  band-structure tuning of graphene moir{\'{e}} superlattices with pressure},\
  }\href {https://doi.org/10.1038/s41586-018-0107-1} {\bibfield  {journal}
  {\bibinfo  {journal} {Nature}\ }\textbf {\bibinfo {volume} {557}},\ \bibinfo
  {pages} {404} (\bibinfo {year} {2018})}\BibitemShut {NoStop}%
\bibitem [{\citenamefont {Brown}(1964)}]{Brown1964s}%
  \BibitemOpen
  \bibfield  {author} {\bibinfo {author} {\bibfnamefont {E.}~\bibnamefont
  {Brown}},\ }\bibfield  {title} {\bibinfo {title} {Bloch electrons in a
  uniform magnetic field},\ }\href {https://doi.org/10.1103/physrev.133.a1038}
  {\bibfield  {journal} {\bibinfo  {journal} {Physical Review}\ }\textbf
  {\bibinfo {volume} {133}},\ \bibinfo {pages} {A1038} (\bibinfo {year}
  {1964})}\BibitemShut {NoStop}%
\bibitem [{\citenamefont {Zak}(1964)}]{Zak1964s}%
  \BibitemOpen
  \bibfield  {author} {\bibinfo {author} {\bibfnamefont {J.}~\bibnamefont
  {Zak}},\ }\bibfield  {title} {\bibinfo {title} {Magnetic translation group},\
  }\href {https://doi.org/10.1103/physrev.134.a1602} {\bibfield  {journal}
  {\bibinfo  {journal} {Physical Review}\ }\textbf {\bibinfo {volume} {134}},\
  \bibinfo {pages} {A1602} (\bibinfo {year} {1964})}\BibitemShut {NoStop}%
\bibitem [{\citenamefont {Bistritzer}\ and\ \citenamefont
  {MacDonald}(2011)}]{Bistritzer2011s}%
  \BibitemOpen
  \bibfield  {author} {\bibinfo {author} {\bibfnamefont {R.}~\bibnamefont
  {Bistritzer}}\ and\ \bibinfo {author} {\bibfnamefont {A.~H.}\ \bibnamefont
  {MacDonald}},\ }\bibfield  {title} {\bibinfo {title} {Moire bands in twisted
  double-layer graphene},\ }\href {https://doi.org/10.1073/pnas.1108174108}
  {\bibfield  {journal} {\bibinfo  {journal} {Proceedings of the National
  Academy of Sciences}\ }\textbf {\bibinfo {volume} {108}},\ \bibinfo {pages}
  {12233} (\bibinfo {year} {2011})}\BibitemShut {NoStop}%
\bibitem [{\citenamefont {Chebrolu}\ \emph {et~al.}(2019)\citenamefont
  {Chebrolu}, \citenamefont {Chittari},\ and\ \citenamefont
  {Jung}}]{Chebrolu2019s}%
  \BibitemOpen
  \bibfield  {author} {\bibinfo {author} {\bibfnamefont {N.~R.}\ \bibnamefont
  {Chebrolu}}, \bibinfo {author} {\bibfnamefont {B.~L.}\ \bibnamefont
  {Chittari}},\ and\ \bibinfo {author} {\bibfnamefont {J.}~\bibnamefont
  {Jung}},\ }\bibfield  {title} {\bibinfo {title} {Flat bands in twisted double
  bilayer graphene},\ }\href {https://doi.org/10.1103/physrevb.99.235417}
  {\bibfield  {journal} {\bibinfo  {journal} {Physical Review B}\ }\textbf
  {\bibinfo {volume} {99}},\ \bibinfo {pages} {235417} (\bibinfo {year}
  {2019})}\BibitemShut {NoStop}%
\bibitem [{\citenamefont {Jung}\ and\ \citenamefont
  {MacDonald}(2014)}]{Jung2014s}%
  \BibitemOpen
  \bibfield  {author} {\bibinfo {author} {\bibfnamefont {J.}~\bibnamefont
  {Jung}}\ and\ \bibinfo {author} {\bibfnamefont {A.~H.}\ \bibnamefont
  {MacDonald}},\ }\bibfield  {title} {\bibinfo {title} {Accurate tight-binding
  models for the$\pi$bands of bilayer graphene},\ }\href
  {https://doi.org/10.1103/physrevb.89.035405} {\bibfield  {journal} {\bibinfo
  {journal} {Physical Review B}\ }\textbf {\bibinfo {volume} {89}},\ \bibinfo
  {pages} {035405} (\bibinfo {year} {2014})}\BibitemShut {NoStop}%
\bibitem [{\citenamefont {Burg}\ \emph {et~al.}(2019)\citenamefont {Burg},
  \citenamefont {Zhu}, \citenamefont {Taniguchi}, \citenamefont {Watanabe},
  \citenamefont {MacDonald},\ and\ \citenamefont {Tutuc}}]{Burg2019s}%
  \BibitemOpen
  \bibfield  {author} {\bibinfo {author} {\bibfnamefont {G.~W.}\ \bibnamefont
  {Burg}}, \bibinfo {author} {\bibfnamefont {J.}~\bibnamefont {Zhu}}, \bibinfo
  {author} {\bibfnamefont {T.}~\bibnamefont {Taniguchi}}, \bibinfo {author}
  {\bibfnamefont {K.}~\bibnamefont {Watanabe}}, \bibinfo {author}
  {\bibfnamefont {A.~H.}\ \bibnamefont {MacDonald}},\ and\ \bibinfo {author}
  {\bibfnamefont {E.}~\bibnamefont {Tutuc}},\ }\bibfield  {title} {\bibinfo
  {title} {Correlated insulating states in twisted double bilayer graphene},\
  }\href {https://doi.org/10.1103/physrevlett.123.197702} {\bibfield  {journal}
  {\bibinfo  {journal} {Physical Review Letters}\ }\textbf {\bibinfo {volume}
  {123}},\ \bibinfo {pages} {197702} (\bibinfo {year} {2019})}\BibitemShut
  {NoStop}%
\bibitem [{\citenamefont {Shen}\ \emph {et~al.}(2020)\citenamefont {Shen},
  \citenamefont {Chu}, \citenamefont {Wu}, \citenamefont {Li}, \citenamefont
  {Wang}, \citenamefont {Zhao}, \citenamefont {Tang}, \citenamefont {Liu},
  \citenamefont {Tian}, \citenamefont {Watanabe}, \citenamefont {Taniguchi},
  \citenamefont {Yang}, \citenamefont {Meng}, \citenamefont {Shi},
  \citenamefont {Yazyev},\ and\ \citenamefont {Zhang}}]{Shen2020s}%
  \BibitemOpen
  \bibfield  {author} {\bibinfo {author} {\bibfnamefont {C.}~\bibnamefont
  {Shen}}, \bibinfo {author} {\bibfnamefont {Y.}~\bibnamefont {Chu}}, \bibinfo
  {author} {\bibfnamefont {Q.}~\bibnamefont {Wu}}, \bibinfo {author}
  {\bibfnamefont {N.}~\bibnamefont {Li}}, \bibinfo {author} {\bibfnamefont
  {S.}~\bibnamefont {Wang}}, \bibinfo {author} {\bibfnamefont {Y.}~\bibnamefont
  {Zhao}}, \bibinfo {author} {\bibfnamefont {J.}~\bibnamefont {Tang}}, \bibinfo
  {author} {\bibfnamefont {J.}~\bibnamefont {Liu}}, \bibinfo {author}
  {\bibfnamefont {J.}~\bibnamefont {Tian}}, \bibinfo {author} {\bibfnamefont
  {K.}~\bibnamefont {Watanabe}}, \bibinfo {author} {\bibfnamefont
  {T.}~\bibnamefont {Taniguchi}}, \bibinfo {author} {\bibfnamefont
  {R.}~\bibnamefont {Yang}}, \bibinfo {author} {\bibfnamefont {Z.~Y.}\
  \bibnamefont {Meng}}, \bibinfo {author} {\bibfnamefont {D.}~\bibnamefont
  {Shi}}, \bibinfo {author} {\bibfnamefont {O.~V.}\ \bibnamefont {Yazyev}},\
  and\ \bibinfo {author} {\bibfnamefont {G.}~\bibnamefont {Zhang}},\ }\bibfield
   {title} {\bibinfo {title} {Correlated states in twisted double bilayer
  graphene},\ }\href {https://doi.org/10.1038/s41567-020-0825-9} {\bibfield
  {journal} {\bibinfo  {journal} {Nature Physics}\ }\textbf {\bibinfo {volume}
  {16}},\ \bibinfo {pages} {520} (\bibinfo {year} {2020})}\BibitemShut
  {NoStop}%
\bibitem [{\citenamefont {Liu}\ \emph {et~al.}(2020)\citenamefont {Liu},
  \citenamefont {Hao}, \citenamefont {Khalaf}, \citenamefont {Lee},
  \citenamefont {Ronen}, \citenamefont {Yoo}, \citenamefont {Najafabadi},
  \citenamefont {Watanabe}, \citenamefont {Taniguchi}, \citenamefont
  {Vishwanath},\ and\ \citenamefont {Kim}}]{Liu2020s}%
  \BibitemOpen
  \bibfield  {author} {\bibinfo {author} {\bibfnamefont {X.}~\bibnamefont
  {Liu}}, \bibinfo {author} {\bibfnamefont {Z.}~\bibnamefont {Hao}}, \bibinfo
  {author} {\bibfnamefont {E.}~\bibnamefont {Khalaf}}, \bibinfo {author}
  {\bibfnamefont {J.~Y.}\ \bibnamefont {Lee}}, \bibinfo {author} {\bibfnamefont
  {Y.}~\bibnamefont {Ronen}}, \bibinfo {author} {\bibfnamefont
  {H.}~\bibnamefont {Yoo}}, \bibinfo {author} {\bibfnamefont {D.~H.}\
  \bibnamefont {Najafabadi}}, \bibinfo {author} {\bibfnamefont
  {K.}~\bibnamefont {Watanabe}}, \bibinfo {author} {\bibfnamefont
  {T.}~\bibnamefont {Taniguchi}}, \bibinfo {author} {\bibfnamefont
  {A.}~\bibnamefont {Vishwanath}},\ and\ \bibinfo {author} {\bibfnamefont
  {P.}~\bibnamefont {Kim}},\ }\bibfield  {title} {\bibinfo {title} {Tunable
  spin-polarized correlated states in twisted double bilayer graphene},\ }\href
  {https://doi.org/10.1038/s41586-020-2458-7} {\bibfield  {journal} {\bibinfo
  {journal} {Nature}\ }\textbf {\bibinfo {volume} {583}},\ \bibinfo {pages}
  {221} (\bibinfo {year} {2020})}\BibitemShut {NoStop}%
\bibitem [{\citenamefont {Burg}\ \emph {et~al.}(2020)\citenamefont {Burg},
  \citenamefont {Lian}, \citenamefont {Taniguchi}, \citenamefont {Watanabe},
  \citenamefont {Bernevig},\ and\ \citenamefont {Tutuc}}]{Burg2020s}%
  \BibitemOpen
  \bibfield  {author} {\bibinfo {author} {\bibfnamefont {G.~W.}\ \bibnamefont
  {Burg}}, \bibinfo {author} {\bibfnamefont {B.}~\bibnamefont {Lian}}, \bibinfo
  {author} {\bibfnamefont {T.}~\bibnamefont {Taniguchi}}, \bibinfo {author}
  {\bibfnamefont {K.}~\bibnamefont {Watanabe}}, \bibinfo {author}
  {\bibfnamefont {B.~A.}\ \bibnamefont {Bernevig}},\ and\ \bibinfo {author}
  {\bibfnamefont {E.}~\bibnamefont {Tutuc}},\ }\href@noop {} {\bibinfo {title}
  {Evidence of emergent symmetry and valley chern number in twisted
  double-bilayer graphene}} (\bibinfo {year} {2020}),\ \bibinfo {note}
  {2006.14000. arXiv. \url{https://arxiv.org/abs/2006.14000} (date accessed:
  09/10/2021)},\ \Eprint {https://arxiv.org/abs/2006.14000} {arXiv:2006.14000
  [cond-mat.mes-hall]} \BibitemShut {NoStop}%
\bibitem [{\citenamefont {Cao}\ \emph {et~al.}(2020)\citenamefont {Cao},
  \citenamefont {Rodan-Legrain}, \citenamefont {Rubies-Bigorda}, \citenamefont
  {Park}, \citenamefont {Watanabe}, \citenamefont {Taniguchi},\ and\
  \citenamefont {Jarillo-Herrero}}]{Cao2020s}%
  \BibitemOpen
  \bibfield  {author} {\bibinfo {author} {\bibfnamefont {Y.}~\bibnamefont
  {Cao}}, \bibinfo {author} {\bibfnamefont {D.}~\bibnamefont {Rodan-Legrain}},
  \bibinfo {author} {\bibfnamefont {O.}~\bibnamefont {Rubies-Bigorda}},
  \bibinfo {author} {\bibfnamefont {J.~M.}\ \bibnamefont {Park}}, \bibinfo
  {author} {\bibfnamefont {K.}~\bibnamefont {Watanabe}}, \bibinfo {author}
  {\bibfnamefont {T.}~\bibnamefont {Taniguchi}},\ and\ \bibinfo {author}
  {\bibfnamefont {P.}~\bibnamefont {Jarillo-Herrero}},\ }\bibfield  {title}
  {\bibinfo {title} {Tunable correlated states and spin-polarized phases in
  twisted bilayer{\textendash}bilayer graphene},\ }\href
  {https://doi.org/10.1038/s41586-020-2260-6} {\bibfield  {journal} {\bibinfo
  {journal} {Nature}\ }\textbf {\bibinfo {volume} {583}},\ \bibinfo {pages}
  {215} (\bibinfo {year} {2020})}\BibitemShut {NoStop}%
\bibitem [{\citenamefont {He}\ \emph {et~al.}(2020)\citenamefont {He},
  \citenamefont {Li}, \citenamefont {Cai}, \citenamefont {Liu}, \citenamefont
  {Watanabe}, \citenamefont {Taniguchi}, \citenamefont {Xu},\ and\
  \citenamefont {Yankowitz}}]{He2020s}%
  \BibitemOpen
  \bibfield  {author} {\bibinfo {author} {\bibfnamefont {M.}~\bibnamefont
  {He}}, \bibinfo {author} {\bibfnamefont {Y.}~\bibnamefont {Li}}, \bibinfo
  {author} {\bibfnamefont {J.}~\bibnamefont {Cai}}, \bibinfo {author}
  {\bibfnamefont {Y.}~\bibnamefont {Liu}}, \bibinfo {author} {\bibfnamefont
  {K.}~\bibnamefont {Watanabe}}, \bibinfo {author} {\bibfnamefont
  {T.}~\bibnamefont {Taniguchi}}, \bibinfo {author} {\bibfnamefont
  {X.}~\bibnamefont {Xu}},\ and\ \bibinfo {author} {\bibfnamefont
  {M.}~\bibnamefont {Yankowitz}},\ }\bibfield  {title} {\bibinfo {title}
  {Symmetry breaking in twisted double bilayer graphene},\ }\href
  {https://doi.org/10.1038/s41567-020-1030-6} {\bibfield  {journal} {\bibinfo
  {journal} {Nature Physics}\ }\textbf {\bibinfo {volume} {17}},\ \bibinfo
  {pages} {26} (\bibinfo {year} {2020})}\BibitemShut {NoStop}%
\end{thebibliography}%

\end{document}


\renewcommand{\theequation}{S\arabic{equation}}
\renewcommand{\thefigure}{S\arabic{figure}}
\renewcommand{\bibnumfmt}[1]{[S#1]}
\renewcommand{\citenumfont}[1]{S#1}
\renewcommand{\thetable}{S\arabic{table}}
\setcounter{section}{0}
\renewcommand{\thesection}{S\arabic{section}}

\title{Supporting Information: Tailoring the flat bands in twisted double bilayer graphene}
\author{Bálint Szentpéteri}
\affiliation{Department of Physics, Budapest University of Technology and Economics and Nanoelectronics Momentum Research Group of the Hungarian Academy of Sciences, Budafoki ut 8, 1111 Budapest, Hungary}
\author{Peter Rickhaus}
\affiliation{Solid State Physics Laboratory, ETH Zürich, CH-8093 Zürich, Switzerland}
\author{Folkert K. de Vries}
\affiliation{Solid State Physics Laboratory, ETH Zürich, CH-8093 Zürich, Switzerland}
\author{Albin Márffy}
\affiliation{Department of Physics, Budapest University of Technology and Economics and Nanoelectronics Momentum Research Group of the Hungarian Academy of Sciences, Budafoki ut 8, 1111 Budapest, Hungary}
\author{Bálint Fülöp}
\affiliation{Department of Physics, Budapest University of Technology and Economics and Nanoelectronics Momentum Research Group of the Hungarian Academy of Sciences, Budafoki ut 8, 1111 Budapest, Hungary}
\author{Endre Tóvári}
\affiliation{Department of Physics, Budapest University of Technology and Economics and Nanoelectronics Momentum Research Group of the Hungarian Academy of Sciences, Budafoki ut 8, 1111 Budapest, Hungary}
\author{Kenji Watanabe}
\affiliation{Research Center for Functional Materials, National Institute for Materials Science, 1-1 Namiki, Tsukuba 305-0044, Japan}
\author{Takashi Taniguchi}
\affiliation{International Center for Materials Nanoarchitectonics, National Institute for Materials Science, 1-1 Namiki, Tsukuba 305-0044, Japan}
\author{Andor Kormányos}
\affiliation{Department of Physics of Complex Systems, Eötvös Loránd University, Budapest, Hungary}
\author{Szabolcs Csonka}
\email{csonka.szabolcs@ttk.bme.hu}
\affiliation{Department of Physics, Budapest University of Technology and Economics and Nanoelectronics Momentum Research Group of the Hungarian Academy of Sciences, Budafoki ut 8, 1111 Budapest, Hungary}
\author{Péter Makk}
\email{makk.peter@ttk.bme.hu}
\affiliation{Department of Physics, Budapest University of Technology and Economics and Nanoelectronics Momentum Research Group of the Hungarian Academy of Sciences, Budafoki ut 8, 1111 Budapest, Hungary}

\maketitle

\section{Device fabrication}
\begin{figure*}[!ht]
\begin{center}
\includegraphics[width=.7\linewidth]{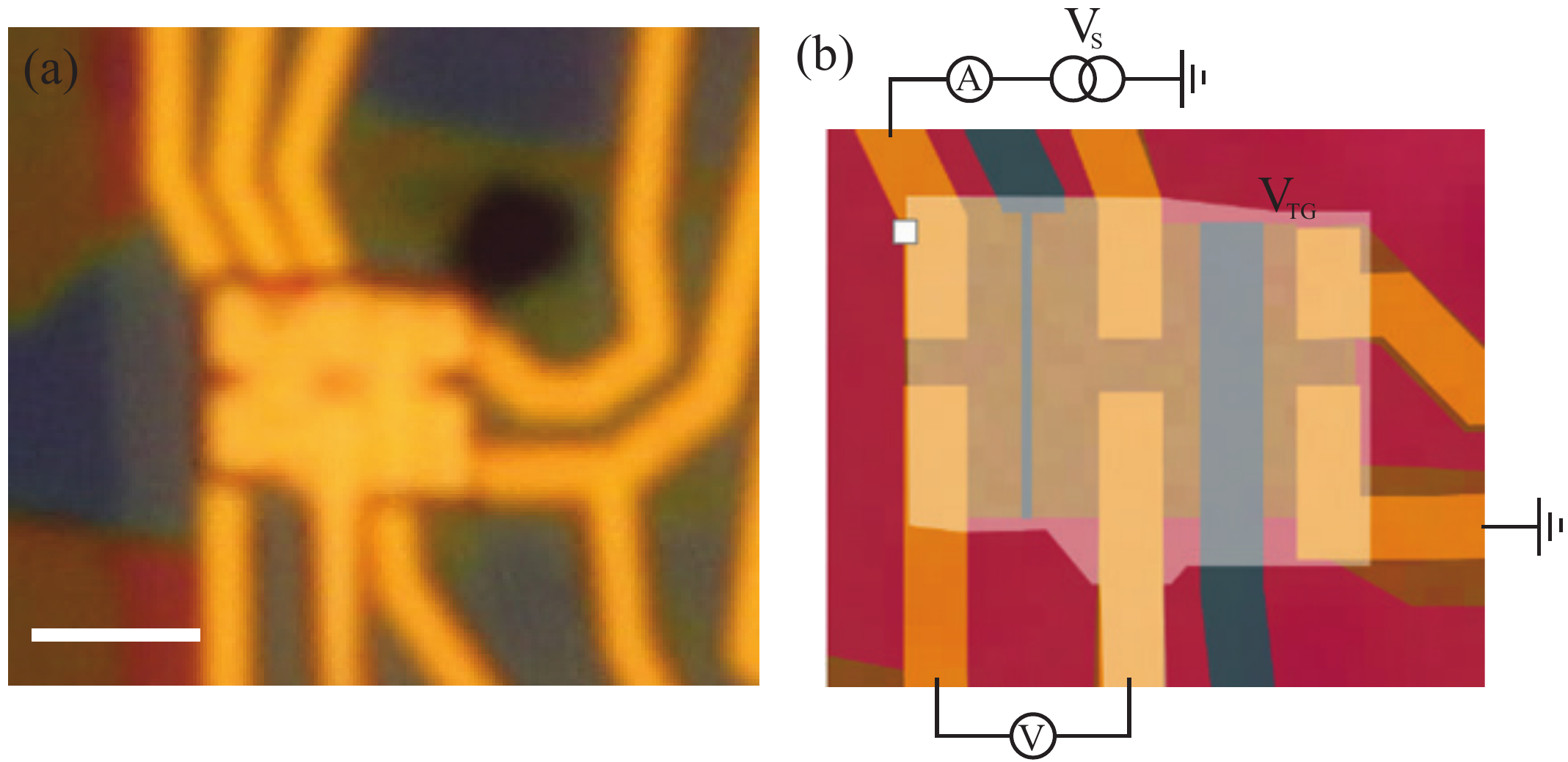}
\caption{(a) Optical microscope image of the measured device. The scale bar is \SI{2}{\micro\metre} (b) Schematic device geometry showing the four-terminal measurement setup.}
\label{supfig1}
\end{center}
\end{figure*}
The van der Waals heterostructure is fabricated using the dry-transfer technique\cite{Kim2016as}. Half of a single BLG is picked up with the top hBN layer ($\sim$\,32\,nm), whereafter the remaining BLG is rotated by 1° and subsequently picked up. Then the bottom hBN ($\sim$\,54\,nm) and graphite layer, that serves as a global bottom gate, are added. The device is fabricated in a cleanroom facility, using standard electron beam lithography techniques to define the different layers. Side contacts to the twisted double bilayer graphene (TDBG) are created by reactive ion etching and evaporation of 10\,nm of chromium (Cr) and 50\,nm of gold (Au). After this the first top gate layer (blue in Fig.~\ref{supfig1}b) is added, consisting of 10/70 nm of Cr/Au, and the device boundaries are defined by reactive ion etching (red in Fig.~\ref{supfig1}b). Finally, atomic layer deposition is used to create a 30\,nm aluminium oxide dielectric layer, which isolates the second top gate layer of 10/110\,nm Cr/Au (grey transparent in Fig.~\ref{supfig1}b) from the device and first top gate layer. The two top gate layers are used as one in the measurements by adjusting their potential such that the density in the TDBG is constant throughout the device.

\section{Transport measurements}
\begin{figure}[!ht]
\begin{center}
\includegraphics[width=.7\linewidth]{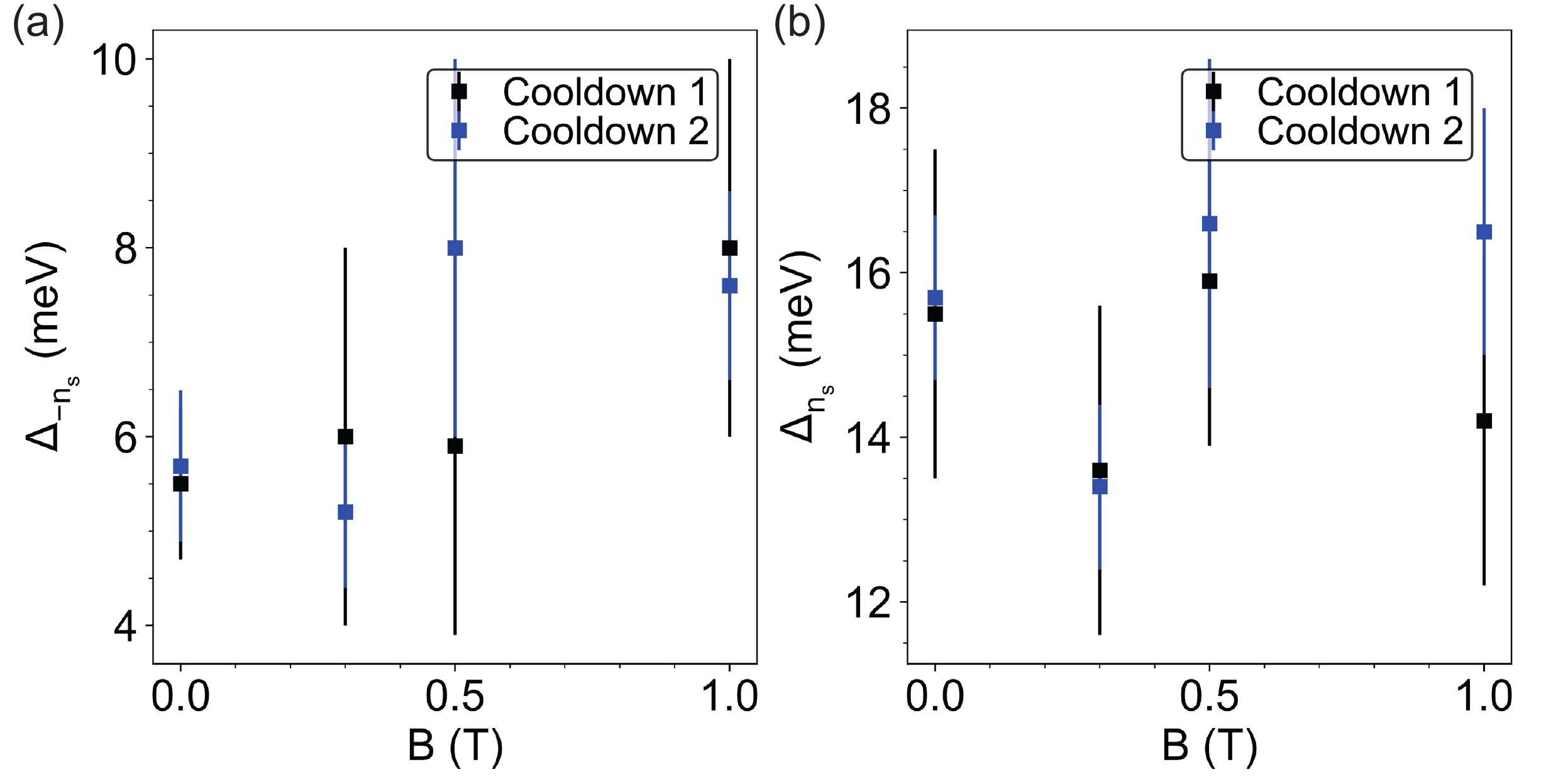}
\caption{Comparison between the measured gap energies at two different cooldowns at ambient pressure. Cooldown 1 is measured before applying the pressure and cooldown 2 is measured after releasing the pressure.}
\label{supfig10}
\end{center}
\end{figure}

Transport measurements were carried out in a four-terminal and two-terminal geometry with typical AC voltage excitation of 0.1\,mV using a standard lock-in technique at 177.13\,Hz. The device was cooled down six times: twice at zero pressure, three times at $p=2$\,GPa and once at $p=1$\,GPa. Twice at $p=2$\,GPa we measured in a two-terminal geometry while for the rest of the cooldowns we measured in a four-terminal geometry as depicted in Fig.~\ref{supfig1}b. In Fig.~3d of the main text the two-terminal measurements are shown with diamond symbols and the four-terminal measurements are shown with rectangular symbols. The measurements in different cooldowns with the same pressure resulted in similar results as it is shown in Fig.~\ref{supfig10} for $p=0$ before applying the pressure and after releasing it.

\subsection{Gate voltage n-D conversion}
\begin{figure}[!ht]
\begin{center}
\includegraphics[width=.9\linewidth]{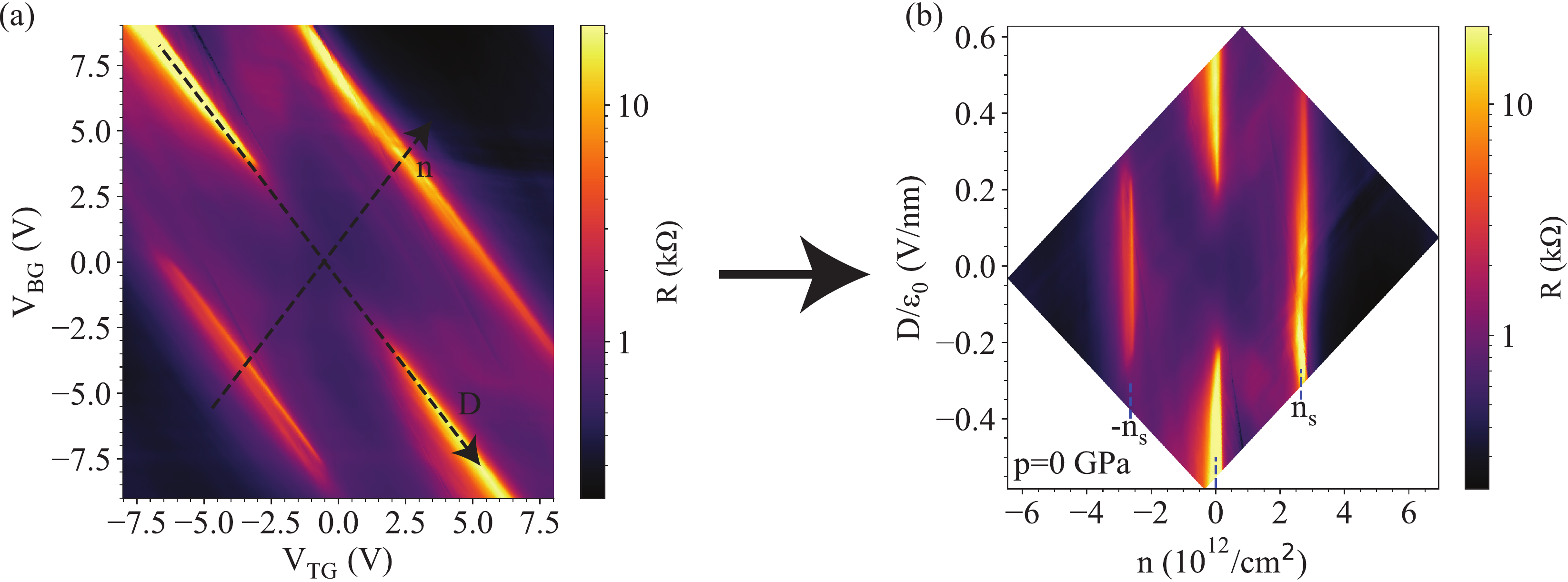}
\caption{Bottom and top gate voltage conversion to $n$ and $D$.}
\label{supfig2}
\end{center}
\end{figure}
The charge density ($n$) and electric displacement field ($D$) is related to the top and bottom gate voltage by
\begin{eqnarray}
n&=&\alpha_{\text{TG}}V_{\text{TG}}+\alpha_{\text{BG}}V_{\text{BG}}+n_0\\
\frac{D}{\epsilon_0}&=&\frac{e}{2\epsilon_0}\left(\alpha_{\text{TG}}V_{\text{TG}}-\alpha_{\text{BG}}V_{\text{BG}}\right)+\frac{D_0}{\epsilon_0},\nonumber
\end{eqnarray}
 where $\alpha_{\text{TG}}$ and $\alpha_{\text{BG}}$ are the lever arm for the top and bottom gate respectively, $\epsilon_0$ is the vacuum permittivity, $V_{\text{TG}}$ and $V_{\text{BG}}$ are the top and bottom gate voltage, respectively, ${n_0=-\alpha_{\text{TG}}V_{\text{TG}0}-\alpha_{\text{BG}}V_{\text{BG}0}}$ is the carrier density when the gate voltages are set to zero, whereas ${D_0=-\frac{e}{2}\left(\alpha_{\text{TG}}V_{\text{TG}0}-\alpha_{\text{BG}}V_{\text{BG}0}\right)}$ is a built-in offset electric field. $V_{\text{TG}0}$ and $V_{\text{BG}0}$ are the values of the top and bottom gate at the zero density and zero displacement field point.

\begin{figure}[!hb]
\begin{center}
\includegraphics[width=.67\linewidth]{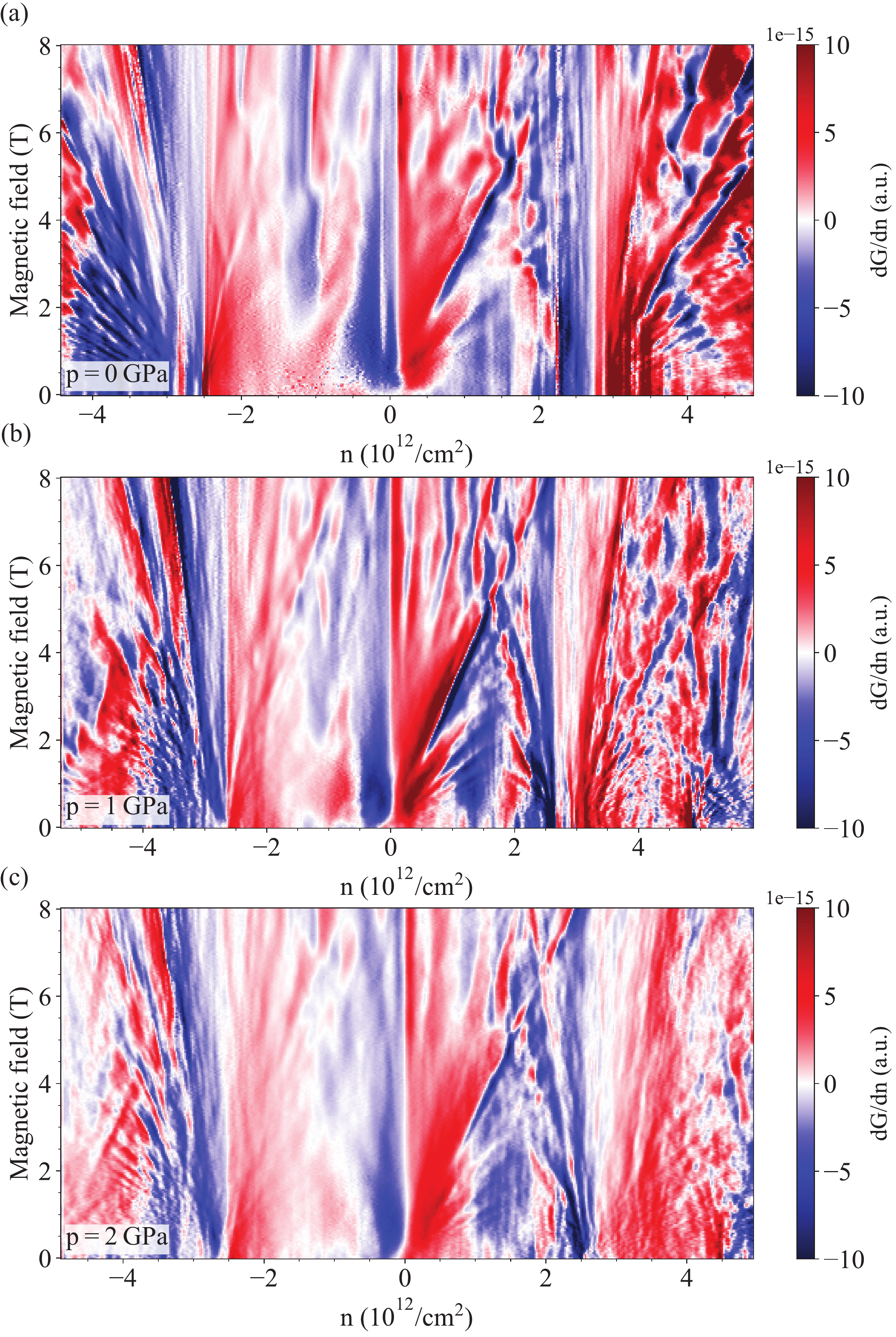}
\caption{Magnetic oscillation measurements of the four-terminal conductance $G$. (a), (b), (c) show d$G$/d$n$ versus $n$ and $B$ at $p=0$, 1\,GPa and 2\,GPa respectively.}
\label{supfig3}
\end{center}
\end{figure}

\begin{table}[!hbt]
\centering
\begin{tabular}{|cccc|}
\hline 
P (GPa)&0&1&2\\
\hline 
$\alpha_{TG}$ $(\frac{10^{15}}{\mathrm{Vm}^2})$&3.38(5)  &3.75(5)  &3.84(5)\\ 
$\alpha_{BG}$ $(\frac{10^{15}}{\mathrm{Vm}^2})$&4.54(6)  &4.94(7)  &4.91(8)\\ 
$n_0$ ($10^{15}/$m) &2.3(2)  &2.5(2)  &2.4(2)\\
$D_0$ (V/nm) &0.016(5)  &0.018(5)  &0.017(5)\\  
\hline 
\end{tabular}
\caption{{The extracted lever arms at different pressures.}}
\label{1tabl}
\end{table}
The lever arms were obtained from the gate-gate maps (Fig.~\ref{supfig2}a) and from the quantum oscillations in magnetoconductance measurements (Fig.~\ref{supfig3}). The extracted lever arms are shown in Table \ref{1tabl}. The pressure dependence of the lever arms is originating from the compression of the dielectrics and pressure-dependent dielectric constant of the hBN as already reported in Refs.\,\citenum{Solozhenko1995s, Yankowitz2018s}. The extracted lever arms within the margin of error were the same at the same pressures at different cooldowns.

\clearpage
\section{Pressure dependence of the twist angle}
\begin{figure*}[!ht]
\begin{center}
\includegraphics[width=1\linewidth]{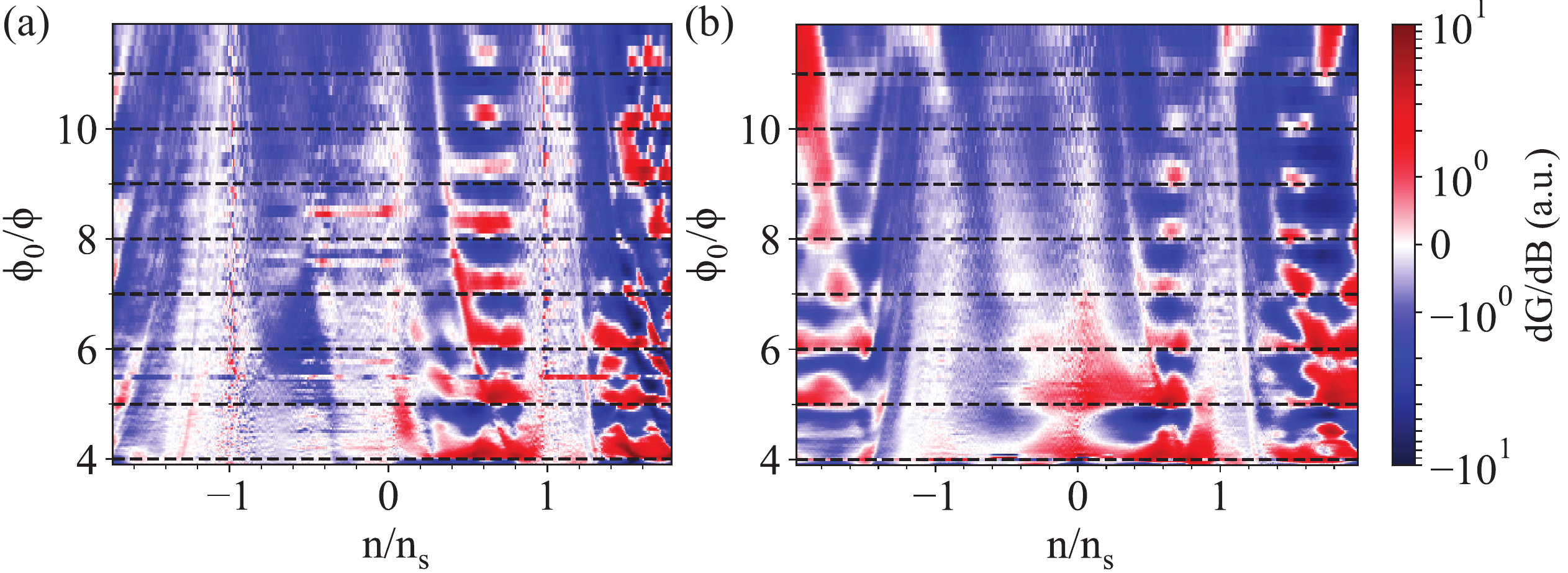}
\caption{Brown-Zak oscillations. (a) 2D color map of d$G$/d$B$ versus $n/n_\mathrm{s}$ and $\Phi_0/\Phi$ at $D=0$ and at ambient pressure. (b) 2D color map of d$G$/d$B$ versus $n/n_\mathrm{s}$ and $\Phi_0/\Phi$ at $D=0$ under 2\,GPa pressure. On the figures, the oscillating pattern is well observable and their periodicity is the same. The dashed lines are a guide to the eye for the oscillations.}
\label{supfig4}
\end{center}
\end{figure*}

At certain magnetic fields, the magnetic length is commensurable with the lattice periodicity which results in the recovery of the translation symmetry thus the electron feels effectively zero magnetic field. This results an oscillation in the resistance called Brown-Zak (BZ) oscillation\cite{Brown1964s, Zak1964s}. A common method to determine the twist angle ($\vartheta$) in superlattice structures such as the TDBG via transport measurements is to calculate it from the BZ oscillations. It can be calculated using that the BZ oscillations have maxima in the conductance at $\phi=BA_\mathrm{s}=\frac{\phi_0}{q}$, where $\phi_0=h/e$ is the flux quantum, $q$ is an integer number and $A_\mathrm{s}$ is the area of the superlattice unit cell which is given by
\begin{equation}
\label{M_area}
A_\mathrm{s}=\frac{\sqrt{3}}{2}\left(\frac{a}{2\sin(\vartheta/2)}\right)^2,
\end{equation} 
where $a$ is the lattice constant of the graphene. We determined the twist angle ($\vartheta=1.067$°$\pm0.003$°) from our data at different pressures and found that they were identical within the uncertainty of our measurements. 

We also estimated the twist angle inhomogeneity from the width of the resistance peaks at $n=\pm n_\mathrm{s}$ in Fig.~\ref{supfig6} panel a d and f at different pressures which varied in a similar range:
at $p=0$\,GPa $\vartheta$ is between 1.06° and 1.09°, at $p=1$\,GPa $\vartheta$ is between 1.06° and 1.1° and at $p=2$\,GPa $\vartheta$ is between 1.05° and 1.08°. In regions containing substantial twist angle inhomogeneity, this inhomogeneity could result in smaller measured gaps values, but it doesn't change our results qualitatively.

\section{Bias measurements}
\begin{figure*}[!ht]
\begin{center}
\includegraphics[width=.9\linewidth]{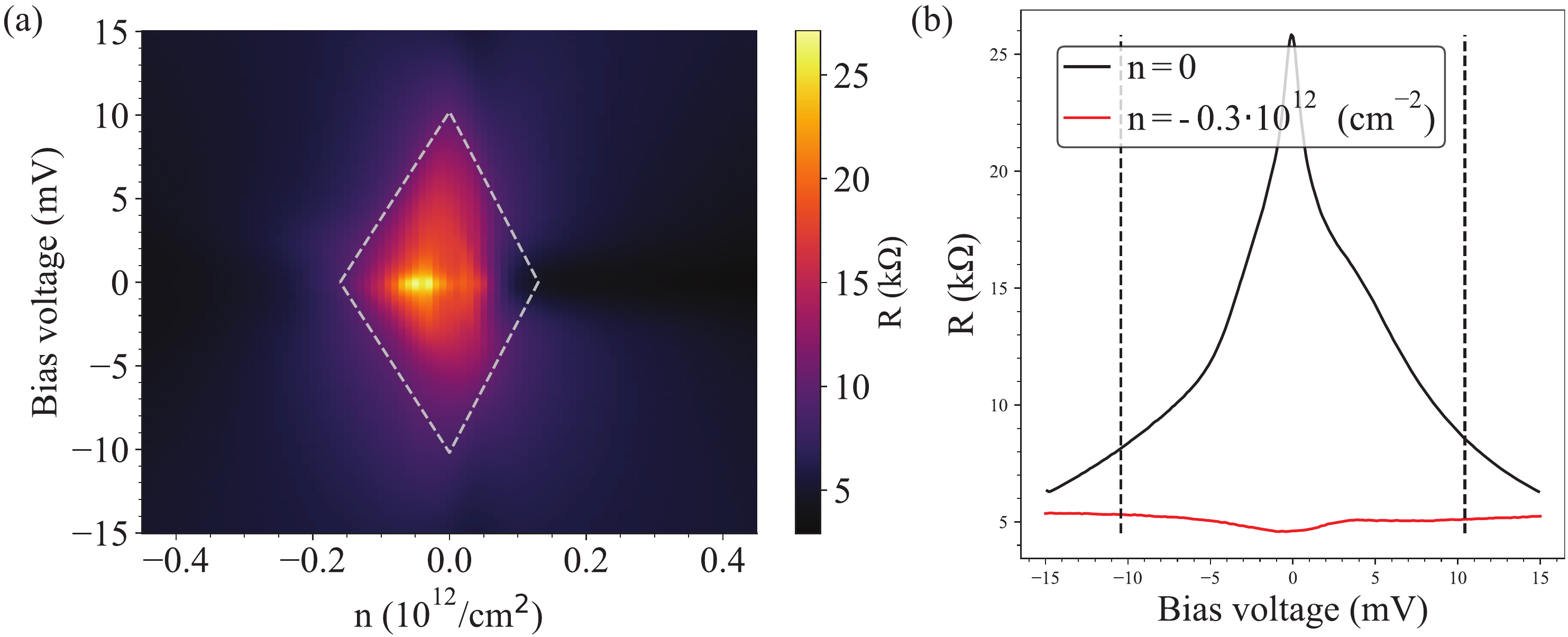}
\caption{Bias measurements. (a) The resistance versus $n$ and the bias voltage ($V_\text{B}$) at $D=-0.3$\,V/nm. (b) Line-cut from (a) at $n=0$ and $n=-3\cdot 10^{11}$ 1/cm$^2$. The gaps were obtained by taking the bias voltage where the resistance drops to 15\% of the highest value at $V_\text{B}=0$, after subtracting a background.}
\label{supfig_bias}
\end{center}
\end{figure*}

At the CNP we also estimated the gaps with bias measurements. We measured the two-terminal resistance of the device with the lock-in technique at low frequency and applied a DC voltage bias ($V_\text{B}$) to the sample. If we apply a higher or equal bias voltage to the gap, transport processes become available. We performed bias measurements over a small range of $n$ at fix $D$ (see Fig.~\ref{supfig_bias}a). First the maximum resistance ($R_\text{max}$) was determined (occurring in Fig.~\ref{supfig_bias}a at $n=0$ and $V_\text{B}=0$). We defined the gap value as the point where the resistance drops to 15\% of the maximum resistance value, also taking into account a background value. To determine the gap size, the following procedure was followed.  At the gap ($n=0$, $V_\text{B}=0$) we took the peak value ($R_\text{max}$), while at a slightly different density value, where the resistance $R$ is virtually independent of $V_\text{B}$, we took the background value $R_\text{gapless}$. We approximated the gap energy with $e|V_\text{B}|$ where $V_\text{B}=(V_\text{B+}+V_\text{B-})/2$ with $R(\pm V_\mathrm{B\pm})=R_\text{gapless}+0.15(R_\text{max}-R_\text{gapless})$ as it is shown in Fig.~\ref{supfig_bias}b. 

\section{Band structure calculation}
The band structure of TDBG is calculated using the non-interacting Bistritzer-Macdonald model\,\cite{Bistritzer2011s} and applied to the TDBG using the parameters from Ref.\,\citenum{Chebrolu2019s}. The matrix elements in the model can be written as
\begin{eqnarray}
\label{Hinterintra}
\braket{ls\mathbf{k}|H|l's'\mathbf{k}'}&=&\delta_{l,l'} H_{ls,l's'}(\mathbf{k})\delta_{\mathbf{k},\mathbf{k}'}+
(1-\delta_{l,l'}) H_{ls,l's'}(\mathbf{K}) \left[\delta_{\mathbf{k},\mathbf{k}'}\right.+\\
&&e^{-i(\xi\mathbf{b}_1\mathbf{\tau}_{s}-\xi\mathbf{b}_1' \mathbf{\tau}_{s'})} \delta_{\mathbf{k}-\xi\mathbf{b}_1,\mathbf{k}'-\xi\mathbf{b}_1'}+ 
\left.e^{-i(\xi\mathbf{b}_2\mathbf{\tau}_{s}-\xi\mathbf{b}_2'\mathbf{\tau}_{s'})} \delta_{\mathbf{k}-\xi\mathbf{b}_2,\mathbf{k}'-\xi\mathbf{b}_2'}\right]\nonumber,
\end{eqnarray} 
where $\delta_{l,l'}$ is the Kronecker delta fuction, $l=\{\mathrm{top},\mathrm{bottom}\}$ is the layer index of the BLGs, $s=\{A,B,A',B'\}$ is the site index within the BLG, $\mathbf{k}$ is a wave-vector in the $\mathbf{k}$-space, $H_{ls,l's'}(\mathbf{k})$ is the low-energy Hamiltonian of the BLG, $\mathbf{K}$ is the Dirac-point wave vector, $\xi=\pm1$ for $\mathbf{K}$ and $\mathbf{K'}$, $\mathbf{b}_i$ and $\mathbf{b}_i'$ are the primitive reciprocal lattice vectors for the top and bottom BLGs and $\mathbf{\tau}_{s}$ are the sublattice positions. 

In our calculations we used a configuration space with cutoff in momentum space with a radius up to $4|b_\text{s}|=32\sin(\vartheta/2)/(\sqrt{3}a)$ using Hamiltonian matrices with a size of $648\times648$. In the model for the low-energy Hamiltonian of the BLG ($\delta_{l,l'} H_{ls,l's'}(\mathbf{k})$) we included the remote hoppings and used the following parameters: $\gamma_0=3.1$ eV for the intralayer nearest-neighbor hopping, $\gamma_1=3\omega(p)$ eV for the interlayer coupling between orbitals on the dimer sites, $\gamma_3=0.283$ eV for the interlayer coupling between orbitals on the non-dimer sites, $\gamma_4= 0.138$ eV for the interlayer coupling between dimer and non-dimer orbitals and $\delta = 0.015$ eV onsite energy difference between dimer and non-dimer sites\cite{Jung2014s,Chebrolu2019s}. For the matrix elements between the BLGs ($[1-\delta_{l,l'}] H_{ls,l's'}(\mathbf{K})$) which are given by
\begin{equation}
H_{ls,l's'}(\mathbf{K})=\left(\begin{array}{cc}
\omega'(p)&\omega(p)\\
\omega(p)&\omega'(p)
\end{array}\right),
\end{equation}
 the pressure dependent interlayer couplings $\omega(p)$ and $\omega'(p)$ were taken from Table I. in \mbox{Ref.\,\citenum{Chebrolu2019s}} as
\begin{equation}
\omega(p)=A+\sqrt{B+C\cdot p},
\end{equation}
where $A=0.0546$ $B=0.0044$ $C=0.0031$ for $\omega$ and $A=0.0561$ $B=0.0018$ $C=0.0018$ for $\omega'$.

The external electric field modifies the layer potentials ($u_i$), where $i=\{1,2,3,4\}$ marks the 4 layers of graphene from top to bottom. The effect of the external electric field is modelled with $u_1=-u_4$, $u_2=-u_3$, and we introduce $u$ as the potential difference within each BLG and $2u'$ as the potential difference between the BLGs as $u_1=(u+u')/2$ and $u_2=(-u+u')/2$. In the calculations, we modeled the external electric field with equal interlayer potential drops ($u'=2u$) and neglected the quantum capacitance corrections and the electron-electron interactions, which could modify the low-energy flat bands.
\section{Magnetic field dependence}
\begin{figure*}[!ht]
\begin{center}
\includegraphics[width=1\linewidth]{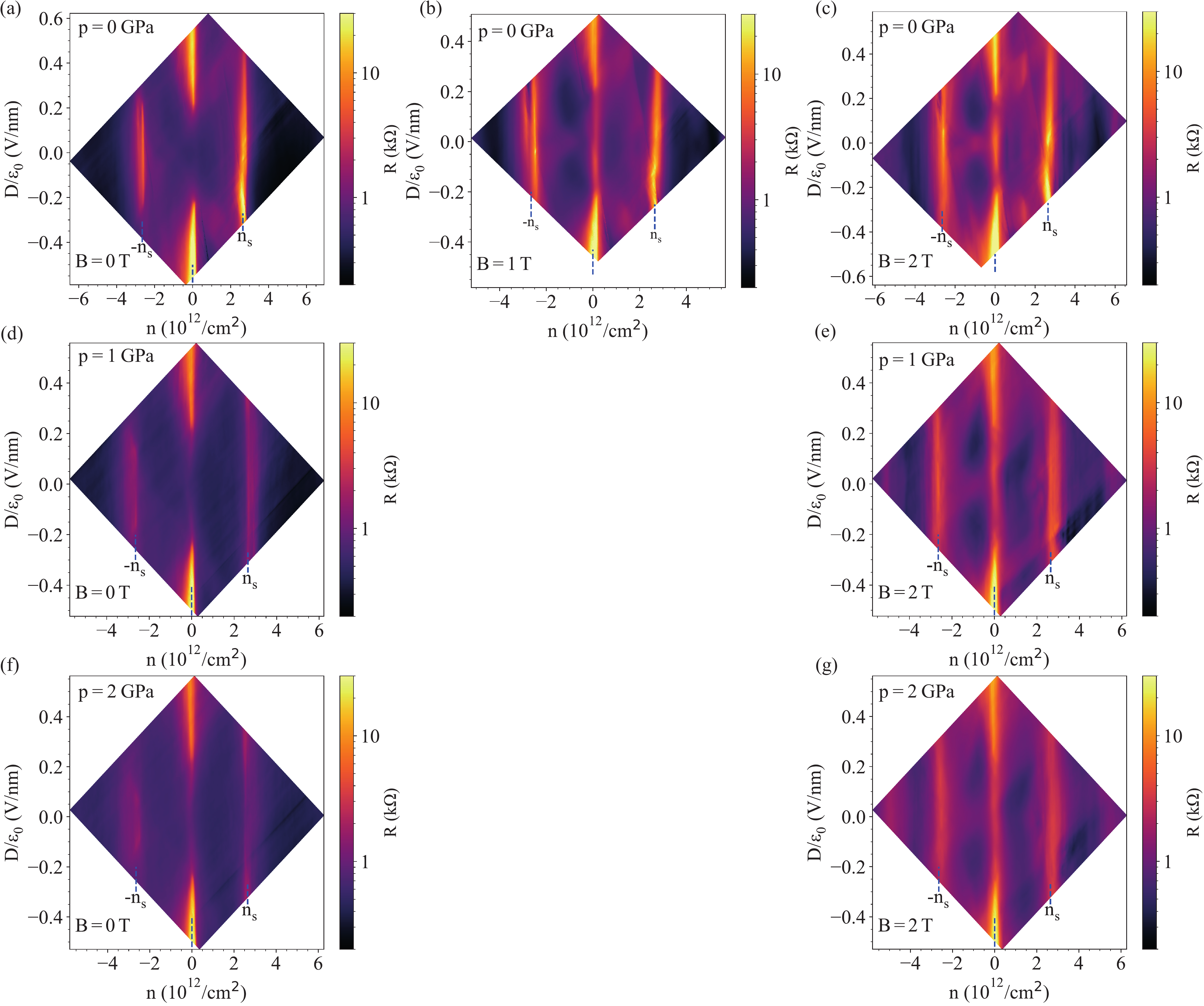}
\caption{Four-probe resistance of the TDBG as a function of  the charge density ($n$) and electric displacement field ($D$) measured in (a-c) at ambient pressure, in (d) and (e) at $p=1$\,GPa and in (f) and (g) at $p=2$\,GPa at different magnetic fields. }
\label{supfig6}
\end{center}
\end{figure*}
In Fig.~\ref{supfig6} we present n-D resistance maps at $B_\mathrm{z}=0$\,T, 1\,T and 2\,T at $1.5$\,K. At ambient pressure a gap opens at the charge density of $n=n_\mathrm{s}/2$ which corresponds to a correlated insulating phase\cite{Burg2019s,Shen2020s,Liu2020s}. In panel d,e and f,g the measurements are shown for 0 and 2\,T at 1\,GPa and 2\,GPa, respectively. Under pressure this gap disappears for all applied $B_\mathrm{z}$ fields. 

\begin{figure*}[!ht]
\begin{center}
\includegraphics[width=.8\linewidth]{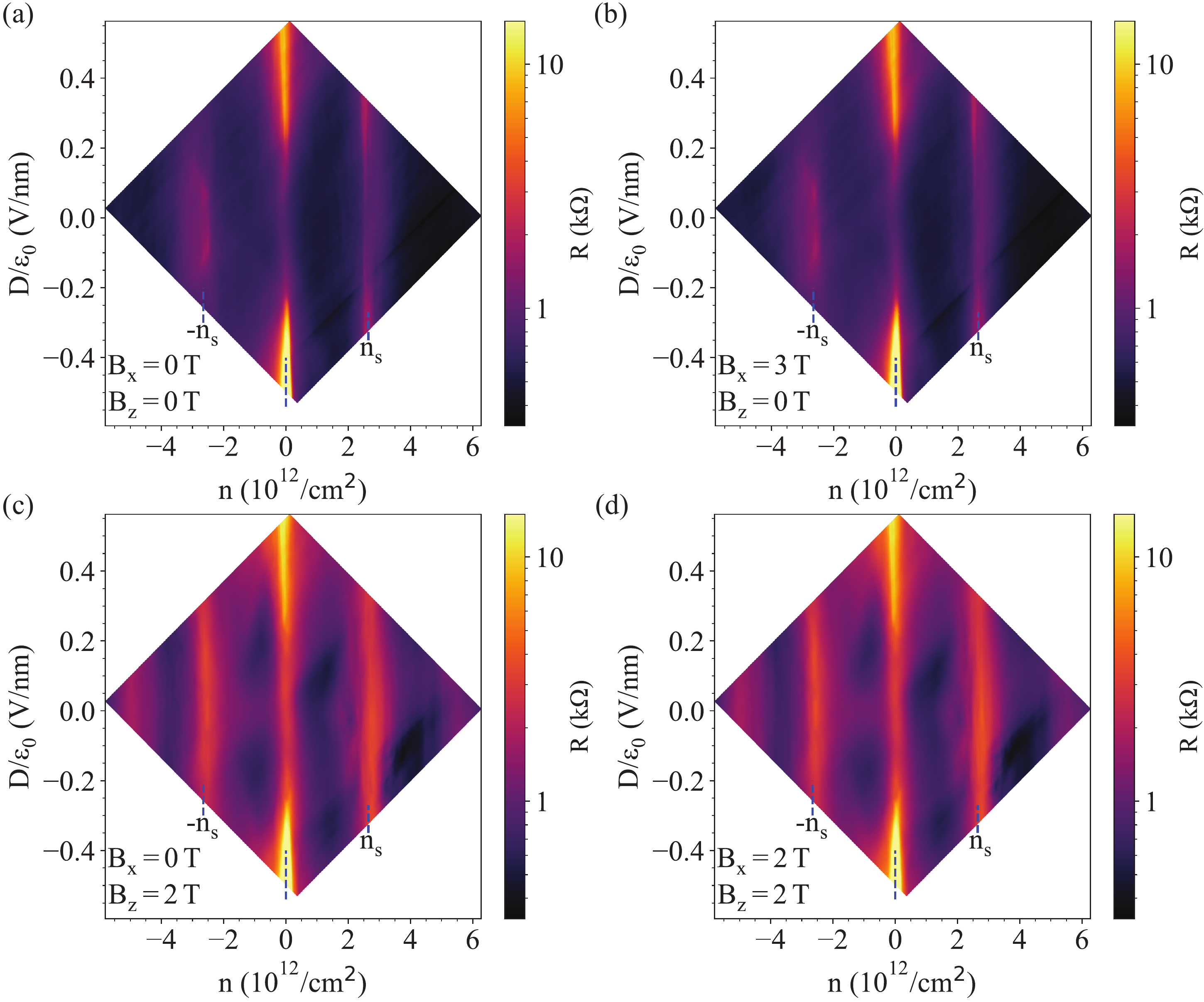}
\caption{Four-probe resistance of the TDBG as a function of $n$ and $D$ measured at 2\,GPa in different in- and out-plane magnetic fields. (a) presents the data for zero magnetic field, (b) for $B_\mathrm{x}=3$\,T in-plane magnetic field, (c) for $B_\mathrm{z}=2$\,T vertical magnetic field whereas (d) shows the measurement in $B_\mathrm{z}=2$\,T and $B_\mathrm{x}=2$\,T.}
\label{supfig7}
\end{center}
\end{figure*}

In Fig.~\ref{supfig7} we show resistance maps at various in-plane magnetic fields ($B_\mathrm{x}$) at $p=2$\,GPa. Comparing $B_\mathrm{x}=0$ (panel a) and $B_\mathrm{x}=3$\,T (panel b) it is visible that the effect of the in-plane magnetic field is negligible. At a finite perpendicular magnetic field, the effect of applying an in-plane field is also negligible. In thermal activation measurements, $B_\mathrm{x}$ also had a negligible effect on the extracted gaps.

\begin{figure*}[!ht]
\begin{center}
\includegraphics[width=.6\linewidth]{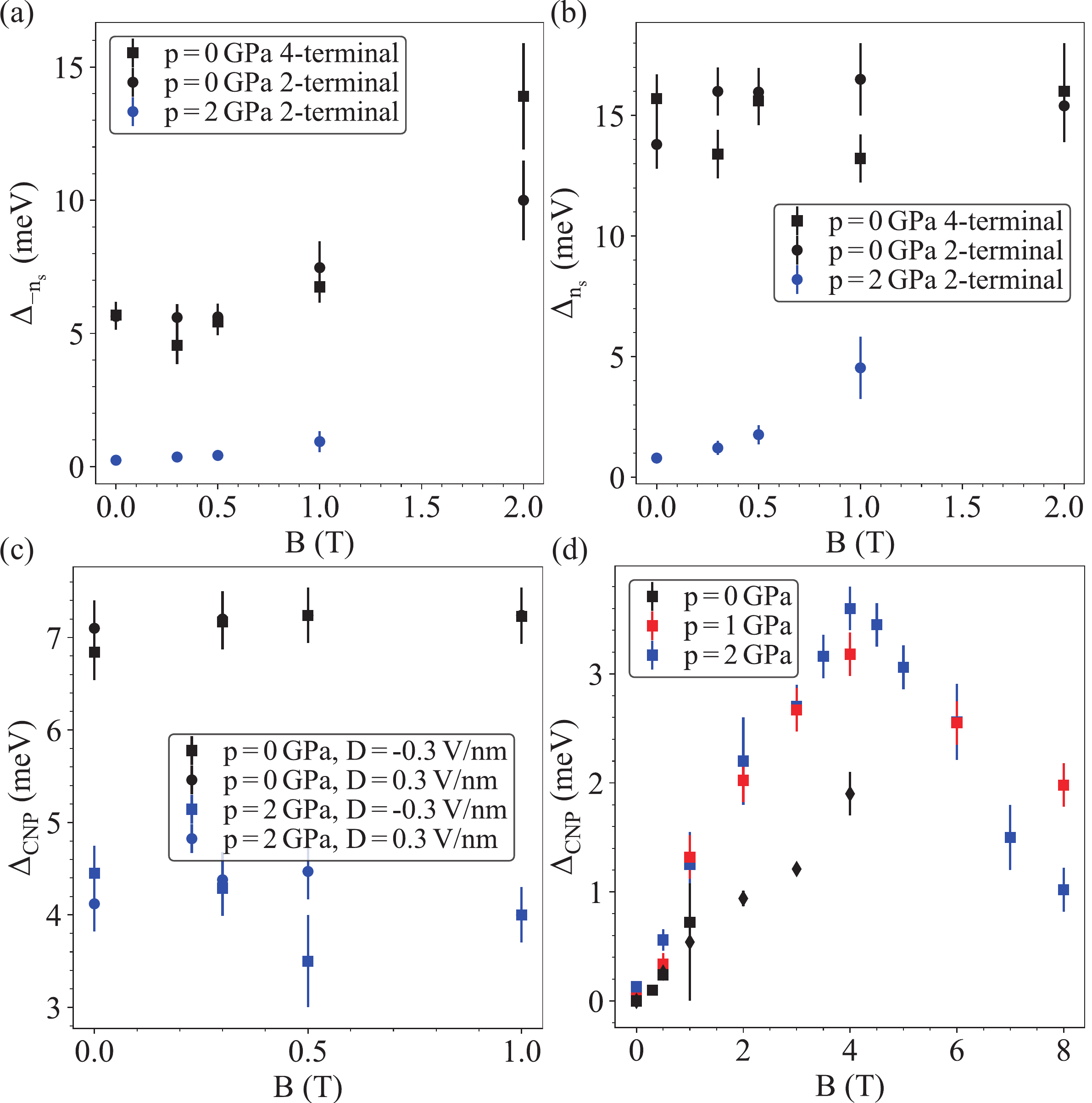}
\caption{Out-of-plane magnetic field dependence of the band gaps. (a) $\Delta_\mathrm{-n_{s}}$ versus $B$ at $D=0$ at $p=0$ and at $p=2$\,GPa. (b) $\Delta_\mathrm{n_{s}}$ versus $B$ at $D=0$ at $p=0$ and at $p=2$\,GPa. (c) $\Delta_\mathrm{CNP}$ at $D/\epsilon_0$=0.3\,V/nm and $-0.3$\,V/nm shown with squares and circles respectively. (d) $\Delta_\mathrm{CNP}$ at $D=0$.}
\label{supfig8}
\end{center}
\end{figure*}

In Fig.~\ref{supfig8} panel a, b the vertical magnetic field dependence of the measured moiré band gaps is shown. $\Delta_\mathrm{-n_{s}}$ increases with $B$ at $D=0$ for both $p=0$ and $p=2$\,GPa (shown in panel a). The same tendency is visible for $\Delta_\mathrm{n_{s}}$ at 2\,GPa and at $D=0$, but at $p=0$ $\Delta_\mathrm{n_{s}}$ seems independent of $B$ in a limited magnetic field range as depicted in panel b. $\Delta_\mathrm{CNP}$ at finite $D$ is independent of $B$ at small magnetic fields as it is shown in Fig.~\ref{supfig8}c. Fig.~\ref{supfig8}d shows the magnetic field dependence of the gaps at the CNP for different pressures for $D=0$. At $p=0$ only measurements up to 4\,T were available. These findings are similar to Ref.\,\citenum{Burg2020s} and are discussed in the main text. We note that our model does not include the effect of magnetic fields and more advanced model capable of handling that goes beyond the scope of this manuscript.

\clearpage

\section{Temperature dependence at half and three-quarter fillings}
\begin{figure*}[!ht]
\begin{center}
\includegraphics[width=.6\linewidth]{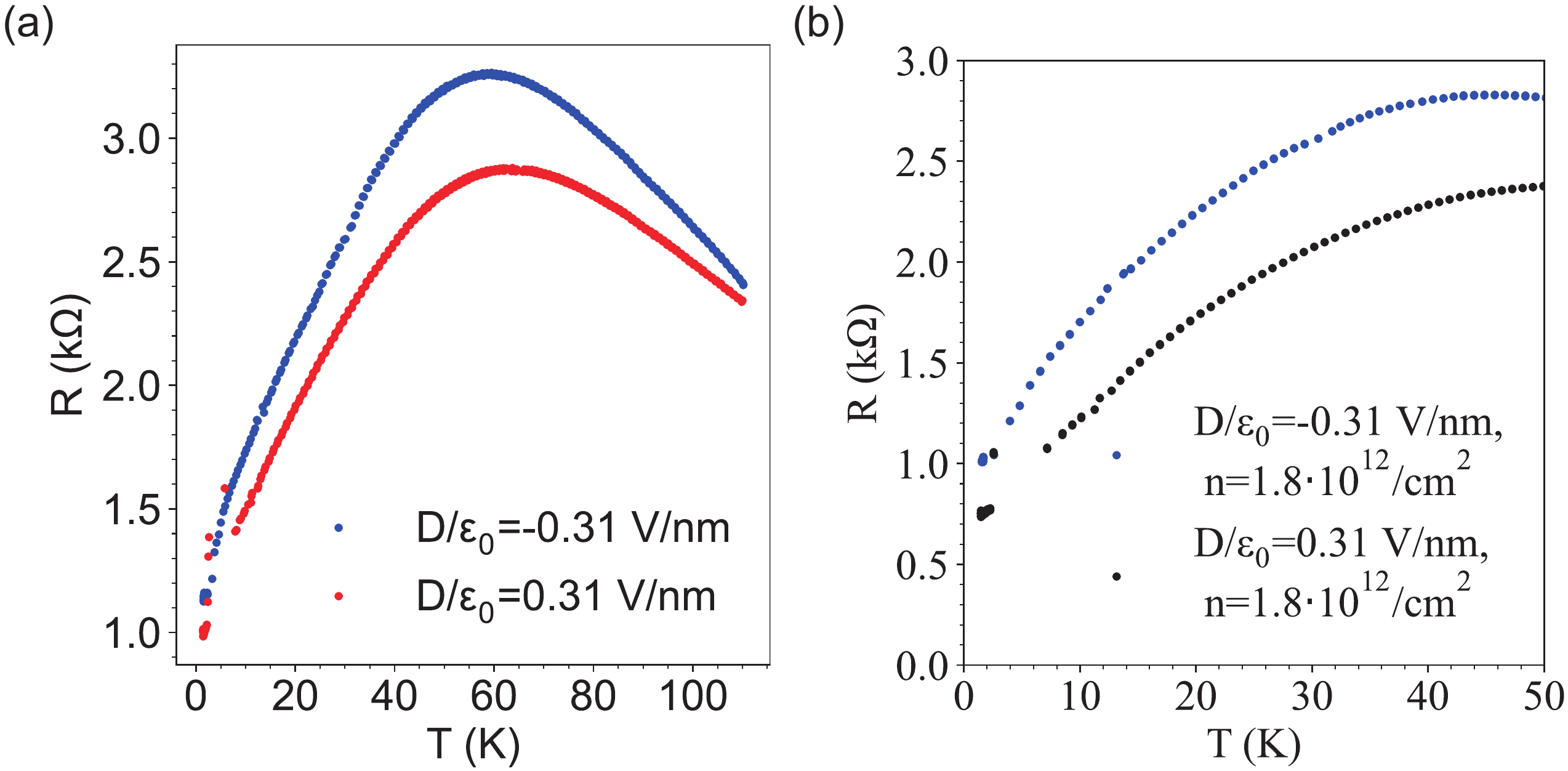}
\caption{Temperature dependence of the resistance at the correlated phases. (a) $R(T)$ at $n=\mathrm{n_{s}}/2$ at two different displacement fields. (b) $R(T)$ near three-quarter filling. }
\label{corr}
\end{center}
\end{figure*}

Fig.~\ref{corr} show the temperature dependence of the correlated states at half and near three-quarter fillings which are similar what was observed in Ref.\,\citenum{Cao2020s,Liu2020s,He2020s,Shen2020s}. At finite pressure the correlated states disappear as shown in Fig.~\ref{supfig6}.

\section{Comparison between the measurement and the model}
\begin{figure*}[!ht]
\begin{center}
\includegraphics[width=.6\linewidth]{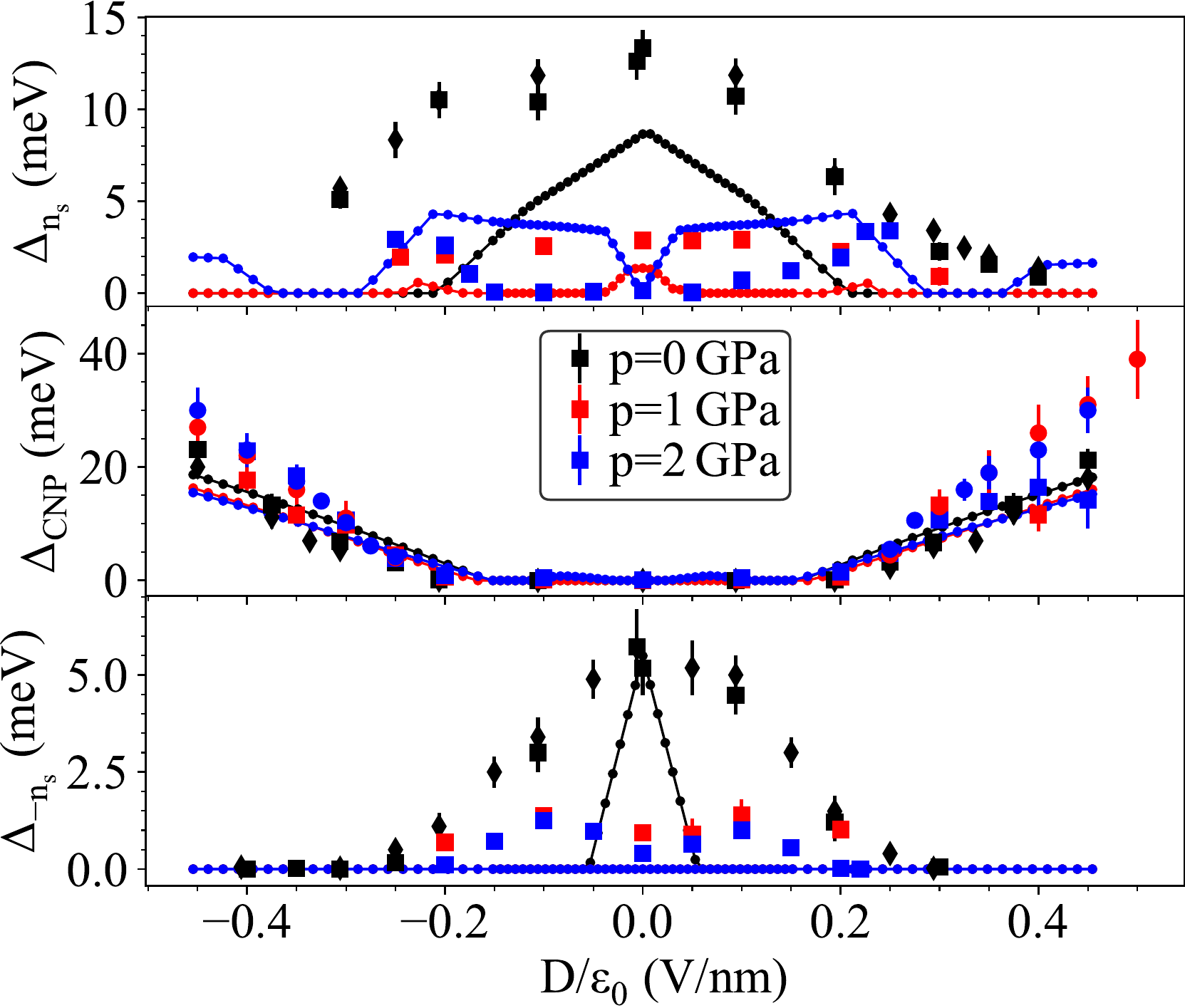}
\caption{Comparison between the measurement and the model. The displacement field was calculated from the interlayer potential with $\frac{D}{\epsilon_0}= eu\frac{\epsilon}{d}$ with $\epsilon=5$.}
\label{supfig9}
\end{center}
\end{figure*}

To compare experimental findings with our calculations we used the relation of $\frac{D}{\epsilon_0}= eu\frac{\epsilon}{d}$, where $d=0.33$\,nm is the interlayer distance of bilayer graphene and $\epsilon$ is the relative dielectric constant of bilayer graphene. We also used the same conversion between the top layer of the bottom BLG and the bottom layer of the twisted, top BLG. A qualitatively good agreement at the CNP is achieved using $\epsilon=5$ which is depicted in Fig.~\ref{supfig9} where the experiments and the theory are shown in the same figure. 

\bibliography{Szentpeteri_etal_2021_SupMat}